\documentclass[11pt]{article}
\usepackage{verbatim,amsmath,amssymb}
\usepackage{epsfig,float,color}
\usepackage[font=small,labelfont=bf]{caption}
\usepackage{geometry}
\usepackage{appendix}
\usepackage{setspace}
\usepackage{natbib}
\usepackage{amsthm,mathrsfs,amsfonts,dsfont} 
\usepackage{bm}
\usepackage{hyperref}
\usepackage[labelfont=bf]{caption}
\usepackage{tikz}
\usetikzlibrary{arrows,decorations.markings}\usepackage{tikz}
\usetikzlibrary{arrows,decorations.markings}
\usepackage{multirow}
\usepackage[labelfont=normalfont]{subcaption}
\usepackage{array}
\usepackage{adjustbox}
\usepackage{pifont}
\newcommand{\xmark}{\ding{55}}%

\definecolor{darkblue}{rgb}{0,0,1}
\definecolor{col1}{rgb}{1,0,1}
\definecolor{col2}{rgb}{0,0.5,0}
\definecolor{col3}{rgb}{0.5,0,1}
\definecolor{col4}{rgb}{0.1,.75,0}

\hypersetup{pdftex=true, colorlinks=true, breaklinks=true, linkcolor=darkblue, menucolor=darkblue, pagecolor=darkblue, citecolor=darkblue, urlcolor=darkblue}

\newtheoremstyle{rem}%name
{6pt}%Space above
{6pt}%Space below
{\small}%Body font
{}%Indent amount
{\bf}%Theorem head font
{:}%Punctuation after theorem head
{.5em}%Space after theorem head
{}%Theorem head spec(can be left empty, meaning ‘normal’)

\theoremstyle{rem}
\newtheorem{remark}{Remark}[section]

\newcommand{\bitm}{\begin{itemize}}
\newcommand{\eitm}{\end{itemize}}
\newcommand{\bnumr}{\begin{enumerate}}
\newcommand{\enumr}{\end{enumerate}}

\newcommand {\aab}{a^{\alpha\beta}}
\newcommand {\auab}{a_{\alpha\beta}}

\newcommand {\Aab}{A^{\alpha\beta}}
\newcommand {\Auab}{A_{\alpha\beta}}
\newcommand {\Agd}{A^{\gamma\delta}}

\newcommand {\Mab}{M^{\alpha\beta}}
\newcommand {\Nab}{N^{\alpha\beta}}

\newcommand {\bab}{b^{\alpha\beta}}

\newcommand {\buab}{b_{\alpha\beta}}

\newcommand {\bugd}{b_{\gamma\delta}}

\newcommand {\tauab}{\tau^{\alpha\beta}}

\newcommand {\eqb}[1]{\begin{equation}\begin{array}{#1}}
\newcommand {\eqe}{\end{array}\end{equation}}

\newcommand {\esb}[1]{\begin{equation*}\begin{array}{#1}}
\newcommand {\ese}{\end{array}\end{equation*}}

\newcommand {\pa}[2]{\frac{\partial{#1}}{\partial{#2}}}

\newcommand {\norm}[1]{\|#1\|}

\newcommand {\dif}{\mathrm{d}}

\newcommand {\II}{{I\kern-.3em I}}
\newcommand {\III}{{I\kern-.3em I\kern-.3em I}}

\newcommand {\intooe}{\int_{\Omega^e_0}}

\newcommand {\mra}{\mathrm{a}}
\newcommand {\mrb}{\mathrm{b}}
\newcommand {\mrc}{\mathrm{c}}

\newcommand {\mf}{\mathbf{f}}

\newcommand {\mk}{\mathbf{k}}

\newcommand {\mx}{\mathbf{x}}

\newcommand {\ba}{\boldsymbol{a}}
\newcommand {\bb}{\boldsymbol{b}}

\newcommand {\be}{\boldsymbol{e}}

\newcommand {\bg}{\boldsymbol{g}}

\newcommand {\bn}{\boldsymbol{n}}

\newcommand {\bt}{\boldsymbol{t}}
\newcommand {\bu}{\boldsymbol{u}}

\newcommand {\bx}{\boldsymbol{x}}

\newcommand {\mM}{\mathbf{M}}
\newcommand {\mN}{\mathbf{N}}

\newcommand {\mX}{\mathbf{X}}

\newcommand {\bA}{\boldsymbol{A}}

\newcommand {\bG}{\boldsymbol{G}}

\newcommand {\bK}{\boldsymbol{K}}
\newcommand {\bL}{\boldsymbol{L}}

\newcommand {\bN}{\boldsymbol{N}}

\newcommand {\bS}{\boldsymbol{S}}

\newcommand {\bX}{\boldsymbol{X}}

\newcommand {\eps}{\varepsilon}
\newcommand {\sig}{\sigma}

\newcommand {\btau}{\mbox{\boldmath$\tau$}}

\newcommand {\IR}{{\rm\kern.24em
   \vrule width.02em height1.53ex depth-.05ex
   \kern-.3em R}}
\newcommand {\ic}{{\rm\kern.20em
   \vrule width.02em height1.0ex depth-.05ex
   \kern-.22em c}}
\newcommand {\ia}{{\rm\kern.20em
   \vrule width.02em height1.05ex depth-.0ex
   \kern-.25em a}}
\newcommand {\IC}{{\rm\kern.24em
   \vrule width.02em height1.4ex depth-.05ex
   \kern-.26em C}}
\newcommand {\ID}{{\rm\kern.34em
   \vrule width.02em height1.5ex depth-.05ex
   \kern-.36em D}}
\newcommand {\IS}{{\rm\kern.24em
   \vrule width.02em height1.6ex depth.05ex
   \kern-.26em S}}
\newcommand {\IT}{{\rm\kern.50em
   \vrule width.02em height1.55ex depth-.05ex
   \kern-.52em T}}

\newcommand {\IE}{{\rm\kern.24em
   \vrule width.02em height1.55ex depth-.05ex
   \kern-.33em E}}
\newcommand {\IEa}{{\rm\kern.24em
   \vrule width.02em height1.55ex depth-.05ex
   \kern-.33em E}^{1}_{ijkl}}
\newcommand {\IEb}{{\rm\kern.24em
   \vrule width.02em height1.55ex depth-.05ex
   \kern-.33em E}^{2}_{ijkl}}

\newcommand {\sS}{\mathcal{S}}

\newcommand {\Ass}[2]{\kern 0.9ex \vrule width0.45em height0.2ex depth0ex \kern -2.1ex \bigwedge_{#1}^{#2}}
\newcommand {\ASS}[2]{\kern 1.45ex \vrule width0.5em height0.2ex depth0ex \kern -2.65ex \bigwedge_{#1}^{#2}}

\newcommand {\Cabgd}{{c}^{\alpha\beta\gamma\delta}}
\newcommand {\Dabgd}{{d}^{\alpha\beta\gamma\delta}}
\newcommand {\Eabgd}{{e}^{\alpha\beta\gamma\delta}}

\newcommand {\cabgd}{{c}^{\alpha\beta\gamma\delta}}
\newcommand {\dabgd}{{d}^{\alpha\beta\gamma\delta}}
\newcommand {\eabgd}{{e}^{\alpha\beta\gamma\delta}}
\newcommand {\fabgd}{{f}^{\alpha\beta\gamma\delta}}

\begin{document}

\begin{center}
\Large{\bf A new anisotropic bending model for nonlinear shells: Comparison with existing models and isogeometric finite element implementation}\\
\end{center}

\renewcommand{\thefootnote}{\fnsymbol{footnote}}

\begin{center}
\large{Eshwar J.~Savitha$^\mra$ and Roger A.~Sauer$^{\mra,\mrb,\mrc,}$\footnote[1]{corresponding author, email: roger.sauer@pg.edu.pl, sauer@aices.rwth-aachen.de}
}\\
\vspace{4mm}

\small{\textit{
$^\mra$Aachen Institute for Advanced Study in Computational Engineering Science (AICES), \\ 
RWTH Aachen University, Templergraben 55, 52056 Aachen, Germany \\[1.1mm]
$^\mrb$Faculty of Civil and Environmental Engineering, Gda\'{n}sk University of Technology, ul.~Narutowicza 11/12, 80-233 Gda\'{n}sk, Poland \\[1.1mm]
$^\mrc$Dept.~of Mechanical Engineering, Indian Institute of Technology Guwahati, Assam 781039, India}}

\end{center}

\vspace{-4mm}

\renewcommand{\thefootnote}{\arabic{footnote}}
\begin{center}
	\small{Published\footnote{This pdf is the personal version of an article whose journal version is available at \href{https://doi.org/10.1016/j.ijsolstr.2023.112169}{https:/\!/sciencedirect.com}}
		in \textit{International Journal of Solids and Structures}, \href{https://doi.org/10.1016/j.ijsolstr.2023.112169}{DOI: 10.1016/j.ijsolstr.2023.112169} \\
		Submitted on 11 October 2022; Revised on 30 Jan 2023; Accepted on 14 February 2023}
\end{center}
\rule{\linewidth}{.15mm}
{\bf Abstract:} A new nonlinear hyperelastic bending model for shells formulated directly in surface form is presented, and compared to four existing prominent bending models. Through an essential set of elementary nonlinear bending test cases, the membrane and bending stresses of each model are examined analytically. Only the proposed bending model passes all the test cases, while the other bending models either fail or only pass the test cases for small deformations. The proposed new bending model can handle large deformations and initially curved surfaces. It is based on the principal curvatures and their directions in the initial configuration, and it thus can have different bending moduli along those directions. These characteristics make it flexible in modeling a given material, while it does not suffer from the pathologies of existing bending models. Further, the bending models are compared computationally through four classical benchmark examples and one contact example. As the underlying shell theory is based on Kirchhoff-Love kinematics, isogeometric NURBS shape functions are used to discretize the shell surface. The linearization and efficient finite element implementation of the proposed new model are also provided.


\vspace{-5mm}
\rule{\linewidth}{.15mm}

\section{Introduction}\label{s:intro}

% Introduction
Shells are curved thin-walled structures appearing in nature and engineering designs. The curvature enables shells to be designed with high load-bearing capacity at minimal use of materials. This high strength-to-weight ratio makes them ubiquitous in many applications. Extensive efforts have been made to accurately describe the load-carrying behavior of shells. The finite element method (FEM) is the predominantly used numerical technique to solve shell problems. It is common practice in FEM formulations of slender structures such as plates, membranes, and shells, to reduce the dimension from volume to surface. This significantly simplifies the numerical discretization, reduces the degrees-of-freedom (dofs), and condenses the 3D kinematics to 2D. Savings in computational time naturally follow.

% Types of shell modeling
 The three mainly used ways of dimensionality reduction are derived, degenerate, and direct surface approaches \citep{bischoff2004models}. In the derived approach, an approximate shell theory is derived by asymptotic (analytical or numerical) integration of the 3D equations. Some examples of the early works, which mainly focused on linear theories, are those by \cite{gol1963derivation}, \cite{reissner1963derivation} and \cite{cicala1965systematic}. In the degenerate solid approach, the shell behavior is obtained by reducing or degenerating 3D continuum mechanics through kinematic assumptions. This was first developed by \cite{ahmad1970analysis} and a comprehensive presentation of the methodology can be found in \cite{hughes1987finite}. In the direct surface approach, only the shell surface is considered ab initio and well-defined constitutive laws are proposed to obtain membrane and bending stresses. The underlying theory of this method goes back to \cite{cosserat1909cosserat}. This was followed by prominent works of \cite{ericksen1957exact}, \cite{green1965general}, \cite{naghdi1973theory} and \cite{simo1990stress}. 

% Nonlinear shells
In solid mechanics, nonlinearities arise from the material behavior and the geometry of large deformations. Some early works on the theory of nonlinear analysis of shells were given by \cite{novozhilov1953foundations}, \cite{naghdinonlinear}, \cite{Simmondsnonlinear}, \cite{bathe1980geometric} and \cite{pietraszkiewicz1989geometrically}. Computational aspects of nonlinear shell modeling were presented in \cite{hughes1981nonlinear,dvorkin1984continuum,simo1990stress,betsch19964} among others. There are also several nonlinear shell formulations put forth for the modeling of soft materials such as rubbers \citep{Chroscielewski,Basar}, tissues \citep{itskov2001generalized,pandolfi2006model,prot2007transversely}, red blood cells \citep{DAO,mills2004nonlinear}, lipid bilayers \citep{steigmann1999fluid,FENG} and viscoelastic materials \citep{EVANS,neff2005geometrically}.  

% Shell theory 
In addition to the dimensionality reduction, the underlying shell kinematics also plays a vital role in constructing shell theories. The two widely used shell theories are those of Reissner-Mindlin (RM) \citep{reissner,mindlin} and Kirchhoff-Love (KL) \citep{kirchhoff,love1888xvi}. In KL theory the cross-section remains normal to the mid surface during deformation, whereas in RM shell theory, a shear angle can appear. Therefore, RM theory has both displacement and rotational degrees-of-freedom (dofs), and $ C^0 $-continuous shape functions suffice for discretization. On the other hand, KL theory has only displacement dofs. It therefore accommodates bending in the governing equations through the displacement field itself. This results in a fourth-order strong form equation for KL shells. The corresponding principle of virtual work contains second order derivatives which necessitates $ C^1 $-continuous shape functions for discretization. There are several methods developed to enforce $ C^1 $-continuity for Lagrange shape functions like rotation-free elements \citep{onate2000rotation,brunet2006analysis}, discontinuous Galerkin formulation \citep{NOELS} and mesh-free methods \citep{KRYSL}. However, these methods are usually complex and/or expensive. Therefore RM based shells are commonly used in commercial finite element (FE) codes as they can be used with simple classical Lagrange shape functions.

A new approach to obtain $ C^1 $-continuous shape function called Isogeometric analysis (IGA) was introduced by \cite{hughes2005isogeometric}. There the spline-based basis functions used to create the geometry, such as B-splines or Non-Uniform Rational B-Splines (NURBS), themselves are used for the FE analysis. The method can also be used with other spline-based formulations such as subdivision surfaces \citep{Cirak} or T-splines \cite{Tsplines}. The Bézier extraction operator developed by \cite{Borden} enables a NURBS surface to be decomposed into Bézier elements to seamlessly incorporate isogeometric analysis into existing FE code. The advantages of IGA over conventional approaches in shell problems were shown by \cite{kiendl2009isogeometric} for KL and \cite{benson2010isogeometric} for RM shells. A KL shell formulation does not just benefit from having fewer dofs compared to a RM shell, but it is also shear locking-free. However, both KL and RM shells suffer from membrane-bending locking. This type of locking arises due to the undesirable coupling of membrane and bending modes. One of the early works on alleviating locking in the context of isogeometric shells is by \cite{ECHTER2013170}. There intrinsically shear-locking-free RM formulations are used and membrane locking is alleviated in both RM and KL shells using either the Discrete Strain Gap method or a mixed method. \cite{BOUCLIER201386} used the $\bar{B}$-method to alleviate membrane locking in 3D isogeometric shells. Shell locking is still actively researched and some of the recent remedies use reduced quadrature \citep{ADAM,ZOU,ZOU2022114722}, mixed formulations \citep{Bieber,ZOU2020113283} and assumed natural strain methods \citep{caseiro2014assumed,casquero2022removing}.

The concept of using 3D constitutive models for shells \citep{DeBorst,Klinkel2002} is adapted to isogeometric KL shell formulation in \cite{kiendl2015isogeometric}. Along similar lines, \cite{duong2017new} proposed a formulation that admits constitutive laws obtained by thickness-integration of 3D material models as well as those constructed directly in surface energy form. IGA shell formulations have been applied to the study of rubbers \citep{Taylor,ELGUEDJ}, tissues \citep{TEPOLE,roohbakhshan2017efficient}, red blood cells \citep{CASQUERO,BARTEZZAGHI}, lipid bilayers \citep{sauer2017stabilized,auddya2021biomembranes} and viscoelastic materials \citep{DORTDIVANLIOGLU,paul2022isogeometric}. IGA shells have also been used in inverse problems such as shape optimization \citep{KIENDL2014148} and material reconstruction \citep{BORZESZKOWSKI}.

% Current focus and highlights
One of the material models for the direct surface approach was proposed by \cite{koiter1966nonlinear}. It is a linear relation between certain strains and stresses that can be derived from thickness integration of the 3D St.Venant-Kirchhoff material model. The Koiter model was later extended by \cite{STEIGMANN2012} to materials exhibiting arbitrary symmetries. A comprehensive derivation can also be found in \cite{steigmann2013koiter}. Two other direct surface material model were proposed by \cite{CANHAM}, to study red blood cells, and \cite{helfrich1973elastic}, to study the elastic properties of lipid bilayers. The Helfrich model, which includes the Canham model as a special case, is one of most widely used material models in morphological studies of vesicles, which are closed bilayer films.

KL shell formulations are primarily used with bending models derived from 3D or directly proposed for the surface such as the Koiter and Helfrich model. There aren't many examples of direct bending models besides those of Koiter and Helfrich. As is shown here, these suffer from an inability to handle nonlinear deformations or initially stress-free curved surfaces. We address these shortcomings by proposing a new nonlinear anisotropic bending model for the direct surface formulation of shells. It is motivated by eliminating the \textit{physical} stretch-dependency of bending that is affecting existing bending models at large deformations. This is different from treating membrane-bending locking, which is caused by a \textit{numerical} stretch-dependency of bending of very thin shells that is already a problem at small deformations. Membrane-bending locking is due to shortcomings in the underlying finite element discretization and therefore needs to be treated at that level, as was noted above. Our concern is purely physical and therefore needs to be treated at the constitutive level. Our description uses a direct surface formulation for KL shells in curvilinear coordinates together with an isogeometric surface discretization, which is very general and accurate, yet straightforward to implement \citep{duong2017new}.  

The salient features of the proposed new bending model are:
\vspace*{-4mm}
\begin{itemize}
	\item It is objective, admits large deformations and captures anisotropic bending.
	\item It allows to describe initially curved stress-free shells. 
	\item It offers great flexibility in modeling a given material. 
	\item It is compared to existing bending models and it agrees with those at small deformations.
	\item In contrast to existing models, it passes a proposed set of nonlinear bending test cases.
	\item It allows for an efficient implementation within isogeometric shell FE.
\end{itemize}  

The remainder of this paper is organized as follows: Sec.~\ref{sec:2} provides an overview of general thin shell theory in curvilinear coordinates. In Sec.~\ref{sec:3}, existing bending models are presented along with the proposed new bending model. In Sec.~\ref{sec:4}, these models are then compared and their shortcomings are illustrated using five analytical test cases. In Sec.~\ref{sec:5}, the new bending model is examined in several numerical examples. The paper concludes with Sec.~\ref{sec:6}.

\section{Thin-shell formulation}\label{sec:2}

This section summarizes the Kirchhoff-Love thin-shell theory formulation of \cite{sauer2017theoretical} and \cite{duong2017new} that is based on curvilinear coordinates and isogeometric finite elements. First, the essential kinematical relations, governing strong and weak form equations, and hyperelastic constitutive equations are introduced in Sec.~\ref{ssec:kin}-\ref{ssec:selas}. Then, the linearization and finite element  approximation of the resulting nonlinear equations are presented in Sec.~\ref{ssec:lin} and \ref{ssec:fe}.

%----------------------------------------------------------------------------------------%
\subsection{Surface kinematics}\label{ssec:kin}

The current and initial configuration of a shell surface embedded within 3D space can be parametrized respectively as 
\begin{align}
\bx\,=\,\bx(\xi^\alpha)\,,\quad\,\text{and}\quad \bX\,=\,\bX(\xi^\alpha)\,,
\end{align} 
with parameters $ \xi^1 $ and $ \xi^2 $. 
A basis in the curvilinear setting can then be defined by a pair of covariant tangent vectors ($ \ba_{\alpha} $, $ \alpha\,=\,1,\,2 $) and normal ($ \bn $) as
\begin{gather}
\ba_{\alpha}\,:=\,\frac{\partial\bx}{\partial \xi^\alpha}\,, \quad \bn\,:=\,\frac{\ba_1\,\times\, \ba_2}{\norm{\ba_1\,\times\,\ba_2}}\,.
\end{gather}
A similar set of basis vectors is defined for the initial configuration as 
\begin{gather}
\bA_{\alpha}\,:=\,\frac{\partial\bX}{\partial \xi^\alpha}\,, \quad \bN\,:=\,\frac{\bA_1\,\times\, \bA_2}{\norm{\bA_1\,\times\,\bA_2}}\,.
\label{eq:A_a}
\end{gather}
Since the covariant tangent vectors are not orthonormal, i.e.~the covariant surface metric $ \auab\,=\,\ba_{\alpha}\,\cdot\,\ba_{\beta} $ does not correspond to the identity matrix,	 their dual vectors are introduced by
\begin{gather}
\ba^{\alpha}\,:=\,\aab\,\ba_{\beta}\,,
\end{gather} 
such that $ \ba_{\alpha}\,\cdot\ba^{\beta}\,=\,\delta^{\beta}_{\alpha} $ and $ [\aab]\,:=\,\left[\auab\right]^{-1} $, where $ [\delta^{\beta}_{\alpha} ] $ is the identity. The surface stretch is given by $ J\,=\,\sqrt{J_a/J_A} $ where $J_a\,:=\,\norm{\ba_1\,\times\ba_2}  $ and $J_A\,:=\,\norm{\bA_1\,\times\bA_2}  $ is the area of the parallelogram enclosed by the covariant tangents in the current and initial configuration, respectively. The components of the surface curvature can be defined as
\begin{gather}
\buab\,:=\,\ba_{\alpha,\beta}\,\cdot\,\bn\,,\quad b_{\beta}^{\alpha}\,:=\,a^{\alpha \gamma}\, b_{\gamma \beta}\,,\quad \text{and}\quad b^{\alpha \beta}\,:=\,b_{\gamma}^{\alpha}\, a^{\gamma \beta}\,.\label{eq:b_ab}
\end{gather}
Analogous curvature definitions, denoted $ B^{\alpha\beta}$, $ B^{\alpha}_{\beta} $ and $ B^{\alpha\beta} $, follow for the initial configuration. The two invariants of the curvature tensor, $\bb\,=\,\buab\,\ba^{\alpha}\,\otimes\ba^{\beta}  $, called mean and Gaussian curvature are calculated respectively as
\begin{gather}
H\,:=\,\frac{1}{2}\,\text{tr}~\bb\,=\,\frac{1}{2}\, b_{\alpha}^{\alpha}\,=\,\frac{1}{2} a^{\alpha \beta}\, b_{\alpha \beta}\,,\\
\kappa\,:=\,\det\bb\,=\,\frac{\det[\buab]}{\det [\auab]}\,.
\end{gather}
The principal curvatures of the surface can then be calculated as 
\begin{align}
\kappa^{*}_{1/2}\,=\,H\,\pm\,\sqrt{H^2\,-\,\kappa}\,.\label{eq:Pcurv}
\end{align}
Additionally, we introduce the surface deformation gradient
\begin{gather}
	\boldsymbol{F}\,:=\,\ba_{\alpha}\,\otimes\,\bA_{\alpha},\label{eq:dgrad}
\end{gather}
and two symmetric tensors: The Green-Lagrange surface strain tensor, 
\begin{gather}
\boldsymbol{E}\,=\,\varepsilon_{\alpha \beta}\, \boldsymbol{A}^{\alpha}\, \otimes\, \boldsymbol{A}^{\beta}\,:=\,\frac{1}{2}\,\left(a_{\alpha \beta}\,-\,A_{\alpha \beta}\right) \,\boldsymbol{A}^{\alpha} \,\otimes\, \boldsymbol{A}^{\beta}\,,
\end{gather} 
and the relative surface curvature tensor,
\begin{gather}
\boldsymbol{K}\,=\,K_{\alpha \beta}\, \boldsymbol{A}^{\alpha}\, \otimes\, \boldsymbol{A}^{\beta}\,:=\,\left(b_{\alpha \beta}\,-\,B_{\alpha \beta}\right) \,\boldsymbol{A}^{\alpha} \,\otimes\, \boldsymbol{A}^{\beta}\,.
\end{gather}
These three quantities can be used to characterize the deformation of a thin shell. A more detailed description of thin shell kinematics and its variation can be found for example in \cite{sauer2018computational}.

%----------------------------------------------------------------------------------------%
\subsection{Strong form and weak form}\label{ssec:bal}

The quasi-static shell boundary value problem governs the displacement field $ \bu $ of the surface $ \mathcal{S} $ through the fourth order partial differential equation
\begin{gather}
\boldsymbol{T}_{; \alpha}^{\alpha}\,+\,\boldsymbol{f}\,=\,\mathbf{0} \quad \forall\, \boldsymbol{x} \,\in\, \mathcal{S}\,,
\label{eq:strf}
\end{gather}
where $ \boldsymbol{f} $ is a body force on $ \sS $, and the prescribed displacement ($ \bar{\bu} $), traction ($ \bar{\bt} $), and bending moment ($ \bar{m}_{\tau} $) boundary conditions
\begin{alignat}{2}
\boldsymbol{u}\,&=\,\bar{\boldsymbol{u}}\quad &&\text { on } \partial_{u} \mathcal{S}\,, \\
\boldsymbol{t}\,&=\,\bar{\bt}\quad &&\text { on } \partial_{t} \mathcal{S}\,,\\
m_{\tau}\,&=\,\bar{m}_{\tau}\quad &&\text { on } \partial_{m} \mathcal{S}\,.
\end{alignat}
In Eq.~\eqref{eq:strf},
\begin{gather}
\boldsymbol{T}^{\alpha}\,=\,N^{\alpha \beta}\, \boldsymbol{a}_{\beta}\,+\,S^{\alpha}\, \boldsymbol{n}
\end{gather}
is the stress vector. Here, $ N^{\alpha\beta} $ and $ S^{\alpha} $ are the in-plane membrane and the out-of-plane shear stress components defined via Cauchy's theorem.
From angular momentum balance follows
\begin{align}
\begin{array}{l}
\sigma^{\alpha \beta}\,:=\,N^{\alpha \beta}\,-\,b_{\gamma}^{\beta}\, M^{\gamma \alpha}\,=\,\sigma^{\beta\alpha}, \\
S^{\alpha}\,=\,-M_{; \beta}^{\beta \alpha}\,,
\end{array}
\end{align}
where $ \Mab $ are the stress couples caused by out-of-plane bending. Tab.~\ref{tab:Strs} gives an overview of the different stress components. As shown, they can be either expressed per current surface area or per reference surface area.
\begin{table}[H]
	\small
	\centering
	\begin{tabular}{|c|c|c|}
		\hline
		& per current area & per reference area           \\ \hline
		& & \\[-2mm]
		physical membrane stresses & $ N^{\alpha\beta}\,=\,\sigma^{\alpha \beta}\,+\,b_{\gamma}^{\alpha}\, M^{\gamma \beta} $                     & $ N^{\alpha\beta}_0\,=\,J\,N^{\alpha\beta} $                  \\[3mm] \hline
		& & \\[-2mm]
		effective membrane stresses&$ \sigma^{\alpha\beta}\,=\,\dfrac{2}{J} \dfrac{\partial\,W}{\partial a_{\alpha \beta}} $ & $ \tau^{\alpha\beta}\,=\,J\,\sigma^{\alpha\beta} $ \\[3mm]\hline
		& & \\[-2mm]
		bending stress couples & $ M^{\alpha\beta}\,=\,\dfrac{1}{J} \dfrac{\partial\,W}{\partial b_{\alpha \beta}} $ & $ M^{\alpha\beta}_0\,=\,J\,M^{\alpha\beta} $                   \\[3mm] \hline
		& & \\[-2mm]
		out-of-plane shear stresses & $ S^{\alpha}\,=\,-M^{\alpha\beta}_{;\beta} $ & $ S^{\alpha}_0\,=\,J\,S^{\alpha} $                   \\[3mm] \hline
	\end{tabular}
	\caption{Different stress components w.r.t.~the current or reference configuration. $ N^{\alpha\beta} $ are the Cauchy membrane stresses appearing in the strong form equilibrium equation \eqref{eq:strf}, while the effective stresses $ \sig^{\alpha\beta} $ appear in the weak form. Their transformation to the reference configuration correspond to Kirchhoff membrane stresses.}\label{tab:Strs}
\end{table}
Multiplying strong form Eq.~\eqref{eq:strf} with a suitable variation $ \delta\bx \in \mathcal{V}$ and integrating it 	over surface $ \mathcal{S} $, leads to the weak form
\begin{align}
G_{\text {int }}\,-\,G_{\text {ext }}\,=\,0 \quad \forall\, \delta x\, \in\, \mathcal{V}\,,
\label{eq:virt}
\end{align}
where
\begin{align}
G_{\mathrm{int}} &\,=\,\int_{\mathcal{S}_{0}}\,\delta \varepsilon_{\alpha \beta}\, \tau^{\alpha \beta} \,\mathrm{d} A\,+\,\int_{\mathcal{S}_{0}}\, \delta b_{\alpha \beta} \,M_{0}^{\alpha \beta}\, \mathrm{d} A\,, \label{eq:weakfrm}\\
G_{\mathrm{ext}} &\,=\,\int_{\mathcal{S}} \delta \boldsymbol{x} \,\cdot \,\boldsymbol{f} \,\mathrm{d} a\,+\,\int_{\partial_{t} \mathcal{S}} \delta \boldsymbol{x} \,\cdot \,\boldsymbol{t}\, \mathrm{d} s\,+\,\int_{\partial_{m} \mathcal{S}} \delta \boldsymbol{n}\, \cdot\, m_{\tau} \,\boldsymbol{\nu}\,\mathrm{d} s\,.\label{eq:ext}
\end{align}
Eq.~\eqref{eq:weakfrm} contains the virtual work of the in-plane membrane deformations and out-of-plane bending deformations. It can be extended to out-of-plane shear and strain deformations \citep{simo1990stress} and in-plane bending deformations \citep{steigmann2018equilibrium,duong2020general} in the context of more general shell theories. The last part of Eq.~\eqref{eq:ext} represents the virtual work of moment $ m_{\tau} \,\boldsymbol{\nu} $, with $\boldsymbol{\nu}$ being the normal to the boundary where bending moment $ m_\tau $ is applied.

\begin{remark}
The membrane stresses and bending stress couples are also referred to as membrane forces and bending moments in the literature. 	
\end{remark}
%----------------------------------------------------------------------------------------%
\subsection{Surface Hyperelasticity}\label{ssec:selas}

For hyperelastic materials, the membrane and bending stress components in weak form~\eqref{eq:virt} are calculated from the surface energy density $ W $ (with units $ [\text{J/m}^2] $) by
\begin{equation}\label{eq:TauMo}
\begin{aligned}
\tauab\,&=\,2\,\pa{W}{a_{\alpha \beta}},\\
\Mab_{0}\,&=\,\,\pa{W}{\buab}\,.
\end{aligned}
\end{equation}
It is convenient to decompose the surface energy density into two parts,
\begin{align}
W\,=\,W_{\mathrm{m}}\,+\,W_{\mathrm{b}}\,,
\end{align}
associated with membrane and bending deformations. Ideally, the former should only generate membrane stresses, while the latter only generates bending stresses. However this is not possible in the case of coupled membrane-bending material behavior. We proceed by introducing some simple hyperelastic membrane models below. Hyperelastic bending models are then discussed in detail in Sec.~\ref{sec:3}.

An example of a linear elastic membrane strain energy is the Koiter model. Its membrane strain energy is of the form
\begin{equation}\label{eq:kmem1}
	W_{\mathrm{mKo}}\,=\,\dfrac{1}{2} \,\boldsymbol{E}\,:\, \mathbb{C}\,:\, \boldsymbol{E}\,,
\end{equation}
where the fourth order tensor $ \mathbb{C} $ is defined as 
\begin{equation}\label{eq:4c}
	\mathbb{C}\,=\,\Lambda \,\boldsymbol{I} \,\odot\, \boldsymbol{I}\,+\,2 \,\mu\,(\boldsymbol{I}\, \otimes\, \boldsymbol{I})^{\mathrm{s}}\,.
\end{equation}
Here, $ \Lambda $ and $ \mu $ are the surface Lam\'e parameters (with units [N/m]). Inserting Eq.~\eqref{eq:4c} into \eqref{eq:kmem1} leads to
\begin{equation}
W_{\mathrm{mKo}}\,=\,\dfrac{1}{2}\left(\Lambda\,\left(\mathrm{tr}\boldsymbol{E}\right)^2\,+\,2\,\mu\,E_{\alpha\beta}\,E^{\alpha\beta}\right)\,,\label{eq:mKo}
\end{equation}
where 
\begin{equation}\label{eq:kmem2}
\operatorname{tr} \boldsymbol{E}\,=\,E^{\alpha \beta} A_{\alpha \beta}\,, \quad \text { and } \quad E^{\alpha \beta}\,:=\,A^{\alpha \gamma} E_{\gamma \delta} A^{\delta \beta}.
\end{equation}
The membrane and bending stress contributions of the Koiter membrane model are obtained from Eq.~\eqref{eq:TauMo} as
\begin{equation}\label{eq:memkoi}
\begin{aligned}
	\tau_{\mathrm{mKo}}^{\alpha\beta}\,&=\,\Lambda \operatorname{tr} \boldsymbol{E} A^{\alpha \beta}+2 \mu E^{\alpha \beta}\,,\\
	M^{\alpha\beta}_{0\,\mathrm{mKo}}\,&=\,0\,.
\end{aligned}
\end{equation}
The membrane stresses $ \tau_{\mathrm{mKo}}^{\alpha\beta} $ can also be expressed as 
\begin{align}
	\tau_{\mathrm{mKo}}^{\alpha\beta}\,=\,c^{\alpha\beta\gamma\delta}_{\mathrm{Ko}}\,(a_{\gamma\delta}\,-\,A_{\gamma\delta})/2\,,
\end{align}
with
\begin{align}\label{eq:memkoic}
	c^{\alpha\beta\gamma\delta}_{\mathrm{Ko}}\,=\,\Lambda \,A^{\alpha \beta}\, A^{\gamma \delta}\,+\,\mu\left(A^{\alpha \gamma}\, A^{\beta \delta}\,+\,A^{\alpha \delta}\, A^{\beta \gamma}\right)\,.
\end{align}
$ c^{\alpha\beta\gamma\delta}_{\mathrm{Ko}} $ is equivalent to the in-plane components of the material tensor of the St.~Venant-Kirchhoff law \citep{ciarlet2005introduction}. Eq.~(\ref{eq:memkoi}.1) is thus equal to the membrane constitutive model in \cite{kiendl2009isogeometric}, which is obtained through thickness integration. 

Another commonly used membrane model, which can be derived from 3D elasticity, is the Neo-Hookean surface model. Its strain energy function can be written as
\begin{align}
W_{\mathrm{mNH}}\,=\,\frac{\Lambda}{4}\,\left(J^2\,-\,1-2\,\text{ln}~J\right)\,+\,\frac{\mu}{2}\,\left(I_1\,-\,2\,-\,2\,\text{ln}~J\right).
\label{eq:neohook}
\end{align}  
Here, $ I_1 \,=\,A^{\alpha\beta}\,\auab$ is the first invariant of the surface Cauchy-Green tensors. The Neo-Hookean model only produces the membrane stress components
\begin{equation}\label{eq:neohookstr}
\tauab_{\mathrm{mNH}}\,=\,\frac{\Lambda}{2}\,\left(J^2\,-\,1\right)\,a^{\alpha\beta}\,+\,\mu\,\left(A^{\alpha\beta}\,-\,a^{\alpha\beta}\right)\,,
\end{equation}
i.e.~$M^{\alpha\beta}_{0\,\mathrm{mNH}}\,=\,0$. 

%----------------------------------------------------------------------------------------
\subsection{Linearization}\label{ssec:lin}

In order to solve the nonlinear Eq.~\eqref{eq:virt} using the Newton-Raphson method, its linearization is necessary. This leads to the increment for the internal virtual work
\begin{align}
\Delta G_{\text{int}}\, = \,\int_{\mathcal{S}_o} \bigg(&\cabgd\,\delta\varepsilon_{\alpha \beta}\,\Delta\varepsilon_{\gamma \delta} \,+\, \dabgd \,\delta\varepsilon_{\alpha \beta}\, \Delta\bugd\, +\, \tauab\, \Delta\delta \varepsilon_{\alpha \beta}\,\nonumber \,+\, \\
&\eabgd\, \delta\buab\,\Delta\varepsilon_{\gamma \delta} \,+\, \fabgd\, \delta\buab\,\Delta\bugd \,+\, M^{\alpha\beta}_{0}\,\Delta\delta\buab\bigg)\,\dif A\,,
\label{eq:wkfrm}
\end{align}
where the material tangents are defined as 
\begin{align}
&\cabgd \,:=\, \frac{\partial \tauab}{\partial \varepsilon_{\gamma \delta}}\,, && \dabgd \,:=\, \frac{\partial \tauab}{\partial \bugd}\,,\nonumber\\
&\eabgd \,:=\, \frac{\partial M^{\alpha\beta}_{0}}{\partial \varepsilon_{\gamma \delta}}\,, && \fabgd \,:= \,	\frac{\partial M^{\alpha\beta}_{0}}{\partial \bugd}\,.
\label{eq:1.31}
\end{align}
Examples for these, along with $ \Delta\delta\eps_{\alpha\beta} $ and $ \Delta\delta b_{\alpha\beta} $ can be found in \cite{sauer2018computational}.

%----------------------------------------------------------------------------------------%
\subsection{Isogeometric FE approximation}\label{ssec:fe}

The geometry within an undeformed finite element $ \Omega_{0}^e $ and its deformed counterpart $ \Omega^e $ is interpolated from the positions of control points $ \mX_{e} $ and $ \mx_{e} $, respectively, as
\begin{align}
\bX \,=\, \mN_e\,\mX_{e}\,,\quad\bx\, =\, \mN_e\,\mx_{e}\,,\label{eq:inter}
\end{align}
with
\begin{align}
\mN_e(\xi^\alpha)\,:=\,\left[N_1\,\boldsymbol{1},\,N_2\,\boldsymbol{1},\,\hdots,\,N_{n_e}\,\boldsymbol{1}\right]\,.
\end{align}
Here $ \{N_A\,(\xi^\alpha)\}^{n_e}_{A\,=\,1} $ are the $ n_e $ $ C^1-$continuous NURBS basis functions of $ \Omega^e $. Such an arrangement results in a direct correspondence between the discretized equations and their implementation in a computer code. Discretizing Eq.~\eqref{eq:virt} gives 
\begin{align}
\sum_{e =1}^{n_{\text{el}}} \big( G^e_{\text{int}} \,+\, G^e_{\text{ext}}\big) \,=\, 0 \qquad \forall \,\delta \mx_{e} \,\in\, \mathcal{V}\,,
\end{align}
where $ n_{\mathrm{el}} $ is the number of elements. Based on Eq.~\eqref{eq:weakfrm} and \eqref{eq:inter}, the internal virtual work due to membrane stresses $ \tauab $ and the bending stress couples $ M^{\alpha\beta}_{0} $ will be
\begin{align}
G^e_{\text{int}}\, =\, \delta\mx^{\mathrm{T}}_e \,\big(\mf^e_{\text{int}\tau} \,+\, \mf^e_{\text{int}M}\big)\,,
\end{align}
where
\begin{align}
\mf^e_{\text{int}\tau} &\,:=\, \intooe \tauab \,\mN^\mathrm{T}_{e,\alpha}\,\ba_{\beta}\,\dif A\,,\nonumber\\
\mf^e_{\text{int}M} &\,:=\, \intooe M_{0}^{\alpha\beta}\,\mN^\mathrm{T}_{e;\alpha\beta}\,\bn\,\dif A\,,
\end{align}
and 
\begin{align}
\mN_{e;\alpha\beta}\,:=\,\mN_{e,\alpha\beta}\,-\,\Gamma^{\gamma}_{\alpha\beta}\,\mN_{e,\gamma}\,.
\end{align}
Here, $ \mN_{e,\alpha} $ and $ \mN_{e,\alpha\beta} $ denote the first and second derivatives of $\mN_{e}$ w.r.t.~parameter $ \xi^{\alpha} $, while $ \Gamma^\gamma_{\alpha\beta}\, := \,\ba^\gamma\,\cdot\,\ba_{\alpha,\beta} $ denote the Christoffel symbols. Discretizing Eq.~\eqref{eq:ext}, the external virtual work follows as
\begin{align}
G^e_{\text{ext}} \,=\, \delta \mx^{\mathrm{T}}_{e}\,\big(\mf^{e}_{\text{ext}o} \,+\, \mf^{e}_{\text{ext}t} \,+\, \mf^{e}_{\text{ext}m}\,+\, \mf^{e}_{\text{ext}p}\big)\,,\label{eq:virtualwork}
\end{align}
where the external FE force vectors are 
\begin{equation}
\begin{aligned}
\mf^{e}_{\text{ext}o} &\,:=\, \intooe \mN^{\mathrm{T}}_{e}\,\boldsymbol{f}_0\,\dif A\,,\\
\mf^{e}_{\text{ext}t} &\,:=\, \int_{\partial_t\Omega^{e}} \mN^{\mathrm{T}}_{e}\,\bt\, \dif s\,,\\
\mf^{e}_{\text{ext}p} &\,:=\, \int_{\Omega^{e}} \mN^{\mathrm{T}}_{e}\,p\,\bn\, \dif a\,,\\
\mf^{e}_{\text{ext}p} &\,:=\, -\int_{\partial_m\Omega^{e}} \mN^{\mathrm{T}}_{e,\alpha}\,\nu^{\alpha}\,m_{\tau}\,\bn\, \dif s\,,
\label{eq:fext}
\end{aligned}
\end{equation}
 for the body force $ \boldsymbol{f}\, = \,\boldsymbol{f}_0\, +\, p \,\bn $. Here $ \boldsymbol{f}_0 $ denotes dead loading, while $ p $ denotes an external pressure always acting normal to the surface, and $ \boldsymbol{\nu} \,=\, \nu^a \,\ba_{\alpha} $ denotes the boundary normal on $ \partial\sS $.

Discretizing Eq.~\eqref{eq:wkfrm} gives
\begin{align}
\Delta G_{\text{int}} \,=\, \delta \mx^T_e \big[
\mk^e_{\tau\tau} \,+\, \mk^e_{\tau M} \,+\, \mk^e_{\tau} \,+\, \mk_{M \tau}^{e} \,+\, \mk_{\tau\tau}^e \,+\, \mk^e_{M}\big] \Delta \mx_e\,,
\end{align}
where 
\begin{align}
\mk_{\tau\tau}^e &\,:= \,\intooe \cabgd\, \mN^\mathrm{T}_{e,\alpha}\,\left(\ba_{\beta} \,\otimes\, \ba_{\gamma}\right)\,\mN_{e,\delta}\, \dif A\,,\nonumber\\
\mk_{\tau M}^e &\,:=\, \intooe \dabgd \,\mN^\mathrm{T}_{e,\alpha}\,\left(\ba_{\beta}\, \otimes\, \bn\right)\,\mN_{e;\gamma\delta}\, \dif A\,,\nonumber\\
\mk_{M \tau}^e &\,:=\, \intooe \eabgd \,\mN^\mathrm{T}_{e;\alpha\beta}\left(\bn \,\otimes\, \ba_{\gamma}\right)\,\mN_{e,\delta}\, \dif A\,,\nonumber\\
\mk_{M M}^e &\,:=\, \intooe \fabgd \,\mN^\mathrm{T}_{e;\alpha\beta}\,\left(\bn \,\otimes\, \bn\right)\,\mN_{e;\gamma\delta} \,\dif A\,,
\label{eq:1.42}
\end{align}
are the material stiffness matrices and 
\begin{align}
\mk_{\tau}^{e} &\,:=\, \intooe \mN^\mathrm{T}_{e,\alpha}\,\tauab\,\mN_{e,\beta}\,\dif A\,,\nonumber\\
\mk^e_{M} &\,:=\, \mk^e_{M1} \,+\, \mk^e_{M2} \,+\, (\mk^e_{M2})^\mathrm{T}\,,
\end{align} 
with
\begin{align}
\mk^e_{M1} &\,:=\,-\intooe \buab \,\Mab_{0}\,  a^{\gamma\delta}\,\mN^\mathrm{T}_{e,\gamma}\,\left(\bn\,\otimes\,\bn\right)\,\mN_{e,\delta}\,\dif A\,,\\
\mk_{M2}^e &\,:=\, -\intooe \Mab_0 \,\mN^\mathrm{T}_{e,\gamma}\,\left(\bn\,\otimes\,\ba^{\gamma}\right)\,\tilde{\mN}_{e;\alpha\beta}\,\dif A\,,  
\end{align}
are the geometric stiffness matrices. The linearization of Eq.~\eqref{eq:virtualwork} leads to 
\begin{align}
\Delta G_{\mathrm{ext}}^{e}\,=\,\delta \mx_{e}^{\mathrm{T}}\,\left(\mk_{\mathrm{ext}p}^{e}\,+\,\mk_{\mathrm{ext}m}^{e}\right)\, \Delta \mathbf{x}_{e}\,,
\end{align}
where $ \mk_{\mathrm{ext}p}^{e} $ and $ \mk_{\mathrm{ext}m}^{e} $ can be found in \cite{sauer2014computational} and \cite{duong2017new}, respectively.

%%%%%%%%%%%%%%%%%%%%%%%%%%%%%%%%%%%%%%%%%%%%%%%%%%%%%%%%%%%%%%%%%%%%%%%%%%%%%%%%%%%%%%%%%%
\section{Bending constitution}\label{sec:3}

This section presents hyperelastic bending models for thin shells. We first summarize existing bending models and then propose a new nonlinear, anisotropic bending model. In each subsection we state the surface energy per unit area and then define the membrane and bending stresses calculated using Eq.~\eqref{eq:TauMo}.

%----------------------------------------------------------------------------------------%
\subsection{Koiter bending model}\label{koiBend}

The surface energy density for the Koiter bending model \citep{ciarlet2005introduction,steigmann2013koiter} is
\begin{align}
W_{\mathrm{bKo}}\,&=\,\frac{1}{2}\, \boldsymbol{K}\,:\, \mathbb{F}\,:\, \boldsymbol{K}\,,\label{eq:koi}
\end{align}
where the fourth order tensor $ \mathbb{F} $ is defined w.r.t. $ \mathbb{C} $ from Eq.~\eqref{eq:4c} as
\begin{align}
\mathbb{F}\,=\,\frac{T^2}{12}\,\mathbb{C}\, .
\end{align}
Here, $ T $ is the shell thickness. Inserting $ \mathbb{F} $ into Eq.~\eqref{eq:koi} leads to
\begin{align}
W_{\mathrm{bKo}}\,=\,\frac{T^2}{24}\,\left(\Lambda\,(\text{tr}\bK)^2\,+\,2\,\mu\,K_{\alpha\beta}\,K^{\alpha\beta}\right)\,,\label{eq:koiw}
\end{align}
where
\begin{align}
\text{tr}~\bK\,=\,K^{\alpha\beta}\,A_{\alpha\beta}\,,\quad\text{and}\quad K^{\alpha\beta}:=\,A^{\alpha\gamma}K_{\gamma\delta}\,A^{\delta\beta}\,,
\end{align} 
similar to Eq.~\eqref{eq:mKo}-\eqref{eq:kmem2}. According to Eq.~\eqref{eq:TauMo}, the Koiter bending model only causes the bending stress couples
\begin{align}
\Mab_{0\,\mathrm{bKo}}\,=\,\frac{T^2}{12}\,\left(\Lambda\,\text{tr}\bK\,\Aab\,+\,2\,\mu\,K^{\alpha\beta}\right)\,,\label{eq:koiMo}
\end{align} 
while $ \tau^{\alpha\beta}_{\mathrm{bKo}} = 0 $. The surface Lam\'e parameters can be derived from Young's modulus ($ E $) and Poisson's ratio ($ \nu $) as \citep{ciarlet2005introduction}
\begin{align}
\mu\,:=\, \frac{T\,E}{2\,\left(1\,+\,\nu\right)}\,, \quad \Lambda\,:=\,\frac{2\,\mu\,\nu}{1-\nu}\,,
\label{eq:surflam}
\end{align} 
that admit the special case $ \nu = 0.5 $.
From these, one can identify the bending stiffness \citep{landau1986course}
\begin{align}
c_{\mathrm{Koi}}\,=\,\dfrac{E\,T^3}{12\,\left(1\,-\,\nu^2\right)}\,=\,\frac{T^2}{12}\,\left(\Lambda\,+\,2\,\mu\right)\,.\label{eq:ckoi}
\end{align}
Using Eq.~\eqref{eq:memkoic}, the Koiter bending stress couples can be expressed as 
\begin{align}
	\Mab_{0\,\mathrm{bKo}}\,=\,\frac{T^2}{12}\,c^{\alpha\beta\gamma\delta}_{\mathrm{Ko}}\,(b_{\gamma\delta}\,-B_{\gamma\delta})\,,
\end{align}
which is equivalent to the bending model in \cite{kiendl2009isogeometric} that is obtained from thickness integration of the St.~Venant-Kirchhoff model.
%----------------------------------------------------------------------------------------%
\subsection{Canham bending model}

The surface energy density for the bending model of \cite{CANHAM} is of the form
\begin{gather}
W_{\mathrm{Can}}\,=\,c\,J\,\left(2 \,H^{2}\,-\,\kappa\right),
\end{gather}
where $ c $ is the bending modulus with units [Nm]. According to Eq.~\eqref{eq:TauMo}, the Canham bending model contributes to both membrane and bending stresses as
\begin{equation}\label{eq:caneq}
\begin{aligned}
\tauab_{\mathrm{Can}}\,&=\,c \,J\,\left(2\, H^{2}\,+\,\kappa\right) \,a^{\alpha \beta}\,-\,4 \,c\, J \,H \,b^{\alpha \beta}\,,\\
\Mab_{0\,\mathrm{Can}}\,&=\,c\,J\,\bab\,.
\end{aligned} 
\end{equation}
As seen, $ M^{\alpha\beta}_{0\,\mathrm{Can}} $ is proportional to the total curvature $ \bab $ instead of the relative curvature $  K^{\alpha\beta} $ as for the Koiter model in Eq.~\eqref{eq:koiMo}. This makes the Canham model unsuitable for initially curved stress-free shells. Comparing their stiffness $ \fabgd  $, further shows that the Canham and Koiter model are equivalent for small deformations of initially flat shells when
\begin{align}
c\,=\,c_{\mathrm{Can}} \,= \,\dfrac{\mu\, T^2}{6}\quad \text{and}\quad  \Lambda\, = \,0\,,
\end{align}
implying
\begin{align}
c_{\mathrm{Can}}\,=\,\dfrac{E\,T^3}{12}\,
\end{align}  
according to Eq.~\eqref{eq:ckoi}.

%----------------------------------------------------------------------------------------%
\subsection{Helfrich bending model}

The surface energy density for the bending model of \citet{helfrich1973elastic} is defined by 
\begin{gather}
W_{\mathrm{Hel}}\,=\,J\,\left(k\,\,(H\,-\,\bar{H}_0)^2\,+\,\bar{k}\,\kappa\right)\,,
\end{gather}
where $ \bar{H}_0 $ is the spontaneous curvature of the material, which can be an externally prescribed quantity, or taken as the initial mean curvature of the shell. Further, $ k $ and $ \bar{k} $  are material constants. The membrane and bending stresses follow from Eq.~\eqref{eq:TauMo} as
\begin{equation}
\begin{aligned}
\tau^{\alpha \beta}_{\mathrm{Hel}}\,&=\,J\,\left(k\, \Delta H^{2}\,-\,\bar{k} \,\kappa\right)\, a^{\alpha \beta}\,-\,2 \,k \,J\, \Delta H\, b^{\alpha \beta}\,,\\
M_{0\,\mathrm{Hel}}^{\alpha \beta}\,&=\,J\,\left(k \,\Delta H\,+\,2\, \bar{k}\, H\right) \,a^{\alpha \beta}\,-\,\bar{k}\, J\, b^{\alpha \beta}\,,\label{eq:helfeq}
\end{aligned}
\end{equation}
with $ \Delta H\,:=\,H\,-\,\bar{H}_0 $. This model reduces to the Canham model for $ \bar{H}_{0}\,=\,0 $, $ k\,=\,2c $, and $ \bar{k}\,=\,-c $. Comparing their stiffness $ \fabgd $ \citep{sauer2017theoretical}, shows that the Helfrich and Koiter bending model are equivalent for flat shells in the linear regime if
\begin{align}
k\,=\,\frac{T^2}{6}\,\left(\Lambda\,+\,2\,\mu\right)\,,\quad\text{and}\quad\bar{k}\,=\,-\frac{T^2\,\mu}{6}\,,
\label{eq:HelP}
\end{align}
which then implies $ c_{\mathrm{Hel}} = c_{\mathrm{Koi}} $.

%----------------------------------------------------------------------------------------%
\subsection{Shell models derived from 3D constitutive laws}
In derived shell formulations, the constitutive law is usually obtained by projecting 3D material models onto the surface through thickness integration. A general derivation of this for isogeometric shells is provided in \cite{kiendl2015isogeometric} and \cite{duong2017new}. Given the Kirchhoff stress tensor ($ \tilde{\btau} $) or the second Piola-Kirchhoff stress tensor ($ \tilde{\bS} $) of a 3D constitutive model, their in-plane components can be obtained from
\begin{align}
	\tilde{\tau}^{\alpha\beta}\,=\,\bg^{\alpha}\,\cdot\,\tilde{\btau}\,\bg^{\beta}\,=\,\bG^{\alpha}\,\cdot\,\tilde{\bS}\,\bG^{\beta}\,,
\end{align}
where $ \bg^{\alpha} $ denotes the contra-variant tangent vectors at point $ \bx\,+\,\xi_{0}\,\bn $ described by thickness coordinate $ \xi_{0} \,\in[-T/2,\,T/2]$. Similarly, $ \bG^{\alpha} $ denotes the contra-variant tangent vectors in the initial configuration. The membrane and bending stresses within the shell can then be computed from 
\begin{align}
	\tauab \,\,=\,\int_{-\frac{T}{2}}^{\frac{T}{2}}\,\tilde{\tau}^{\alpha\beta}\,\dif\xi_{0}\,,\quad	\Mab_0 \,\,=\,-\,\int_{-\frac{T}{2}}^{\frac{T}{2}}\,\xi_{0}\,\tilde{\tau}^{\alpha\beta}\,\dif\xi_{0}\,.
\end{align} 
Generally, numerical integration is required for evaluating these expressions. But in some cases analytical integration is possible. An example is the St.Venant-Kirchhoff model discussed in Sec.~\ref{koiBend}. Another example is given in the following section.
\begin{remark}
	For degenerate shells, 3D material models are used with some correction factors \citep{ahmad1970analysis}. The shell finite elements are then directly obtained from 3D kinematics by enforcing certain constraints. 
\end{remark}
%----------------------------------------------------------------------------------------%
\subsection{Analytically projected Neo-Hooke bending model (apH)}

The classical 3D Neo-Hooke material model can be analytically integrated through the shell thickness as presented in \cite{duong2017new}. Assuming incompressibility ($ \nu \,= \,0.5 $) this leads to the membrane and bending stresses \citep{roohbakhshan2017efficient}
\begin{equation}\label{eq:wbapH}
\begin{aligned}
\tauab_{\mathrm{apH}}\,&=\,\mu\,\left(\Aab\,-\,\dfrac{1}{J^2}\,\aab\right)\,,\\
\Mab_{0\,\mathrm{apH}}\,&=\,-\dfrac{\mu\,T^2}{6}\,\left(B^{\alpha\beta}\,-\,\dfrac{1}{J^2}\,\left(\bab\,+\,2\,\left(H\,-\,H_{0}\right)\,\aab\right)\right)\,.	
\end{aligned}
\end{equation} 
Here, $ H_{0} $ refers to the mean curvature in the initial configuration. Model~\eqref{eq:wbapH} becomes a pure membrane model if $ M_{0\,\mathrm{apH}}^{\alpha\beta} \,=\,0$, i.e.~for $ T^2 \rightarrow 0 $. Otherwise it is a complete shell model, meaning, it has both membrane and bending parts, unlike the other models considered in this section. The bending stress parameter, $ c_{\mathrm{apH}} $, is the same as $ c_{\mathrm{Koi}} $ in Eq.~\eqref{eq:ckoi} as long as $ \nu\,=\,0.5 $.

%----------------------------------------------------------------------------------------%
\subsection{Proposed new bending model}

The preceding bending models all have drawbacks, as will be shown in Sec.~\ref{sec:4} and \ref{sec:5}. This motivates the following new bending model defined by the surface energy density per reference area
\begin{align}
W_{\mathrm{new}}\,&=\,\frac{c_1}{2} \, k_1^2\,+\,\frac{c_2}{2} \, k_2^2\,+\,c_{12}\,k_1\,k_2\,+\,\frac{c_3}{2}\,k_{12}^2\,,
\label{e:1}
\end{align}
where the kinematical quantities 
\begin{align}
k_i\,&:=\,\lambda_{i}\,\kappa_{i}\,-\,\kappa_{0i}\,,\quad i = 1,2   \text{ (without summation)}\\
k_{12}\,&:=\,\sqrt{\lambda_{1}\,\lambda_{2}}\,\kappa_{12}\,-\,\kappa_{012}\,,
\end{align}
characterize the relative curvature, and $ c_i $, $ c_{12} $ and $ c_{3} $ are bending moduli. The stretch ($ \lambda_i $) and curvature ($ \kappa_{i} $, $ \kappa_{0i} $, $ \kappa_{12} $ and $ \kappa_{012} $) measures are defined as (without summation on $ i $)
\begin{gather}
\lambda_{i}\,:= \,\sqrt{L^\alpha_i\,\auab\,L^\beta_i}\,,\label{eq:str}\\
%\quad \lambda_{0i}\,:= \sqrt{L^\alpha_i\,\Auab\,L^\beta_i}\,,\\
\kappa_{i} \,:= \,\frac{1}{\lambda_{i}^2}\,L^\alpha_i\,\buab\,L^\beta_i,\quad\kappa_{0i} \,:= \,L^\alpha_i\,B_{\alpha\beta}\,L^\beta_i\label{eq:kurv}\,,
\end{gather}
and
\begin{gather}
\kappa_{12}\,:=\,\dfrac{L^{\alpha}_1\,L^{\beta}_2\,+\,L^{\alpha}_2\,L^{\beta}_1}{\lambda_{1}\,\lambda_{2}}\,b_{\alpha\beta}\,,\quad\kappa_{012}\,:=\,\left(L^{\alpha}_1\,L^{\beta}_2\,+\,L^{\alpha}_2\,L^{\beta}_1\right)\,B_{\alpha\beta}\,.
\end{gather}
 It is emphasized that these kinematic quantities do not depend on the choice of surface parameterization, hence making the new model frame invariant. Here $ \kappa_{0i} $, for $ i \,= \,1,2 $, refers to the two principal surface curvatures in the initial configuration, while
\begin{align}
L^\alpha_i \,=\,\bL_i\,\cdot\, \bA^\alpha,
\label{eq:auxK}
\end{align}
where $ \bL_i $ refers to the corresponding principal curvature directions in the initial configuration. Upon deformation, $ \bL_i $ is transformed to the current direction
\begin{align}
	\boldsymbol{\ell}_i \,=\,\boldsymbol{F}\,\frac{\bL_i}{\lambda_{i}}\,,\label{eq:auxKcurr}
\end{align}
due to Eqs.~\eqref{eq:dgrad} and \eqref{eq:str}.
As a consequence of Eqs.~\eqref{eq:str} and \eqref{eq:kurv}, $ \lambda_i $ and $ \kappa_i $ are generally not equal to the current principal stretches and curvatures. The proposed bending model is a generalization of the 1D fiber bending model of \cite{duong2020general} that appears for $ c_2\, = \,c_{12} \,= \,c_3\, = 0 $. In contrast to the Canham model, the proposed bending model of Eq.~\eqref{e:1} can be employed for initially stress-free curved shells.

The membrane and bending stresses follow from Eqs.~\eqref{eq:TauMo} and \eqref{e:1} as
\begin{equation}
\begin{aligned}
&\tauab_{\mathrm{new}}\,&&=\,-\left(c_1\,k_1\,+\,c_{12}\,k_2\,+\,\dfrac{c_3\,\sqrt{\lambda_{2}}\,\kappa_{12}\,k_{12}}{2\,\sqrt{\lambda_{1}}\,\kappa_{1}}\right)\,\kappa_{1}\,\ell^{\alpha\beta}_{11}\,\\
& &&-\,\left(c_{12}\,k_1\,+\,c_{2}\,k_2+\,\dfrac{c_3\,\sqrt{\lambda_{1}}\,\kappa_{12}\,k_{12}}{2\,\sqrt{\lambda_{2}}\,\kappa_{2}}\right)\,\kappa_{2}\,\ell^{\alpha\beta}_{22}\,,\\
&\Mab_{0\,{\mathrm{new}}}\,&&=\,\left(c_1\,k_1\,+\,c_{12}\,k_{2}\right)\,\ell^{\alpha\beta}_{11}\,+\,\left(c_{12}\,k_1\,+\,c_{2}\,k_{2}\right)\,\ell^{\alpha\beta}_{22}\,+\,c_3\,k_{12}\,\ell^{\alpha\beta}_{12}\,,
\end{aligned}
\end{equation}
\vspace*{-0.5mm}
with
\vspace*{-1.5mm}
\begin{gather}
	\ell^{\alpha\beta}_{11}\,:=\,\dfrac{L^{\alpha}_1\,L^{\beta}_1}{\lambda_{1}}\,,\quad	\ell^{\alpha\beta}_{22}\,:=\,\dfrac{L^{\alpha}_2\,L^{\beta}_2}{\lambda_{2}}\,,\quad
	\ell^{\alpha\beta}_{12}\,:=\,\dfrac{L^{\alpha}_1\,L^{\beta}_2\,+\,L^{\alpha}_2\,L^{\beta}_1}{\sqrt{\lambda_{1}\,\lambda_{2}}}\,.
	\label{eq:auxL}
\end{gather}
The fourth order material tangents and an efficient FE implementation for the new bending model are provided in Appendices~\ref{ap:tan} and \ref{ap:fe}, respectively. Appendix~\ref{ap:2} shows that for small deformations the proposed model is equivalent to the Koiter model, if
\begin{gather}
c_i \,=\,\dfrac{T^2}{12}\left(\Lambda\,+\,2\,\mu\right)\,,\quad c_{12} \,=\,\dfrac{T^2}{12}\Lambda\,,\quad\text{and}\quad c_{3} \,=\,\dfrac{T^2}{12}\mu\,.
\label{eq:newp}
\end{gather} 
$ c_1 $ and $ c_2 $ then play the same role as $ c_{\mathrm{Koi}} $.
\begin{remark}
	The Canham bending model is a special case of the proposed new model when $ \kappa_{0i}\,=\,0 $, $ \lambda_{i}\,=\,\sqrt{J} $, $ c_i\,=\,c $ and $ c_{12}\,=\,c_3\,=\,0 $.
\end{remark} 

%%%%%%%%%%%%%%%%%%%%%%%%%%%%%%%%%%%%%%%%%%%%%%%%%%%%%%%%%%%%%%%%%%%%%%%%%%%%%%%%%%%%%%%%%%
\section{Elementary bending test cases}\label{sec:4}

In this section, three analytical test cases are investigated, and the results for the proposed new bending material model are juxtaposed with the results for the other bending models given in the previous section. In all the test cases considered, the initial configuration of a (half) tube with radius $ R $ and length $ L\,=\,\pi\,R $ is considered. The initial surface of the tube can be parametrized using
\begin{align}
\theta\in\left[-\pi/2,\,\pi/2\right] \,\text{and} \,\,\,\phi\in\left[0,\,\pi\right]\,,
\label{eq:surfX}
\end{align}
to represent any point on it as
\begin{align}
\bX(\theta,\phi)\,=\,R\,\be_r\,+\,R\,\phi\,\be_3\,,
\label{eq:iniX}
\end{align} 
with
\vspace*{-5mm}
\begin{align}
\be_{r}\,&=\,\cos{\theta}\,\be_1\,+\,\sin{\theta}\,\be_2\,,\label{eq:er}\\
\be_{\theta}\,&=\,-\sin{\theta}\,\be_1\,+\,\cos{\theta}\,\be_2\,.\label{eq:etheta}
\end{align}
The tangents and normal for this surface then follow from \eqref{eq:A_a} as
\begin{align}
\bA_1\,&=\,\pa{\bX}{\theta}\,=\,R\,\be_{\theta}\,,\\
\bA_2\,&=\,\pa{\bX}{\phi}\,=\,R\,\be_3\,,
\end{align}
and
\vspace*{-5mm}
\begin{align}
\bN\,&=\,\frac{\bA_1\times\bA_2}{\norm{\bA_1\times\bA_2}}\,=\,\be_{r}\,.
\end{align}
Using these, we can obtain the further quantities
\begin{gather}
[\Auab\,]=\,R^2\,\left[ {\begin{array}{cc}
	1 & 0 \\
	0 & 1 \\
	\end{array} }\right]\,,\quad[\Aab]\,=\,\dfrac{1}{R^2}\left[ {\begin{array}{cc}
	1 & 0 \\
	0 & 1 \\
	\end{array} }\right]\,,
\end{gather}
\vspace*{-5mm}
\begin{gather}
[B_{\alpha\beta}]\,=\,-R\,\left[ {\begin{array}{cc}
	1 & 0 \\
	0 & 0 \\
	\end{array} }\right]\,,\quad [B^{\alpha}_{\beta}]\,=\,-\dfrac{1}{R}\left[ {\begin{array}{cc}
	1 & 0 \\
	0 & 0 \\
	\end{array} }\right]\,,\quad [B^{\alpha\beta}]\,=\,-\dfrac{1}{R^3}\left[ {\begin{array}{cc}
	1 & 0 \\
	0 & 0 \\
	\end{array} }\right]\,,\label{eq:iniBab}
\end{gather}
\vspace*{-5mm}
\begin{gather}
\bL_1\,=\,\be_{\theta}\,,\quad \bL_2\,=\,\be_3\,,\\
L^1_1\,=\,\frac{1}{R}\,,\quad L^2_1\,=\,0,\quad L^1_2\,=\,0\,,\quad L^2_2\,=\,\frac{1}{R}\,,
\end{gather}
and 
\vspace*{-5mm}
\begin{gather}
\kappa_{01}\,=\,-\frac{1}{R}\,,\quad\kappa_{02}\,=\,0\,. 
\end{gather}

In the elementary test cases that follow, we use  $ \Lambda\, = 0$ and $\mu\,T^2/6\,=\,c $ for all the models except the proposed new model. For the new bending model, the bending parameters are considered to be arbitrary but with $ c_1\,=\,c_2\,=\,c $.  This generalization is used so that all the bending models can be compared with the proposed model through a common material parameter. Further, for the Helfrich model, the corresponding material constants are taken from Eq.~\eqref{eq:HelP} and the spontaneous curvature is set to the initial mean curvature value $ \bar{H}_{0}\,=\,-1/(2R) $.

%----------------------------------------------------------------------------------------%
\subsection{Test case 1 -- bending vs.~rigid rotation}\label{subsec:t1}

As shown in Fig.~\ref{fig:test1}, the first test case compares the membrane and bending stresses for two similarly looking but differently obtained final configurations. One is obtained by applying a rigid body rotation and thus no membrane or bending stresses should be induced. The other is obtained by transverse counter bending, wherein both stresses are induced. This test examines whether the principal curvatures are assigned properly during deformation. This is essential for bending models directly defined on the principal curvatures, such as our proposed model.

%~~~~~~~~~~~~~~~~~~~~~~~~~~~~~~~~~~~~~~~~~~~~~~~~~~~~~~~~~~~~~~~~~~~~~~~~~~~~~~~~~~~~~~~~%
\subsubsection{Rigid body rotation}

\begin{figure}[H]
	\small
	\centering
	\begin{tikzpicture}
		\node[inner sep=0pt] (a) at (-11,3) {\includegraphics[trim={5cm 6cm 5cm 9cm},clip,scale=0.13]{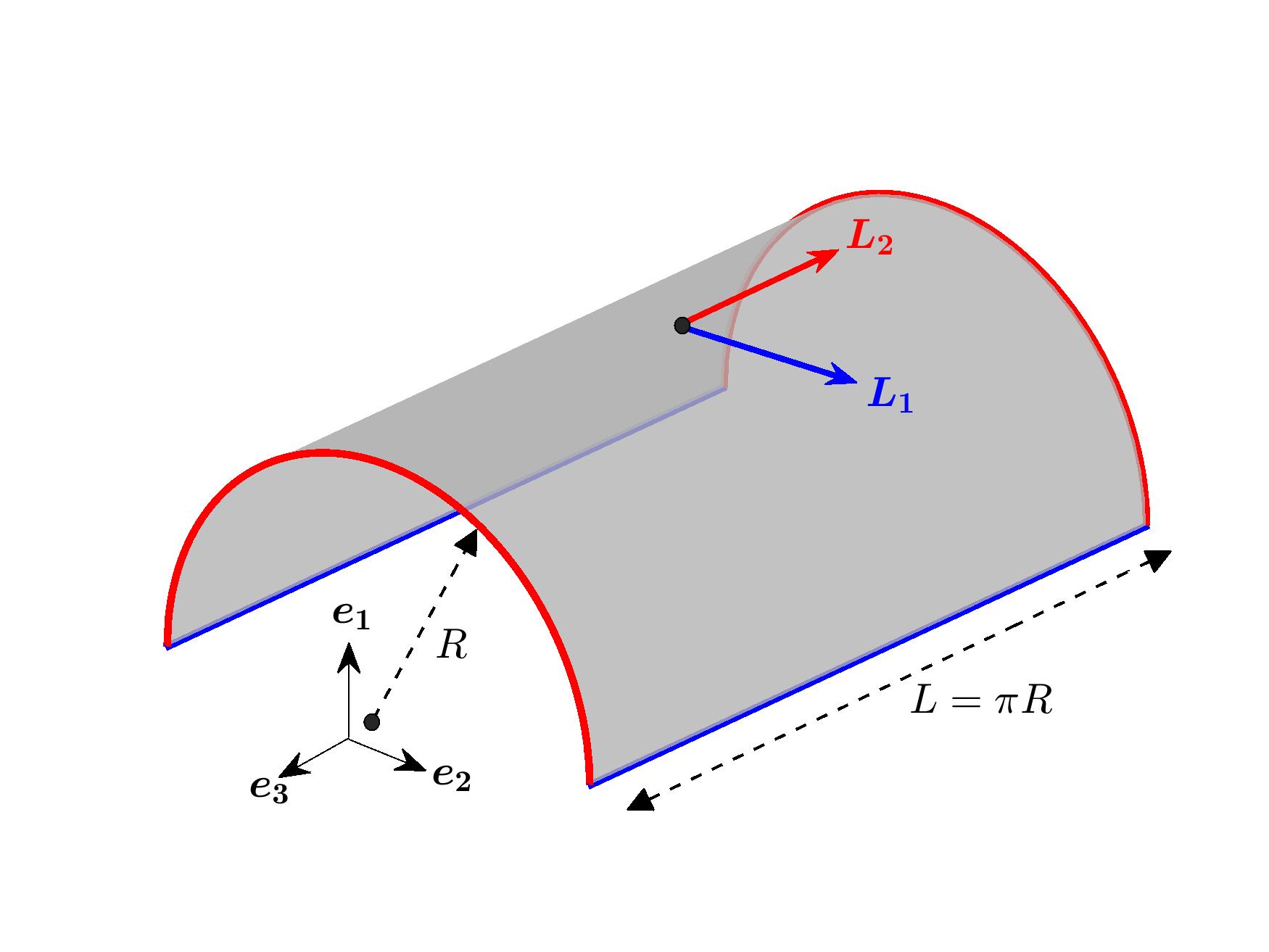}};
		\node[inner sep=0pt] (b) at (-1.9,3) {\includegraphics[trim={5cm 6cm 5cm 9cm},clip,scale=0.13]{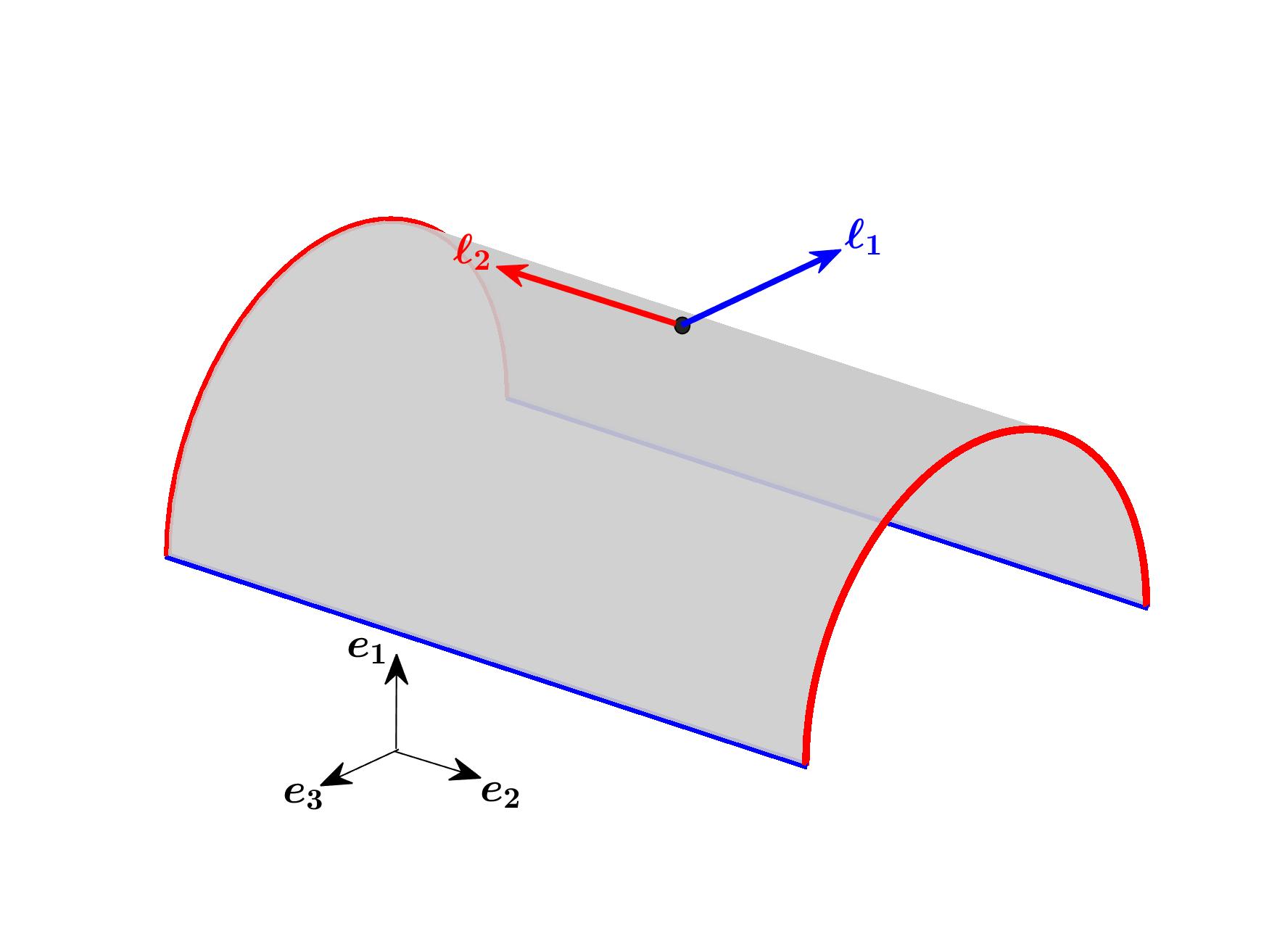}};
		\draw [very thick,  -triangle 60, -triangle 60] plot[smooth, tension=.7] coordinates {(-7.7,3.5) (-4.8,3.5)};
		\node[inner sep=0pt] (a) at (-12,0.8) {Initial configuration};
		\node[inner sep=0pt] (a) at (-0.2,0.8) {Final configuration};
		\node[inner sep=0pt] (a) at (-7.6,4) {(a)};
		\node[inner sep=0pt] (a) at (-6.1,4) {Rigid rotation};		\end{tikzpicture}
	%		\caption{Rigid reformation}
	\centering\\[1cm]
	\begin{tikzpicture}
		\node[inner sep=0pt] (a) at (-11,3) {\includegraphics[trim={5cm 6cm 5cm 9cm},clip,scale=0.13]{fig/test_1ini.jpg}};
		\node[inner sep=0pt] (b) at (-1.9,3) {\includegraphics[trim={5cm 6cm 5cm 9cm},clip,scale=0.13]{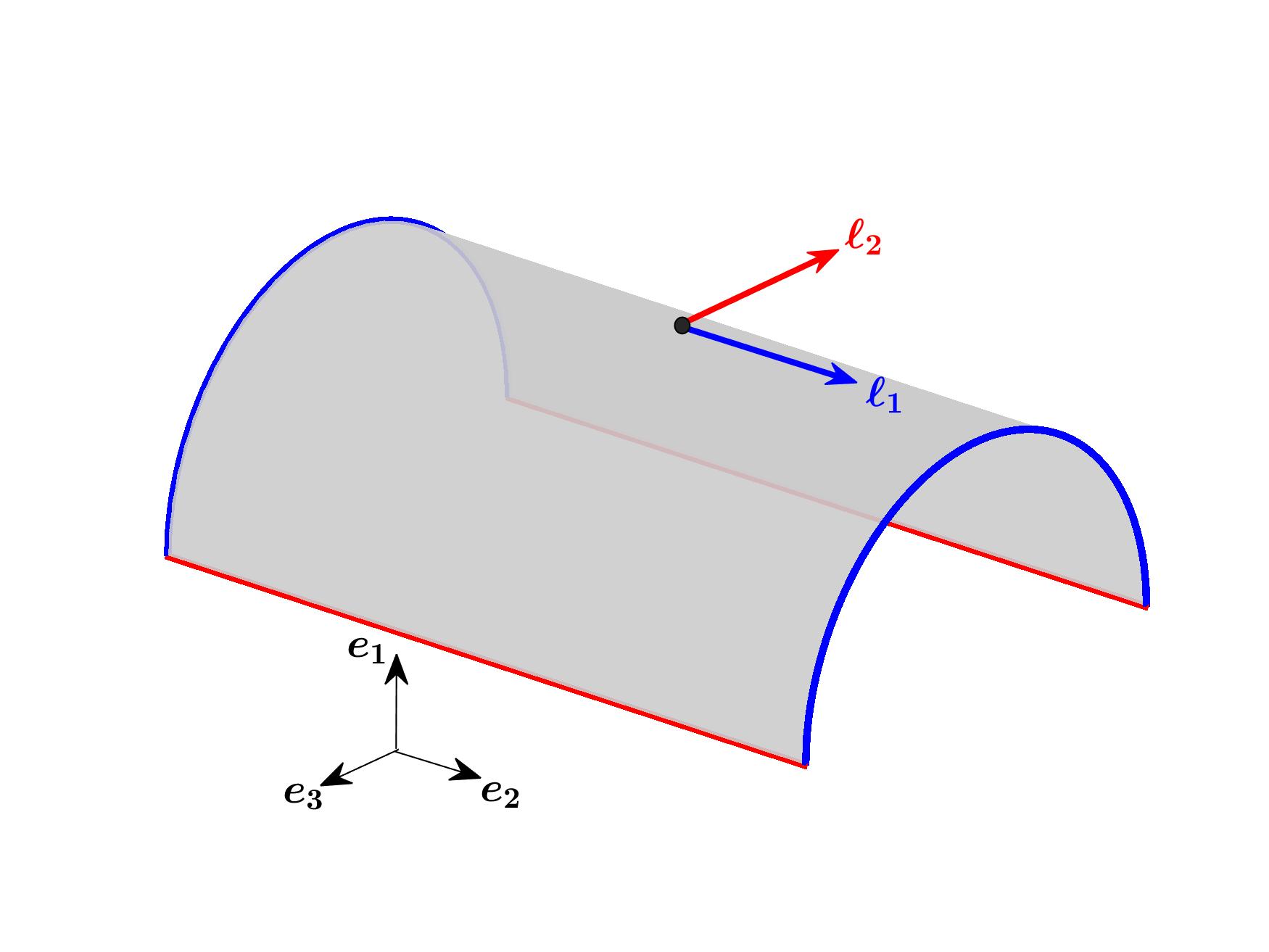}};
		\draw [very thick,  -triangle 60, -triangle 60] plot[smooth, tension=.7] coordinates {(-7.7,3.5) (-4.8,3.5)};
		\node[inner sep=0pt] (a) at (-12,0.5) {Initial configuration};
		\node[inner sep=0pt] (a) at (-0.2,0.5) {Final configuration};
		\node[inner sep=0pt] (a) at (-7.65,4) {(b)};
		\node[inner sep=0pt] (a) at (-6.,4) {Counter bending};
	\end{tikzpicture}
	\caption{Test case 1: Initial and final configuration for (a) rigid body rotation and (b) counter bending.}
	\label{fig:test1}
\end{figure}

Given the parametrization of Eq.~\eqref{eq:surfX}, the current configuration of the half tube rotated by $ 90^{\circ} $ around the $ \be_1 $-axis (Fig.~\ref{fig:test1}a) is described by
\begin{align}
\bx\,=\,R\,\cos \theta\,\be_1\,+\,R\,\phi\,\be_2\,-\,R\,\sin{\theta}\,\be_3\,.
\end{align}
The tangents, normal, surface metric and curvature tensor components thus are 
\begin{gather}
\ba_1\,=\,-R\,\sin{\theta}\,\be_1\,-\,R\,\cos{\theta}\,\be_3\,,\quad \ba_2\,=\,R\,\be_2\,,\quad\bn\,=\,\cos{\theta}\,\be_1\,-\,\sin{\theta}\,\be_3\,,
\end{gather}
and	
\vspace*{-5mm}
\begin{gather}
[\auab]\,=\,R^2\,\left[ {\begin{array}{cc}
	1 & 0 \\
	0 & 1 \\
	\end{array} }\right]\,=\,[\Auab]\,,\quad [b_{\alpha\beta}]\,=\,-R\,\left[ {\begin{array}{cc}
	1 & 0 \\
	0 & 0 \\
	\end{array} }\right]\,=\,[B_{\alpha\beta}]\,.
\end{gather}

\begin{table}[H]
	\small
	\centering
	\rotatebox{-90}{
		\begin{tabular}{|c!{\vrule width 1.1pt} c|c|c|c|c|}
			\hline
			\rotatebox[origin=c]{90}{$ N^{\alpha}_{\beta} $ }   & \rotatebox[origin=c]{90}{$ 0 $}       & \rotatebox[origin=c]{90}{ $\dfrac{c\,\kappa_{1}^2}{2}\left[ {\begin{array}{cc}
					-1 & 0 \\
					0 & 1 \\
					\end{array} }\right]$ }       &  \rotatebox[origin=c]{90}{ $0$ }        & \rotatebox[origin=c]{90}{$ 0 $}        & \rotatebox[origin=c]{90}{$ 0 $}    \\ \hline
			\rotatebox[origin=c]{90}{$ M^{\alpha}_{\beta\,0} $ }  &   \rotatebox[origin=c]{90}{$ 0 $}     & \rotatebox[origin=c]{90}{ $c\,\kappa_{1}\left[ {\begin{array}{cc}
					1 & 0 \\
					0 & 0 \\
					\end{array} }\right]$ }        & \rotatebox[origin=c]{90}{ $c\,\kappa_{1}\,{\begin{bmatrix}
					0 & 0 \\
					0 &  -1 \\
					\end{bmatrix}}$ }         &    \rotatebox[origin=c]{90}{$ 0 $}     & \rotatebox[origin=c]{90}{$ 0 $}    \\ \hline
			\rotatebox[origin=c]{90}{$ \tau^{\alpha}_{\beta} $ } &  \rotatebox[origin=c]{90}{$ 0 $}      &  \rotatebox[origin=c]{90}{ $ \dfrac{c\,\kappa_{1}^2}{2}\,\left[ {\begin{array}{cc}
					-3 & 0 \\
					0 &  1\\
					\end{array} }\right] $}      &   \rotatebox[origin=c]{90}{ $0$ }       & \rotatebox[origin=c]{90}{$ 0 $}        &  \rotatebox[origin=c]{90}{$ 0 $}   \\ \noalign{\hrule height 1.1pt}
			& \rotatebox{90}{\rotatebox{90}{Koiter} } & \rotatebox{90}{\rotatebox{90}{Canham} } & \rotatebox{90}{\rotatebox{90}{Helfrich} } & \rotatebox{90}{\rotatebox{90}{apH} } & \rotatebox{90}{ \rotatebox{90}{new} } \\ \hline
	\end{tabular}}
	\caption{Membrane and bending stress components for test case 1(a) -- rigid body rotation -- according to various bending models. For initially stress-free shells, all components should be zero in this test case, which is only satisfied by the Koiter, apH and new model.}
	\label{tab:1}
\end{table}
As the body is not deforming, i.e.~$ \auab\, =\, \Auab $ and $ b_{\alpha\beta}\,=\,B_{\alpha\beta} $, there should be no membrane or bending stresses induced. This can be verified by plugging the current stretch and curvature along the principal curvature directions,
\begin{gather}
	H\,=\,-\dfrac{1}{2\,R}\,,\quad \kappa\,=\,0\,,\\
	\lambda_{1}\,=\,\lambda_{2}\,=\,1\,,\\
	\kappa_{1}\,=\,-\frac{1}{R}\,=\,\kappa_{01}\,,\quad \kappa_{2}\,=\,0\,,
\end{gather}
into the equations of Sec.~\ref{sec:3}. The results are enumerated in Table~\ref{tab:1}. As seen, for the Koiter, apH and proposed new model no membrane or bending stresses are introduced. However, for the Canham model we obtain both non-zero membrane and bending stresses and for the Helfrich model, a non-zero bending stress. This is because for the considered initial configuration, these two models are not completely stress-free. The Helfrich bending model only satisfies this test for $ \bar{k}\,=\,0 $ as otherwise the bending stresses $ M^{\alpha\beta}_{0\,\mathrm{Hel}} $ becomes non-zero even when $ \Delta H \,=\,0$ as seen in Eq.~(\ref{eq:helfeq}.2).
%~~~~~~~~~~~~~~~~~~~~~~~~~~~~~~~~~~~~~~~~~~~~~~~~~~~~~~~~~~~~~~~~~~~~~~~~~~~~~~~~~~~~~~~~%
\subsubsection{Counter bending}

A similarly looking final configuration as before can be obtained by the counter bending shown in Fig.~\ref{fig:test1}b. In this case the current configuration is given by  
\begin{align}
\bx\,=\,R\,\sin\phi\,\be_1\,-\,R\,\theta\,\be_2\,+\,R\,\cos{\phi}\,\be_3\,.\label{eq:xT1b}
\end{align}
The required kinematic quantities then are
\begin{gather}
\ba_1\,=\,-R\,\be_2\,,\quad\ba_2\,=\,R\,\cos{\phi}\,\be_1\,-\,R\,\sin{\phi}\,\be_3\,,\quad\bn\,=\,\sin{\phi}\,\be_1\,+\,\cos{\phi}\,\be_3\,,\\
[\auab]\,=\,R^2\,\left[ {\begin{array}{cc}
	1 & 0 \\
	0 & 1 \\
	\end{array} }\right]\,=\,[\Auab]\,,\\ [b_{\alpha\beta}]\,=\,-R\,\left[ {\begin{array}{cc}
	0 & 0 \\
	0 & 1 \\
	\end{array} }\right]\,,\quad [b^{\alpha}_{\beta}]\,=\,-\dfrac{1}{R}\,\left[ {\begin{array}{cc}
	0 & 0 \\
	0 & 1\\
	\end{array} }\right]\,,\quad [b^{\alpha\beta}]\,=\,-\dfrac{1}{R^3}\,\left[ {\begin{array}{cc}
	0 & 0 \\
	0 & 1 \\
	\end{array} }\right]\,,\\
H\,=\,-\dfrac{1}{2\,R}\,,\quad \kappa\,=\,0\,,\\
\lambda_{1}\,=\,\lambda_{2}\,=\,1\,,
\end{gather}
and
\vspace*{-5mm}
\begin{gather}
\kappa_{1}\,=\,0\,,\quad\,\kappa_{2}\,=\,-\frac{1}{R}\,=\,\kappa_{01}\,.
\end{gather}
Unlike the rigid body rotation considered before, the bending described by Eq.~\eqref{eq:xT1b} generates non-zero bending stress components. Given the nature of the final configuration in this test case, the bending stresses induced here should satisfy $ M^{1}_1\,=\,-M^{2}_2 $ for isotropic materials, while the Cauchy stress components, $ N^\alpha_\beta $, should still be zero. The membrane and bending stress components of the five bending models are shown in Tab.~\ref{tab:2}. The Canham model neither provides accurate bending stresses nor Cauchy stresses. While the Helfrich model gives accurate Cauchy stresses but erroneous bending stresses, the apH and Koiter models give accurate bending stresses but not Cauchy stresses. Only if $ \kappa_2 $ is small (corresponding to small deformations) $ \kappa_2^2 $ becomes negligible, and hence satisfactory Cauchy stresses are obtained. Thus the Koiter and apH model can be argued to partially pass the test. But only the new model fully passes the test. The test also illustrates its anisotropic nature: If the bending parameter $ c_1\,\neq\,c_2 $ then $ M^1_1/(c_1\,-\,c_{12})\,=\,-\,M^2_2/(c_2\,-\,c_{12})$.

\begin{table}[H]
	\small
	\centering
	\rotatebox{-90}{
		\begin{tabular}{|c!{\vrule width 1.1pt} c|c|c|c|c|}
			\hline
			\rotatebox[origin=c]{90}{$ N^{\alpha}_{\beta} $ }   & \rotatebox[origin=c]{90}{$ c\,\kappa_{2}^2\,\left[ {\begin{array}{cc}
					0 & 0 \\
					0 &  1\\
					\end{array} }\right] $}       & \rotatebox[origin=c]{90}{ $\dfrac{c\,\kappa_{2}^2}{2}\,\left[ {\begin{array}{cc}
					1 & 0 \\
					0 & -1 \\
					\end{array} }\right]$ }       &  \rotatebox[origin=c]{90}{ $0$ }        & \rotatebox[origin=c]{90}{$ c\,\kappa_{2}^2\,\left[ {\begin{array}{cc}
					0 & 0 \\
					0 &  1\\
					\end{array} }\right] $}        & \rotatebox[origin=c]{90}{$ 0 $}    \\ \hline
			\rotatebox[origin=c]{90}{$ M^{\alpha}_{\beta\,0} $ }  &   \rotatebox[origin=c]{90}{$ c\,\kappa_{2}\,\left[ {\begin{array}{cc}
					-1 & 0 \\
					0 & 1 \\
					\end{array} }\right] $}     & \rotatebox[origin=c]{90}{ $c\,\kappa_{2}\,\left[ {\begin{array}{cc}
					0 & 0 \\
					0 & 1 \\
					\end{array} }\right]$ }        & \rotatebox[origin=c]{90}{ $c\,\kappa_{2}\,\left[{\begin{array}{cc}
					-1 & 0 \\
					0 &  0 \\
					\end{array} }\right]$ }         &    \rotatebox[origin=c]{90}{$ c\,\kappa_{2}\,\left[ {\begin{array}{cc}
					-1 & 0 \\
					0 & 1 \\
					\end{array} }\right] $}     & \rotatebox[origin=c]{90}{\hspace*{1mm}$\left(c-c_{12}\right)\kappa_{2}\,\left[ {\begin{array}{cc}
					-1 & 0 \\
					0 & 1\\
					\end{array} }\right]$}    \\ \hline
			\rotatebox[origin=c]{90}{$ \tau^{\alpha}_{\beta} $ } &  \rotatebox[origin=c]{90}{$ 0 $}      &  \rotatebox[origin=c]{90}{ $\dfrac{c\,\kappa_{2}^2}{2}\,\left[ {\begin{array}{cc}
					1 & 0 \\
					0 &  -3\\
					\end{array} }\right] $}      &   \rotatebox[origin=c]{90}{ $0$ }       & \rotatebox[origin=c]{90}{$ 0 $}        &  \rotatebox[origin=c]{90}{\hspace*{1mm}$ (c_{12}-c)\,\kappa_{2}^2\,\left[ {\begin{array}{cc}
					0 & 0 \\
					0 & 1\\
					\end{array} }\right] $}   \\ \noalign{\hrule height 1.1pt}
			& \rotatebox{90}{\rotatebox{90}{Koiter} } & \rotatebox{90}{\rotatebox{90}{Canham} } & \rotatebox{90}{\rotatebox{90}{Helfrich} } & \rotatebox{90}{\rotatebox{90}{apH} } & \rotatebox{90}{ \rotatebox{90}{new} } \\ \hline
	\end{tabular}}
	\caption{Membrane and bending stress components for test case 1(b) -- counter bending -- according to various bending models. In this case, the Cauchy stresses $ N^{\alpha}_{\beta} $ should be zero, while $ M^{1}_1$ should be equal to $-M^{2}_2 $, which is only satisfied by the new model.}
	\label{tab:2}
\end{table}

%----------------------------------------------------------------------------------------%
\subsection{Test case 2 -- bending vs.~stretching}\label{subsec:t2}

\begin{figure}[H]
	\small
	\centering
	\begin{tikzpicture}
	\node[inner sep=0pt] (a) at (-11,3) {\includegraphics[trim={5cm 6cm 5cm 8cm},clip,scale=0.14]{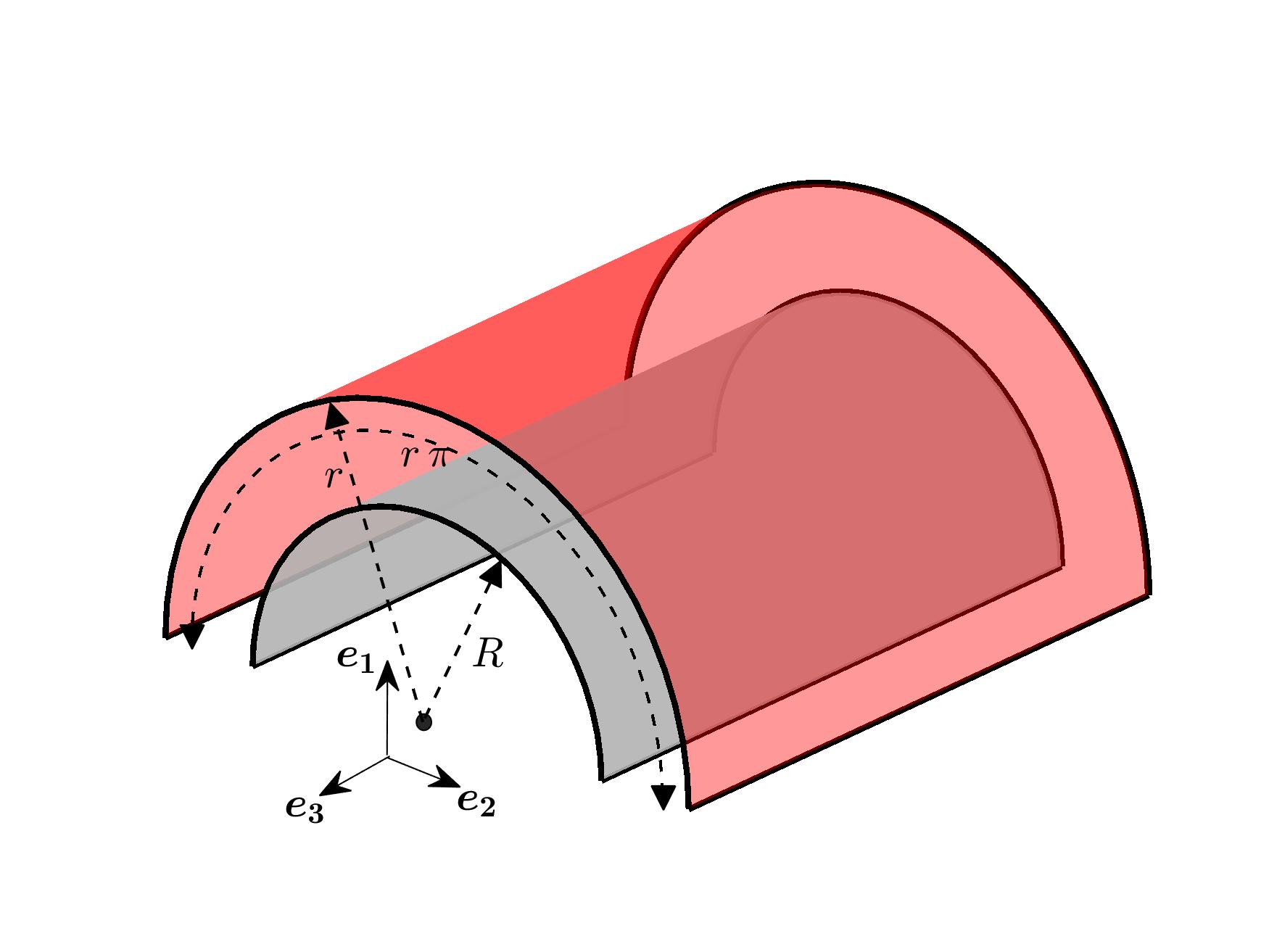}};
	\node[inner sep=0pt] (b) at (-3.5,3) {\includegraphics[trim={5cm 6cm 5cm 7cm},clip,scale=0.14]{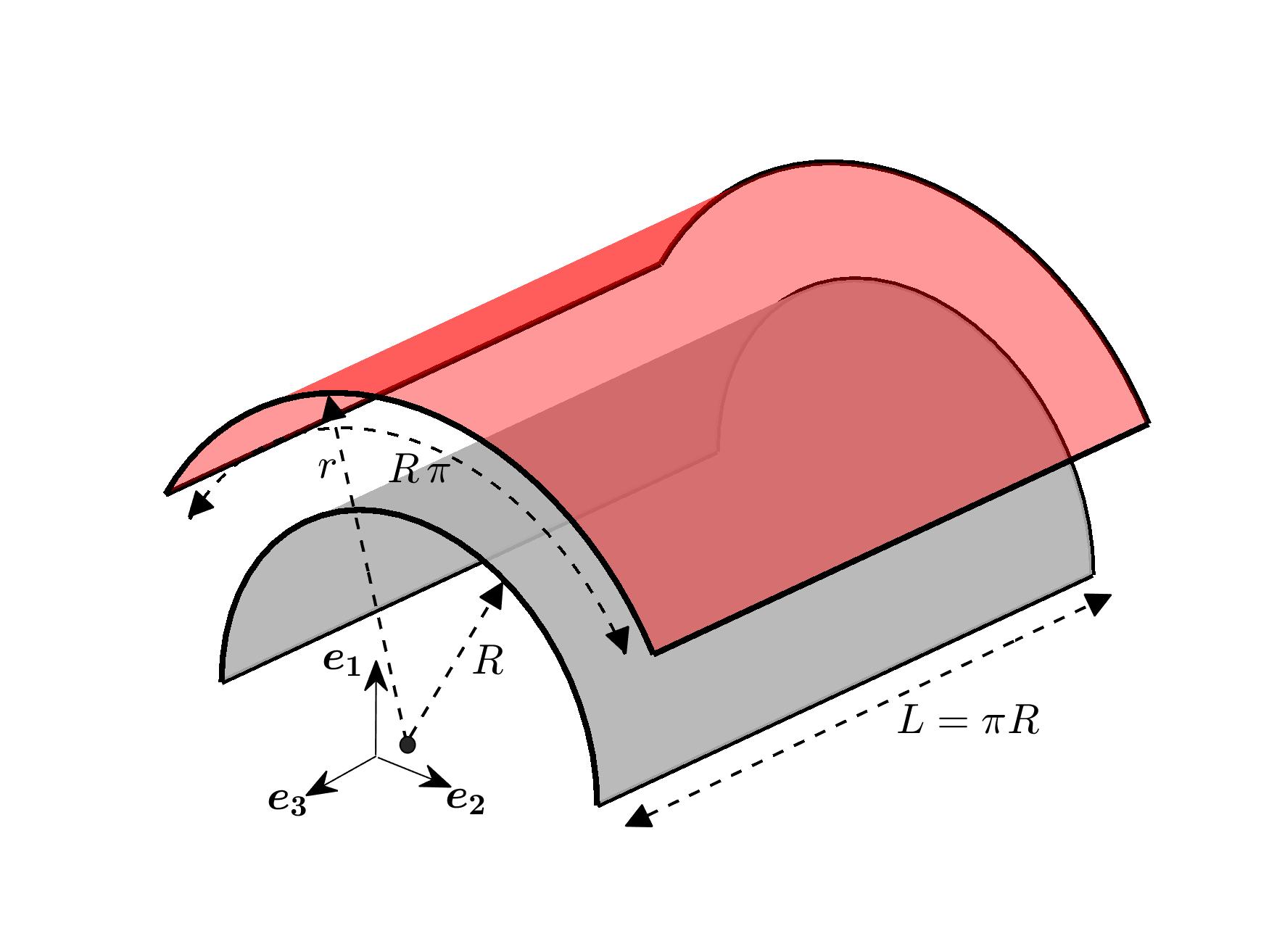}};
	\node[inner sep=0pt] (a) at (-12,0.15) {(a) Inflation};
	\node[inner sep=0pt] (a) at (-3.5,0.15) {(b) Pure bending};
	\end{tikzpicture}
	\caption{Test case 2: initial (grey) and final (red) configurations for (a) inflation and (b) pure bending. }
	\label{fig:test2}
\end{figure}

Test case 2 also examines the behavior of two similarly looking yet different deformations. They are now obtained either by inflation or bending, as Fig.~\ref{fig:test2} shows.

%~~~~~~~~~~~~~~~~~~~~~~~~~~~~~~~~~~~~~~~~~~~~~~~~~~~~~~~~~~~~~~~~~~~~~~~~~~~~~~~~~~~~~~~~%
\subsubsection{Inflation}\label{subsubsec:inf}

If the half tube is inflated to have radius $ r $ (Fig.~\ref{fig:test2}a), the current configuration is given by
\begin{align}
\bx\,=\,r\,\be_r\,+\,R\,\phi\,\be_3\,,
\end{align} 
based on Eqs.~\eqref{eq:surfX} and \eqref{eq:er}.
While the components of the curvature tensor are the same as in Eq.~\eqref{eq:iniBab}, but with $ R $ replaced by $ r $, the surface metric now is
\begin{gather}
[\auab]\,=\,\left[ {\begin{array}{cc}
	r^2 & 0 \vspace*{2mm}\\
	0 &  R^2\\
	\end{array} }\right]\,.
\end{gather} 
The stretch and curvature terms are then of the form
\begin{gather}
H\,=\,-\dfrac{1}{2\,r}\,,\quad\,\kappa\,=\,0\,,\\
\lambda_1\,=\,\frac{r}{R}\,,\quad\lambda_2\,=\,1\,,
\end{gather}
and
\vspace*{-5mm}
\begin{gather}
\kappa_{1}\,=\,-\frac{1}{r}\,,\quad\kappa_2\,=\,0\,.
\end{gather}
The resulting membrane and bending stress components for test case 2(a) are listed in Tab.~\ref{tab:3}. The deformation considered here is a pure membrane action, and hence any stress resultant contribution should only stem from the membrane part but not from the bending part of the material model. This is achieved by our new bending model as it does not introduce any membrane and bending stresses even though the surface curvature changes. Also the apH model achieves this for $ \tauab $ (which comes from the membrane stiffness $ \mu $), but not for $ M^\alpha_\beta $ and $ N^\alpha_\beta $ (which are affected by bending stiffness $ c $). All other models show unphysical stresses in all components. In case of the Koiter and apH model, those vanish in the limit $ r \rightarrow R $ , i.e., for small $ \Delta\kappa_{1}\,:=\,\kappa_{1}\,-\,\kappa_{01} $. In case of the Canham and Helfrich bending models, non-zero membrane and bending stresses remain even for small $ \Delta\kappa_1 $.  
\begin{table}[H]
	\small
	\centering
	\begin{adjustbox}{width=\columnwidth,center}
		\rotatebox{-90}{
			\begin{tabular}{|c!{\vrule width 1.1pt}c|c|c|c|c|}
				\hline
				\rotatebox[origin=c]{90}{$ N^{\alpha}_{\beta} $ }   & \rotatebox[origin=c]{90}{$ c\,\Delta\kappa_{1}\dfrac{\kappa_{01}^2}{\kappa_{1}}\,{\begin{bmatrix}
							1& 0 \\
							0 & 0 \\
					\end{bmatrix} } $}       & \rotatebox[origin=c]{90}{ $\dfrac{c\,\kappa_{1}^2}{2}\,\left[{\begin{array}{cc}
							-1 & 0 \\
							0 & 1 \\
					\end{array}}\right]$ }       &  \rotatebox[origin=c]{90}{ $\dfrac{-c\,\Delta\kappa_{1}}{2}\,\left[{\begin{array}{cc}
							\kappa_{01}\,+\,\kappa_{1} & 0 \\
							0 & -\Delta\kappa_{1} \\
					\end{array} }\right]$ }        & \rotatebox[origin=c]{90}{\begin{tabular}{c}$\dfrac{-\Delta\kappa_{1}}{\kappa_{01}^3\kappa_{1}}\left[{\begin{array}{cc}
								\mu e - cd\kappa_{1} & 0 \\
								0 &  \mu\kappa_{1}^2\left(\kappa_{01}+\kappa_{1}\right)
						\end{array}}\right]$ \end{tabular}}      & \rotatebox[origin=c]{90}{$ 0 $}    \\ \hline
				\rotatebox[origin=c]{90}{$ M^{\alpha}_{\beta\,0} $ }  &   \rotatebox[origin=c]{90}{$ c\,\Delta\kappa_{1}\dfrac{\kappa_{01}^3}{\kappa_{1}^3}\,\left[{\begin{array}{cc}
							-1& 0 \\
							0 & 0 \\
					\end{array}}\right]$}     & \rotatebox[origin=c]{90}{ $c\,J\,\kappa_{1}\,\left[{\begin{array}{cc}
							1 & 0 \\
							0 & 0 \\
					\end{array}}\right]$ }        & \rotatebox[origin=c]{90}{ $J\,c\,\left[{\begin{array}{cc}
							\Delta\kappa_{1} & 0 \\
							0 &  -\kappa_{01}\\
					\end{array} }\right]$ }         &    \rotatebox[origin=c]{90}{ \begin{tabular}{c}$\dfrac{c\,\Delta\kappa_1}{\kappa_{01}^2}\left[{\begin{array}{cc}
								\dfrac{d}{\kappa_{1}^2} & 0 \\
								0 &  \kappa_{1}^2 \\
						\end{array} }\right]$\\ \\ $\text{ with}\quad d:=2\kappa_{1}^4+\kappa_{01}e$
				\end{tabular} }     & \rotatebox[origin=c]{90}{$ 0 $}    \\ \hline
				\rotatebox[origin=c]{90}{$ \tau^{\alpha}_{\beta} $ } &  \rotatebox[origin=c]{90}{$ 0 $}      &  \rotatebox[origin=c]{90}{ $\dfrac{c\,J\,\kappa_{1}^2}{2}\, \left[{\begin{array}{cc}
							-3& 0 \\
							0 &  1\\
					\end{array} }\right] $}      &   \rotatebox[origin=c]{90}{ $\dfrac{-J\,c\,\Delta\kappa_{1}}{2}\, \left[{\begin{array}{cc}
							\kappa_{01}\,+\,3\,\kappa_{1} & 0 \\
							0 &  -\Delta\kappa_{1} \\
					\end{array} }\right]$ }       & \rotatebox[origin=c]{90}{\begin{tabular}{c}$ ~\dfrac{-\mu\,\Delta\kappa_{1}}{\kappa_{01}^2}\, \left[{\begin{array}{cc}
								\dfrac{e}{\kappa_{1}^2}& 0 \\
								0 &  \kappa_{01}\,+\,\kappa_{1} \\
						\end{array} }\right]$\\ \\ $\text{ with}\quad e:=(\kappa_{01}+\kappa_1)(\kappa^2_{01}+\kappa^2_1)$\end{tabular}}        &  \rotatebox[origin=c]{90}{$ 0 $}   \\ \noalign{\hrule height 1.1pt}
				& \rotatebox{90}{\rotatebox{90}{Koiter} } & \rotatebox{90}{\rotatebox{90}{Canham} } & \rotatebox{90}{\rotatebox{90}{Helfrich} } & \rotatebox{90}{\rotatebox{90}{apH} } & \rotatebox{90}{ \rotatebox{90}{new} } \\ \hline
		\end{tabular}}
	\end{adjustbox}
	\caption{Membrane and bending stress components for test case 2(a) -- inflation -- according to various bending models. In this case, all bending induced membrane and bending stresses ($ \tau^{\alpha}_{\beta} $, $ M^{\alpha}_{\beta} $ and $ N^{\alpha}_{\beta} $) should be zero, which is only achieved by the new bending model. $ \Delta\kappa_{1}\,:=\,\kappa_{1}\,-\,\kappa_{01} $ is introduced for simplification.}
	\label{tab:3}
\end{table}
%~~~~~~~~~~~~~~~~~~~~~~~~~~~~~~~~~~~~~~~~~~~~~~~~~~~~~~~~~~~~~~~~~~~~~~~~~~~~~~~~~~~~~~~~%
\subsubsection{Pure bending}\label{subsubsec:ben}

In the second case, the half tube undergoes pure bending to have the radius $ r $ and length $ l $ without any change in the arc length (Fig.~\ref{fig:test2}b). The current configuration, parametrized by Eq.~\eqref{eq:surfX}, is now described by
\begin{align}
\bx\,=\,r\,\be_{\tilde{r}}\,+\,R\,\phi\,\be_{3}\,,
\end{align}
where
\begin{align}
\be_{\tilde{r}}\,&:=\,\cos{\left(\frac{R\theta}{r}\right)}\,\be_1\,+\,\sin{\left(\frac{R\theta}{r}\right)}\,\be_2\,,\\
\be_{\tilde{\theta}}\,&:=\,-\sin{\left(\frac{R\theta}{r}\right)}\,\be_1\,+\,\cos{\left(\frac{R\theta}{r}\right)}\,\be_2\,.
\end{align}  
The corresponding tangents and normal vectors then follow as
\begin{gather}
\ba_1\,=\,R\,\be_{\tilde{\theta}}\,,\quad\ba_2\,=\,R\,\be_3\,,\quad\bN\,=\,\be_{\tilde{r}}\,.
\end{gather}
Further, the  surface metric and curvature tensor components become
\begin{gather}
	[\auab]\,=\,R^2\left[ {\begin{array}{cc}
			1 & 0 \\
			0 & 1 \\
	\end{array} }\right]\,=\,[\Auab]\,,\\
	[\buab]\,=\,-\dfrac{R^2}{r}\,\left[ {\begin{array}{cc}
			1 & 0 \\
			0 & 0 \\
	\end{array} }\right]\,,\quad [b^{\alpha}_{\beta}]\,=\,-\dfrac{1}{r}\,\left[ {\begin{array}{cc}
			1 & 0 \\
			0 & 0 \\
	\end{array} }\right]\,,\quad [b^{\alpha\beta}]\,=\,-\dfrac{1}{r\,R^2}\,\left[ {\begin{array}{cc}
			1 & 0 \\
			0 & 0 \\
	\end{array} }\right]\,.
\end{gather}
With this the stretch and curvature can be calculated as
\begin{gather}
	H\,=\,-\dfrac{1}{2\,r}\,,\quad\,\kappa\,=\,0\,,\\
	\lambda_1\,=\,\lambda_2\,=\,1\,,
\end{gather}
and
\vspace*{-5mm}
\begin{gather}
	\kappa_{1}\,=\, -\frac{1}{r},\quad \kappa_{2}\,=\,0\,.
\end{gather}
The resulting membrane and bending stress components for test case~2(b) are listed in Tab.~\ref{tab:4}. The Cauchy stress components $ N^{\alpha}_{\beta} $ should be zero, and only our new bending model accurately captures this. 
%\vspace*{-10mm}
\begin{table}[H]
	\small
	\centering
	\rotatebox{-90}{
		\begin{tabular}{|c!{\vrule width 1.1pt}c|c|c|c|c|}
			\hline
			\rotatebox[origin=c]{90}{$ N^{\alpha}_{\beta} $ }   & \rotatebox[origin=c]{90}{$ c\,\Delta\kappa_{1}\,\kappa_{1}\,\left[ {\begin{array}{cc}
						1& 0 \\
						0 & 0 \\
				\end{array} }\right] $}       & \rotatebox[origin=c]{90}{ $\dfrac{c\,\kappa_{1}^2}{2}\,\left[ {\begin{array}{cc}
						-1 & 0 \\
						0 & 1 \\
				\end{array} }\right]$ }       &  \rotatebox[origin=c]{90}{ $\dfrac{-c\,\Delta\kappa_{1}}{2}\left[{\begin{array}{cc}
						\kappa_{01}\,+\,\kappa_{1} & 0 \\
						0 & -\Delta\kappa_{1} \\
				\end{array}}\right]$ }        & \rotatebox[origin=c]{90}{  $ 2\,c\,\kappa_{1}\,\Delta\kappa_{1}\left[ {\begin{array}{cc}
						1 & 0 \\
						0 &  0 \\
				\end{array} }\right] $}        & \rotatebox[origin=c]{90}{$ 0 $}    \\ \hline
			\rotatebox[origin=c]{90}{$ M^{\alpha}_{\beta\,0} $ }  &   \rotatebox[origin=c]{90}{$ \hspace*{1mm}c\,\Delta\kappa_{1}\,\left[ {\begin{array}{cc}
						1& 0 \\
						0 & 0 \\
				\end{array} }\right] \hspace*{1mm}$}     & \rotatebox[origin=c]{90}{ $c\,\kappa_{1}\left[ {\begin{array}{cc}
						1 & 0 \\
						0 & 0 \\
				\end{array} }\right]$ }        & \rotatebox[origin=c]{90}{ $c\,\left[{\begin{array}{cc}
						\Delta\kappa_{1} & 0 \\
						0 &  -\kappa_{01} \\
				\end{array} }\right]$ }         &    \rotatebox[origin=c]{90}{$ c\,\Delta\kappa_{1}\left[ {\begin{array}{cc}
						2 & 0 \\
						0 &  1 \\
				\end{array} }\right]$}     & \rotatebox[origin=c]{90}{$\Delta\kappa_{1}\left[ {\begin{array}{cc}
						c & 0 \\
						0 & c_{12} \\
				\end{array} }\right] $}    \\ \hline
			\rotatebox[origin=c]{90}{$ \tau^{\alpha}_{\beta} $ } &  \rotatebox[origin=c]{90}{$ 0 $}      &  \rotatebox[origin=c]{90}{ $ \dfrac{c\,\kappa_{1}^2}{2}\,\left[ {\begin{array}{cc}
						-3 & 0 \vspace*{2mm}\\
						0 &  1\\
				\end{array} }\right] $}      &   \rotatebox[origin=c]{90}{ $\dfrac{-c\,\Delta\kappa_{1}}{2}\,\left[{\begin{array}{cc}
						3\,\kappa_{1}\,+\,\kappa_{01} & 0 \\
						0 &  -\Delta\kappa_{1} \\
				\end{array}}\right]$ }       & \rotatebox[origin=c]{90}{$ 0 $}        &  \rotatebox[origin=c]{90}{$ -c\,\Delta\kappa_{1}\kappa_{1}\left[ {\begin{array}{cc}
						1 & 0 \\
						0 & 0 \\
				\end{array} }\right] $}   \\ \noalign{\hrule height 1.1pt}
			\rotatebox{90}{\rotatebox{90}{} } & \rotatebox{90}{\rotatebox{90}{{Koiter}} } & \rotatebox{90}{\rotatebox{90}{Canham} } & \rotatebox{90}{\rotatebox{90}{Helfrich} } & \rotatebox{90}{\rotatebox{90}{apH} } & \rotatebox{90}{ \rotatebox{90}{new}} \\ \hline
	\end{tabular}}
	\caption{Membrane and bending stress components for test case 2(b) -- pure bending -- according to various bending models. $ N^{\alpha}_{\beta} $ should be zero in this case, which is only achieved by the new model. $ \Delta\kappa_{1}\,:=\,\kappa_{1}\,-\,\kappa_{01} $ is introduced to simplify the results.}
	\label{tab:4}
\end{table}

%----------------------------------------------------------------------------------------%
\subsection{Test case 3 - Torsion}\label{subsec:t3}

\begin{figure}[H]
	\small
	\centering
	\begin{tikzpicture}
	\node[inner sep=0pt] (a) at (-11,3) {\includegraphics[trim={0cm 0cm 0cm 0cm},clip,scale=0.55]{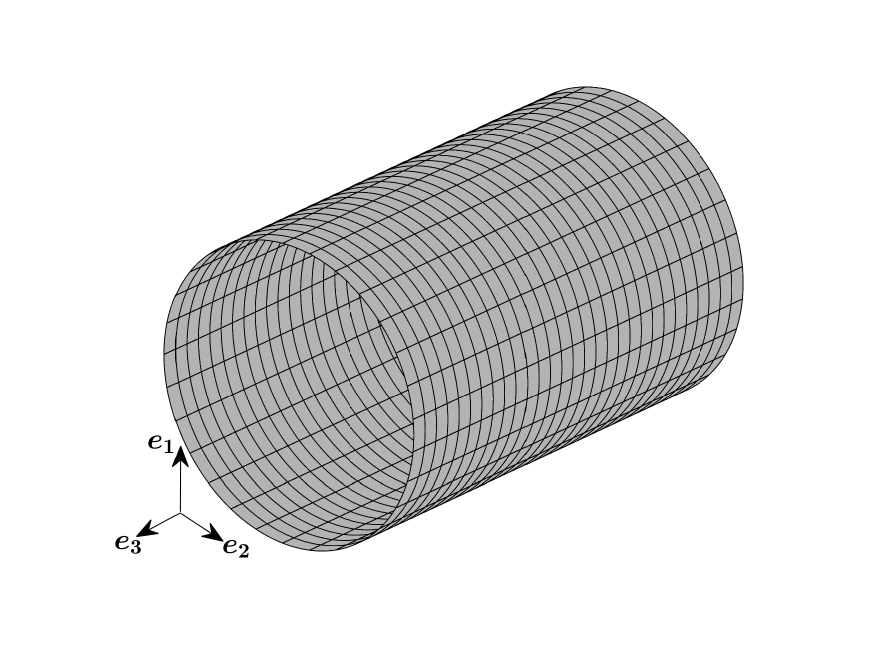}};
	\node[inner sep=0pt] (b) at (-3.5,3) {\includegraphics[trim={0cm 0cm 0cm 0cm},clip,scale=0.55]{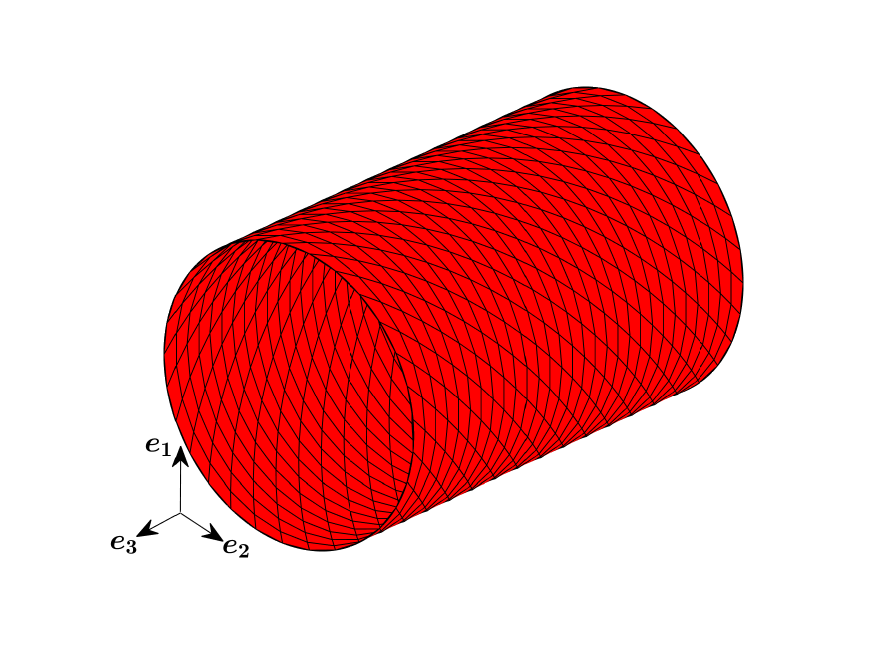}};
	\node[inner sep=0pt] (a) at (-12,0.5) {(a) Initial configuration};
	\node[inner sep=0pt] (a) at (-3.5,0.5) {(b) Current configuration};
	\end{tikzpicture}
	\caption{Torsion: Initial and final configuration.}
	\label{fig:test3}
\end{figure}

Unlike in the previous test cases, we now consider an open ended full cylinder and parametrize it using
\begin{align}
\theta\,\in\,\left[0,\,2\,\pi\right]\,,\,\text{and}\,\,\phi\,\in\,\left[0,\,\pi\right]\,.
\end{align}
The initial surface, shown in Fig.~\ref{fig:test3}a, can still be described using Eq.~\eqref{eq:iniX}. Twisting is then described by (see Fig.~\ref{fig:test3}b)
\begin{align}
\bx(\theta,\phi)\,=\, R\,\be_{\bar{r}}\,+\,R\,\,\phi\,\be_3\,,
\label{eq:tor}
\end{align}
where
\vspace*{-5mm}
\begin{align}
\be_{\bar{r}}\,&:=\,\cos{(\theta\,+\,\gamma\,\phi)}\,\be_1\,+\,\sin{(\theta\,+\,\gamma\,\phi)}\,\be_2\,,\\
\be_{\bar{\theta}}\,&:=\,-\sin{(\theta\,+\,\gamma\,\phi)}\,\be_1\,+\,\cos{(\theta\,+\,\gamma\,\phi)}\,\be_2\,.
\end{align}
Here $ \gamma $ is a constant that defines the extent of twisting. The tangent vectors and normal follow from Eq.~\eqref{eq:tor} as
\vspace*{-5mm}
\begin{align}
	\ba_1\,&=\,R\,\be_{\bar{\theta}}\,,\\
	\ba_2\,&=\,R\,\left(\be_{\bar{\theta}}\,+\,\be_3\right)\,,\\
	\bn\,&=\,\be_{\bar{r}}\,,
\end{align}
while the kinematic quantities become
\begin{gather}
	[\auab]\,=\,R^2\,\left[ {\begin{array}{cc}
			1 & \gamma \\
			\gamma & \left(\gamma^2\,+\,1\right) \\
	\end{array} }\right]\,,\quad [\aab]\,=\,\frac{1}{R^2}\,\left[ {\begin{array}{cc}
			\left(\gamma^2\,+\,1\right) & -\gamma \\
			-\gamma & 1 \\
	\end{array} }\right]\,,\\
	[\buab]\,=\,-R\,\left[ {\begin{array}{cc}
			1 & \gamma \\
			\gamma & \gamma^2 \\
	\end{array} }\right]\,,\quad[b^{\alpha}_{\,\beta}]\,=\,-\dfrac{1}{R}\left[ {\begin{array}{cc}
			1 & \gamma \\
			0 & 0 \\
	\end{array} }\right]\,,\quad 
	[b^{\alpha\beta}]\,=\,-\dfrac{1}{R^3}\,\left[ {\begin{array}{cc}
			1 & 0 \\
			0 & 0 \\
	\end{array} }\right]\,,\\
	H\,=\,-\dfrac{1}{2\,R}\,,\quad\kappa\,=\,0\,,\\
	\lambda_{1}\,=\,1\,,\quad\,\lambda_{2}\,=\,\sqrt{\gamma^2\,+\,1}\,,
\end{gather}
and
\vspace*{-5mm}
\begin{gather}
	\kappa_{1}\,=\,-\frac{1}{R}\,,\quad\kappa_{2}\,=\,-\frac{\gamma^2}{R\,\left(\gamma^2\,+\,1\right)}\,.
\end{gather}
We emphasize that the principal curvatures in the current configuration, $ \kappa^*_i $, which follow from Eq.~\eqref{eq:Pcurv} as
\vspace*{-5mm}
\begin{gather}
	\kappa_{1}^{*}\,=\,-\frac{1}{R}\,,\quad\kappa_{2}^{*}\,=\,0\,,
\end{gather}
will be different from $ \kappa_{i} $.

From the Fig.~\ref{fig:test3}, we know that $ M^{22}_0 $ should be non-zero, as the fibers along $ \xi^2=\phi $ are being bent. In contrast,  $ M^{11}_0 $ should be equal to zero, as there is no change in curvature of the fibers along $ \xi^1=\theta $. As shown in Tab.~\ref{tab:5}, only the Koiter model has $ M^{11}_0\,=\,0 $ and $ M^{22}_0 \,\ne\,0$. For the proposed new model, $ M^{22}_0 \,\ne\,0$ and if we consider $ \Lambda\,=\,0 $, which is also used for all the other bending models, $ M^{11}_0 $ will be zero according to Eq.~\eqref{eq:newp}. 
As torsion, in contrast to the previous test cases, causes in-plane deformation, we need to compare any membrane stresses coming from the bending model to those of the membrane part of the material model in order to further asses the accuracy of the model. For this, we consider the Neo-Hookean membrane model defined in Eq.~\eqref{eq:neohook}. Using Eq.~\eqref{eq:neohookstr}, the Neo-Hookean's membrane stress contribution will be 
\begin{gather}
	[\tau_{\mathrm{mNH}}^{\alpha\beta}]\,=\,\mu\,\kappa_{1}^2\left[ {\begin{array}{cc}
			-\gamma^2 & \gamma \\
			\gamma & 0 \\
	\end{array} }\right].
	\label{eq:torNeo}
\end{gather}
Let $ \Lambda\,=\,0 $, $ \gamma\,=\,1 $ and slenderness ratio $ R/T \,=\,100$. Then, using Eq.~\eqref{eq:newp}, \eqref{eq:torNeo} and $ \tau^{\alpha\beta}_{\mathrm{new}} $ in Tab.~\ref{tab:5}, the maximum absolute value of the effective membrane stress component of the new bending model ($ \tau^{11}_{\mathrm{new}} $) and the Neo-Hookean membrane model ($ \tau^{11}_{\mathrm{mNH}} $) can be compared. We find  
\begin{align}
	\tau^{\alpha\beta\,\mathrm{max}}_{\mathrm{new}}\,=\,10^{-5
	}\,\dfrac{\mu}{R^2}\,\ll\,\tau^{\alpha\beta\,\mathrm{max}}_{\mathrm{mNH}}\,=\,\dfrac{\mu}{R^2}\,.
	\label{eq:compare}
\end{align}
The membrane stress contribution coming from the new material model is thus negligible ($ 10^{5} $ times smaller) in comparison to the membrane stresses from the Neo-Hookean membrane model. Eq.~\eqref{eq:compare} is also true for the new bending model's Cauchy stresses $ \Nab_{\mathrm{new}} $. The Canham bending model's membrane and Cauchy stress contributions are also of the same order as the new model, as Tab.~\ref{tab:5} shows. The membrane stress contribution from the apH model is equal to Eq.~\eqref{eq:torNeo}, which is acceptable as it is from its membrane part Eq.~(\ref{eq:wbapH}.1). Additionally, the Cauchy stresses of the Koiter bending model ($ N^{\alpha\beta}_{\mathrm{bKo}} $) will also be $ 10^{5} $ times smaller than the Cauchy stresses of the Koiter membrane model ($ N^{\alpha\beta}_{\mathrm{mKo}}\,=\,\tau^{\alpha\beta}_{\mathrm{mKo}} $) according to Eq.~\eqref{eq:memkoi}. 
\begin{table}[H]
	\centering
	\scriptsize
	\begin{adjustbox}{width=0.9\linewidth,center}
		\resizebox{12cm}{!}{
			\rotatebox{-90}{
				\begin{tabular}{|c!{\vrule width 1.1pt}c|c|c|c|c|}
					\hline
					\rotatebox[origin=c]{90}{$ N^{\alpha}_{\beta} $ }   & \rotatebox[origin=c]{90}{$c\,\kappa_{1}^2\,\gamma\,\left[{\begin{array}{cc}
								\gamma& \gamma^2 \vspace*{1mm}\\
								\lambda_{2}^2 & \lambda_{2}\,\gamma_2 \\
						\end{array} }\right]$}       & \rotatebox[origin=c]{90}{ $\dfrac{c\,\kappa_{1}^2}{2}\,\left[{\begin{array}{cc}
								-1 & -2\,\gamma \\
								0 & 1 \\
						\end{array}}\right]$ }       &  \rotatebox[origin=c]{90}{ $0$ }        & \rotatebox[origin=c]{90}{$ \mu\,\left[ {\begin{array}{cc}
								0 & \gamma \\
								\gamma &  \gamma^2 \\
						\end{array} }\right] $}        & \rotatebox[origin=c]{90}{\hspace*{1mm}$ \dfrac{\gamma\,\kappa_{1}^2}{\lambda_2^4}\left[ {\begin{array}{cc}
								0 & 0 \\
								c\gamma^2+2c_3\lambda_{2} & 0 \\
						\end{array} }\right] $}    \\ \hline
					\rotatebox[origin=c]{90}{$ M^{\alpha\beta}_{0} $ }  &   \rotatebox[origin=c]{90}{$ c\,\kappa_{1}^3\, \left[{\begin{array}{cc}
								0& \gamma \\
								\gamma & \gamma^2 \\
						\end{array}}\right]$}     & \rotatebox[origin=c]{90}{ $c\,\kappa_{1}^3\,\left[ {\begin{array}{cc}
								1 & 0 \\
								0 & 0 \\
						\end{array} }\right]$ }        & \rotatebox[origin=c]{90}{ $c\,\kappa_{1}^3\,\left[{\begin{array}{cc}
								-\gamma^2& \gamma \\
								\gamma &  -1 \\
						\end{array}}\right]$}         &    \rotatebox[origin=c]{90}{$ 0 $}     & \rotatebox[origin=c]{90}{$\hspace*{1mm}\dfrac{\gamma\,\kappa_{1}^3}{\lambda_{2}}\left[ {\begin{array}{cc}
								c_{12}\,\gamma & 2\,c_3\vspace*{1mm} \\
								2\,c_3 & \dfrac{c\,\gamma}{\lambda_{2}}\\
						\end{array} }\right] $}    \\ \hline
					\rotatebox[origin=c]{90}{$ \tau^{\alpha\beta} $ } &  \rotatebox[origin=c]{90}{$ 0 $}      &  \rotatebox[origin=c]{90}{ $ \dfrac{c\,\kappa_{1}^4}{2}\,\left[ {\begin{array}{cc}
								\gamma^2\,-\,3 & -\gamma \\
								\gamma &  1\\
						\end{array} }\right] $}      &   \rotatebox[origin=c]{90}{ $0$ }       & \rotatebox[origin=c]{90}{$ \mu\,\kappa_{1}^2\left[ {\begin{array}{cc}
								-\gamma^2 & \gamma \\
								\gamma & 0 \\
						\end{array} }\right] $}        &  \rotatebox[origin=c]{90}{$ \hspace*{1mm}-\dfrac{\gamma^2\kappa_{1}^4}{\lambda_2}\left[ {\begin{array}{cc}
								c_{12}\,+\,2c_3 & 0\\
								0 & \dfrac{c\,\gamma^2}{\lambda_{2}^3}+\dfrac{2\,c_3}{\lambda_{2}^2}\\
						\end{array} }\right] $}   \\ \noalign{\hrule height 1.1pt}
					& \rotatebox{90}{\rotatebox{90}{Koiter} } & \rotatebox{90}{\rotatebox{90}{Canham} } & \rotatebox{90}{\rotatebox{90}{Helfrich} } & \rotatebox{90}{\rotatebox{90}{apH} } & \rotatebox{90}{ \rotatebox{90}{new}} \\ \hline
		\end{tabular}}}
	\end{adjustbox}
	\caption{Membrane and bending stress components for test case 3 -- torsion. This test case should satisfy $ M^{11}_0 \,=\,0$ and $ M^{22}_0 \ne\,0$, which is only achieved by the Koiter and the proposed new model (when $ c_{12}\,=\,0 $). The undesired membrane stress contributions from all bending models are much smaller than the stress contributions from a membrane model, even at large deformation.}
	\label{tab:5}
\end{table}
%----------------------------------------------------------------------------------------%
\subsection{Summary}

The preceding results are categorized as passing, failing, or partial passing the tests depending on the results of $ M^{\alpha}_{\beta} $ and $ N^{\alpha}_{\beta} $. This categorization is shown in Tab.~\ref{tab:6}. As seen, only the proposed new bending model passes all test cases. In the table, an additional row is added to report if the models allow for initially stress-free curved surfaces. This is not the case for the Canham and Helfrich model. As a result of not being initially stress-free, the Canham and Helfrich models also fail to provide satisfying results for $ M^{\alpha}_{\beta} $ or $ N^{\alpha}_{\beta} $ in any of the elementary test cases where the body undergoes large deformation. The Koiter and apH models perform similarly well for all the test cases apart from torsion. Both the models pass test case~1(a), but only pass test case~1(b), when $ \kappa_{2} $ is small, such that errors in $ N^{\alpha}_{\beta} $ become negligible. They also only pass test case~2(a), when $ \Delta\,\kappa_{1} $ is negligible, due to errors in $ M^{\alpha}_{\beta} $. In test case~2(b), the Koiter and apH models fail due to errors in both $ M^{\alpha}_{\beta} $ and $ N^{\alpha}_{\beta} $. Test case 3 is passed by the Koiter, but not the apH model due to incorrect bending stresses. 
\begin{table}[H]
	\small
	\centering
	\begin{tabular}{|l|l|c|c|c|c|c|}
		\hline
		&                        & Koiter & Canham & Helfrich & apH & new \\ \hline
		& &  &  &  &  &  \\ [-3.5mm]
		0                  & initially stress-free                 & {\color{blue}\checkmark}    & {\color{red}\xmark}    & {\color{red}\xmark}      & {\color{blue}\checkmark}             & {\color{blue}\checkmark} \\[0.5mm] \hline
		& &  &  &  &  &  \\ [-3.5mm]
		\multirow{2}{*}{1} & a. rigid body rotation & {\color{blue}\checkmark}    & {\color{blue}\checkmark}    & {\color{blue}\checkmark}     & {\color{blue}\checkmark}             & {\color{blue}\checkmark} \\[0.5mm] \cline{2-7} 
		& &  &  &  &  &  \\ [-3.5mm]
		& b. counter bending     & {[\color{black}\checkmark]}    & {\color{red}\xmark}    & {\color{red}\xmark}      & {[\color{black}\checkmark]}             & {\color{blue}\checkmark} \\[0.5mm] \hline
		& &  &  &  &  &  \\ [-3.5mm]
		\multirow{2}{*}{2} & a. inflation           & {\{\color{black}\checkmark\}}    & {\color{red}\xmark}    & {\color{red}\xmark}      & {\{\color{black}\checkmark\}}            & {\color{blue}\checkmark} \\ [0.5mm]\cline{2-7} 
		& &  &  &  &  &  \\ [-3.5mm]
		& b. pure bending        & {\color{red}\xmark}    & {\color{red}\xmark}    & {\color{red}\xmark}      & {\color{red}\xmark}             & {\color{blue}\checkmark} \\[0.5mm] \hline
		3                  & torsion                & {\color{blue}\checkmark}    & {\color{red}\xmark}     & {\color{red}\xmark}       & {\color{red}\xmark}             & {\color{blue}\checkmark} \\ \hline
	\end{tabular}
	\caption{Results of the elementary bending test cases: {\color{red}\xmark} = test failed for either $ M^{\alpha}_{\beta} $ or $ N^{\alpha}_{\beta} $, {\color{blue}\checkmark} = test passed for both $ M^{\alpha}_{\beta} $ and $ N^{\alpha}_{\beta} $, {[\color{black}\checkmark]} = test passed for $ M^{\alpha}_{\beta} $ but with errors in $ N^{\alpha}_{\beta} $ and {\{\color{black}\checkmark\}} = test passed for $ N^{\alpha}_{\beta} $ but with errors in $ M^{\alpha}_{\beta} $. Only the proposed new bending model passes all test cases.}
	\label{tab:6}
\end{table}

%%%%%%%%%%%%%%%%%%%%%%%%%%%%%%%%%%%%%%%%%%%%%%%%%%%%%%%%%%%%%%%%%%%%%%%%%%%%%%%%%%%%%%%%%%
\section{Numerical examples}\label{sec:5}

In this section, the performance of the new bending model is illustrated through two linear and four nonlinear numerical examples. In all examples, unless mentioned otherwise, the in-plane behavior is always modeled by the Koiter membrane model of Eq.~\eqref{eq:kmem1}, while the out-of-plane behavior is modeled by the different bending models of Sec.~\ref{sec:3}, which are then compared. The two linear examples demonstrate that the new bending model is equivalent to existing models in the small deformation regime. The four nonlinear examples illustrate that major differences appear at large deformations. 

%----------------------------------------------------------------------------------------%
\subsection{Simply supported plate under pressure loading}\label{subsec:5.1}

A simply supported square plate of length $ L\,=12\,L_0$ and thickness $ T\,=\,0.375\,L_0 $ under the sinusoidal pressure  
\begin{equation}
p(x,y) \,=\,\sin\left(\dfrac{\pi\,x}{L}\right)\,\sin\left(\frac{\pi\,y}{L}\right)\,E_0\,,
\end{equation} 
is analysed. The plate is considered to have Young's Modulus $ E\,=\,480\,E_0 $ and Poisson's ratio $ \nu\,=\,0.38 $. Tab.~\ref{tab:plate} shows the list of corresponding parameters for the different bending models. As the problem is symmetric, only $ 1/4 $ of the plate is modeled and the symmetry boundary conditions are enforced using the penalty method of \cite{duong2017new} with penalty parameter $\epsilon\,=\,4.8\,n^{q-1}\,E_0\,L_0$. Here, $ n $ refers to the number of elements per side and $ q $ is the order of shape functions used. 

\begin{figure}[H]
	\small
	\centering
	\begin{tikzpicture}
		\node[inner sep=0pt] (a) at (-11,3) {\includegraphics[trim={6cm 7cm 4cm 14cm},clip,scale=0.14]{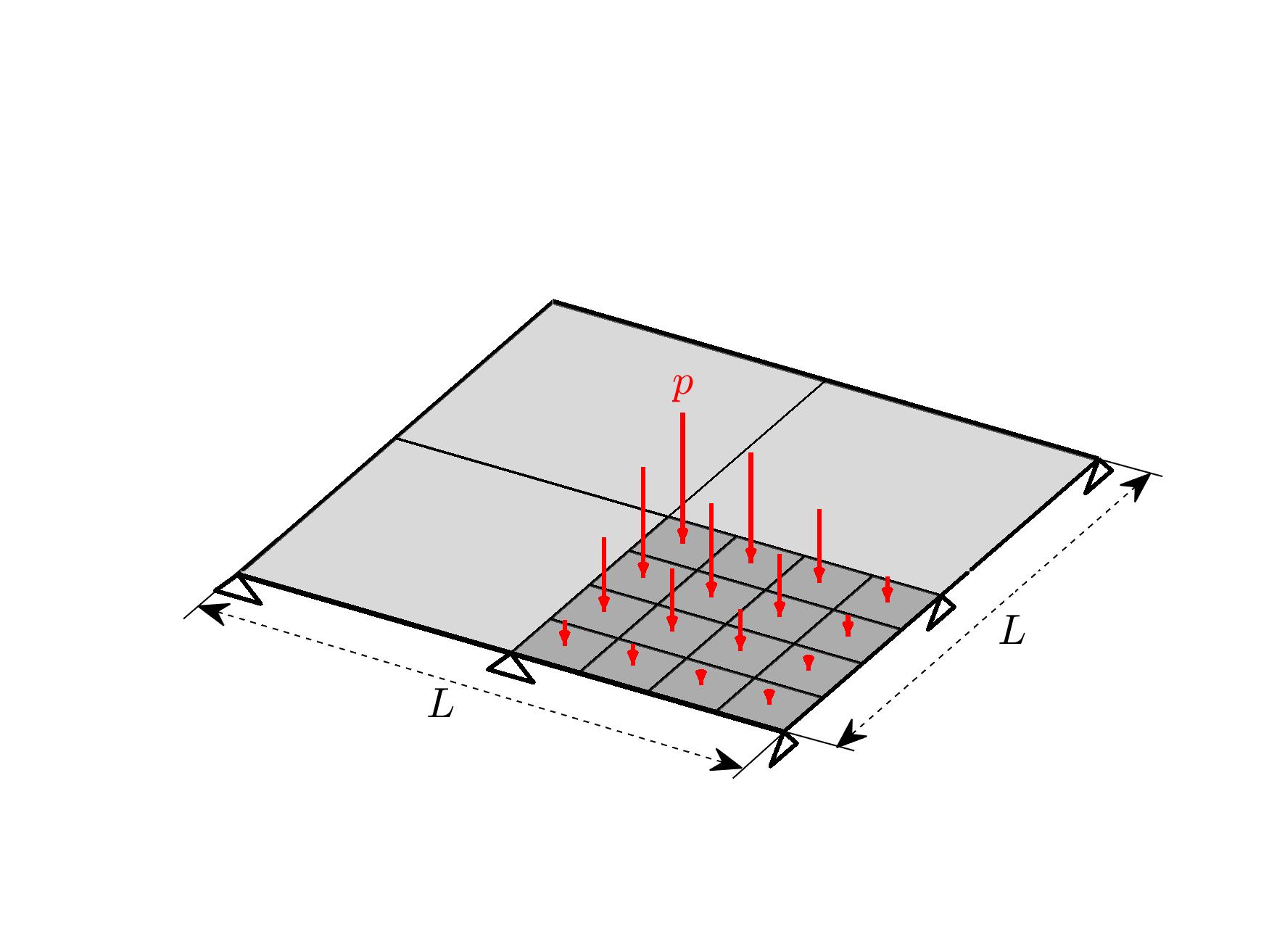}};
		\node[inner sep=0pt] (b) at (-3.5,3) {\includegraphics[trim={6cm 7cm 4cm 14cm},clip,scale=0.14]{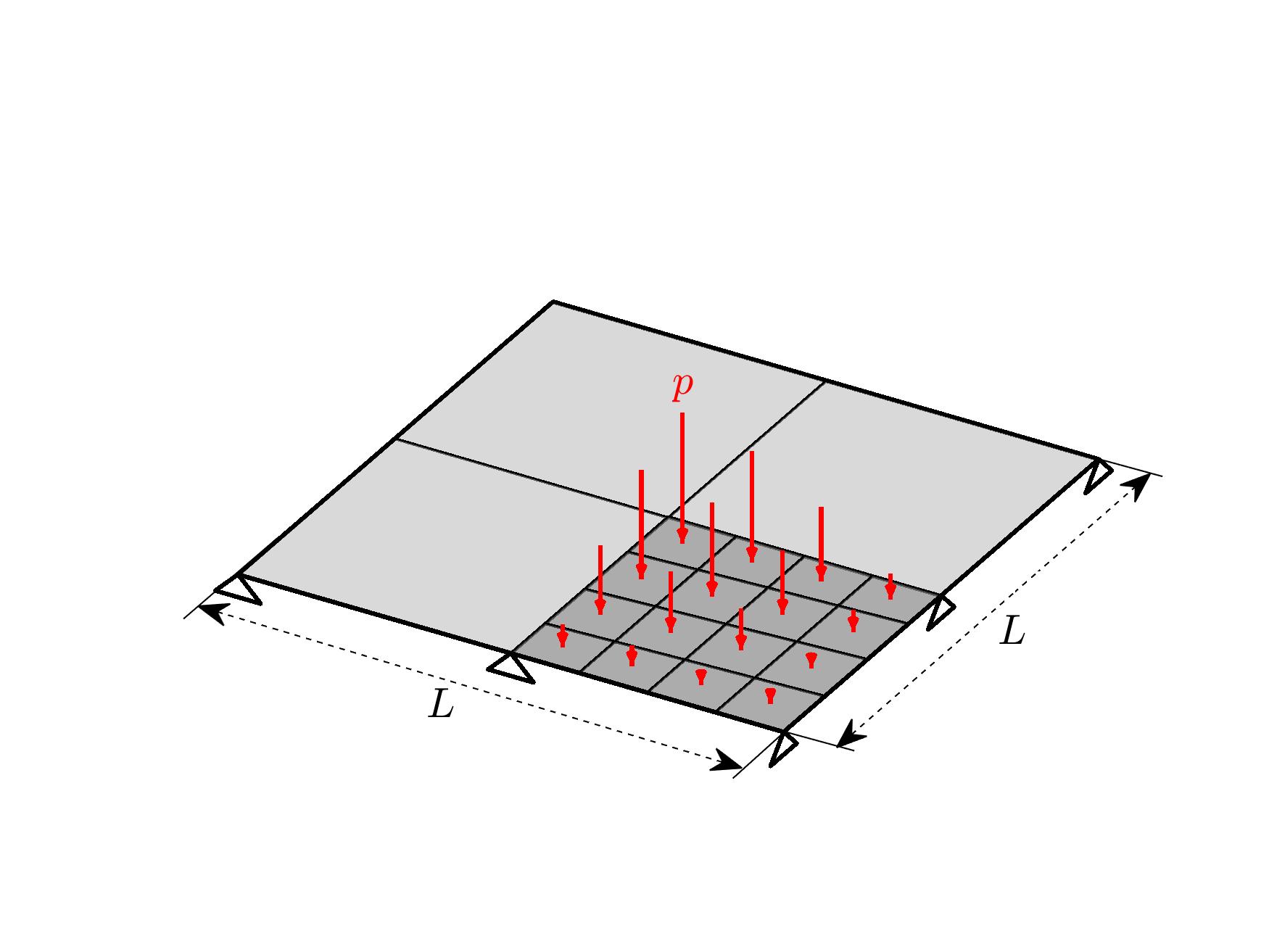}};
		\node[inner sep=0pt] (a) at (-11,1.25) {(a)};
		\node[inner sep=0pt] (a) at (-3.5,1.25) {(b)};
		\node[inner sep=0pt] (a) at (-11.5,-1.5) {\includegraphics[trim={0cm 2.2cm 5.5cm 2cm},clip,scale=0.25]{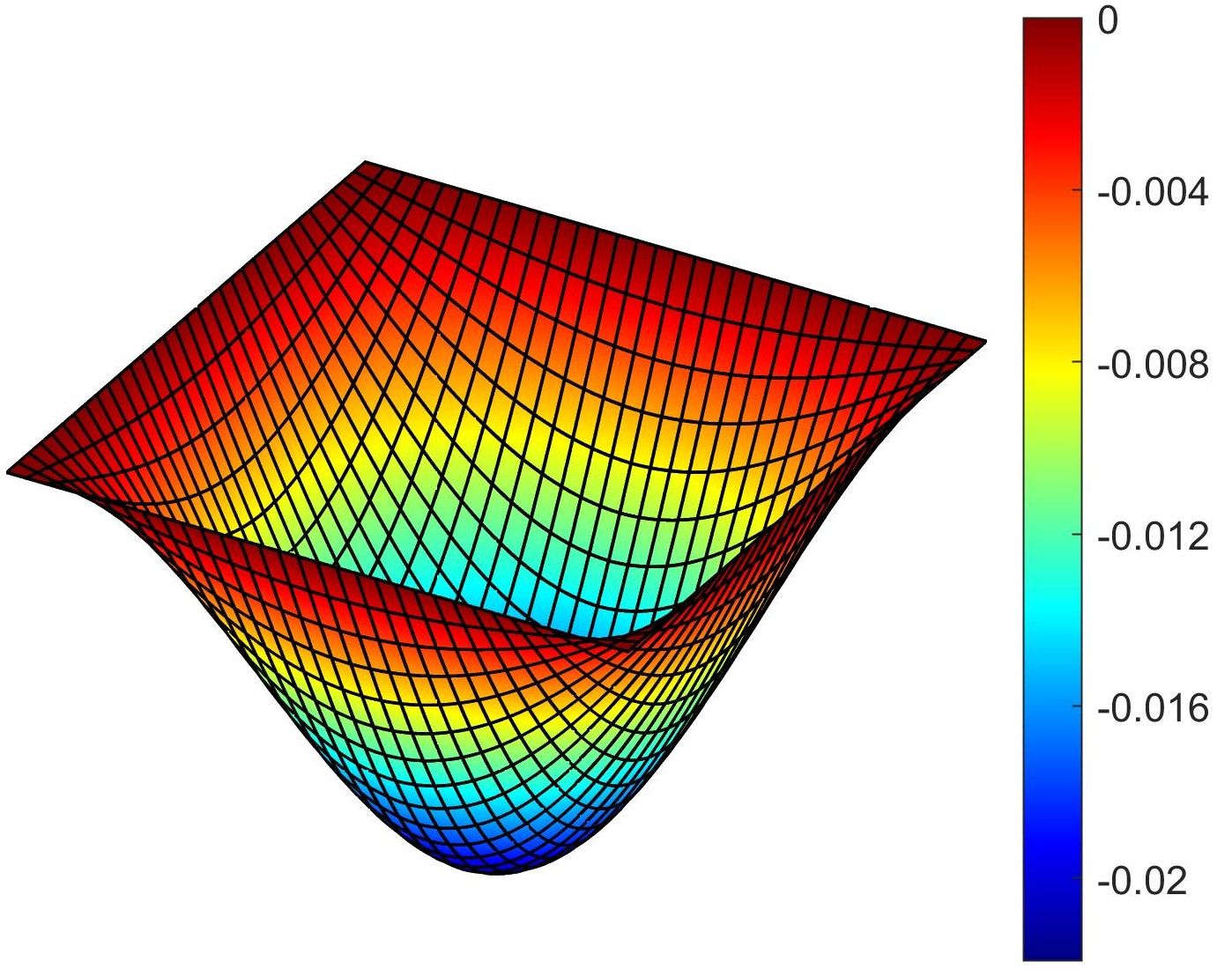}};
		\node[inner sep=0pt] (a) at (-7.8,-1.5) {\includegraphics[trim={24.5cm 0cm 0cm 0cm},clip,scale=0.25]{fig/plate/plate_res}};
		\node[inner sep=0pt] (b) at (-3.5,-1.95) {\includegraphics[scale=0.48]{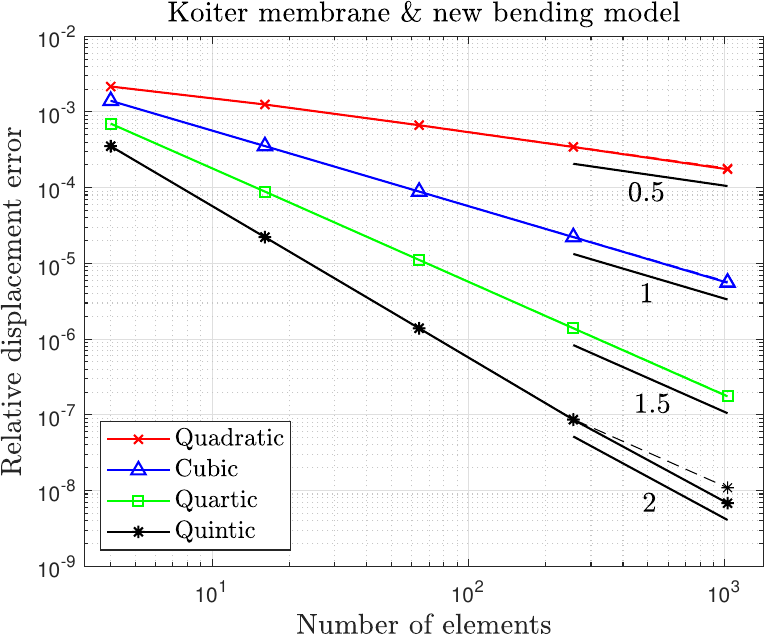}};
		\node[inner sep=0pt] (a) at (-11,-5) {(c)};
		\node[inner sep=0pt] (a) at (-3,-5) {(d)};
	\end{tikzpicture}
	\caption{Simply supported plate: Problem setup with (a) regular mesh and (b) skew mesh. (c) Deformed configuration (scaled by $ 400 $) for the skew mesh colored by the vertical displacement. (d) Difference of the center displacement between analytical and numerical solution for the new bending model. The solid lines show the results of the regular mesh, while the dashed lines show the results of the skew mesh. The displacements obtained for the other bending models are similar to the new bending model and hence omitted here.}
	\label{fig:plate}
\end{figure}

The maximum vertical displacement is compared to the analytical solution from \cite{ugural2009stresses} and the relative error plot is shown in Fig.~\ref{fig:plate}d for the new bending model. All the other bending models show matching error rates and are therefore omitted from the error plot. In this plot, the results obtained with a regular mesh (Fig.~\ref{fig:plate}a) and skew mesh (Fig.~\ref{fig:plate}b) are compared using solid and dashed lines, respectively. The skew mesh is obtained by modifying the knot vectors as elaborated in \cite{duong2017new} using a skewness ratio of 0.6. The two results are very close, confirming that the new bending model works also when $ \bL_{i} $ is not aligned with $ \bA_{\alpha} $. The convergence rates are approximately $ 0.5 $, $ 1 $, $ 1.5 $ and $ 2 $ for quadratic, cubic, quartic and quintic shapes function orders, respectively. Fig.~\ref{fig:plate}c shows the deformed configuration obtained with the skew mesh.
\begin{table}[H]
	\small
	\centering
	\begin{tabular}{|c|c|c|}
		\hline
		Bending model & \multicolumn{2}{c|}{Parameter set} \\ \hline
		& &  \\ [-3.5mm]
		3D linear elasticity & $ E\, = \,480\,E_0$                 &    $ \nu\,=\,0.38 $             \\ [0.5mm] \hline
		& &  \\ [-3.5mm]
		Koiter               & \begin{tabular}{c}$\mu\,=\,65.217\,E_0L_0$\\  $ \Lambda\,=\,79.944\,E_0\,L_0 $ \end{tabular}                 & $ \nu\,=\,0.38 $                 \\ [0.5mm] \hline
		& &  \\ [-3.5mm]
		Helfrich             & \begin{tabular}{c}$ k\,=\,4.9308\,E_0L^3 $ \\$ \bar{k}\,=\,-1.5285\,E_0L^3 $\\ $ H_0\,=\,0 $\end{tabular}                 & $ \nu\,=\,0.38 $                 \\ [0.5mm] \hline
		& &  \\ [-3.5mm]
		apH                  & $ \mu\, =\,60\,E_0L_0 $                  & $ \nu\,=\,0.5 $                 \\ [0.5mm] \hline
		& &  \\ [-3.5mm]
		new                  & \begin{tabular}{c} $ c_{1}\,=\,c_{2}\,=\,2.4654\,E_0\,L_0^3 $\\$ c_{12}\,=\,0.9368\,E_0\,L_0^3 $\\$ c_{3}\,=\,0.7643\,E_0\,L_0^3 $\end{tabular}                 & $ \nu\,=\,0.38 $                \\ [0.5mm] \hline
	\end{tabular}
	\caption{Simply supported plate: Parameter set used for different bending models according to Eqs.~\eqref{eq:surflam}, \eqref{eq:HelP} and \eqref{eq:newp}.}
	\label{tab:plate}
\end{table}

%----------------------------------------------------------------------------------------%
\subsection{Pinched cylindrical shell -- linear case}\label{subsec:5.2}

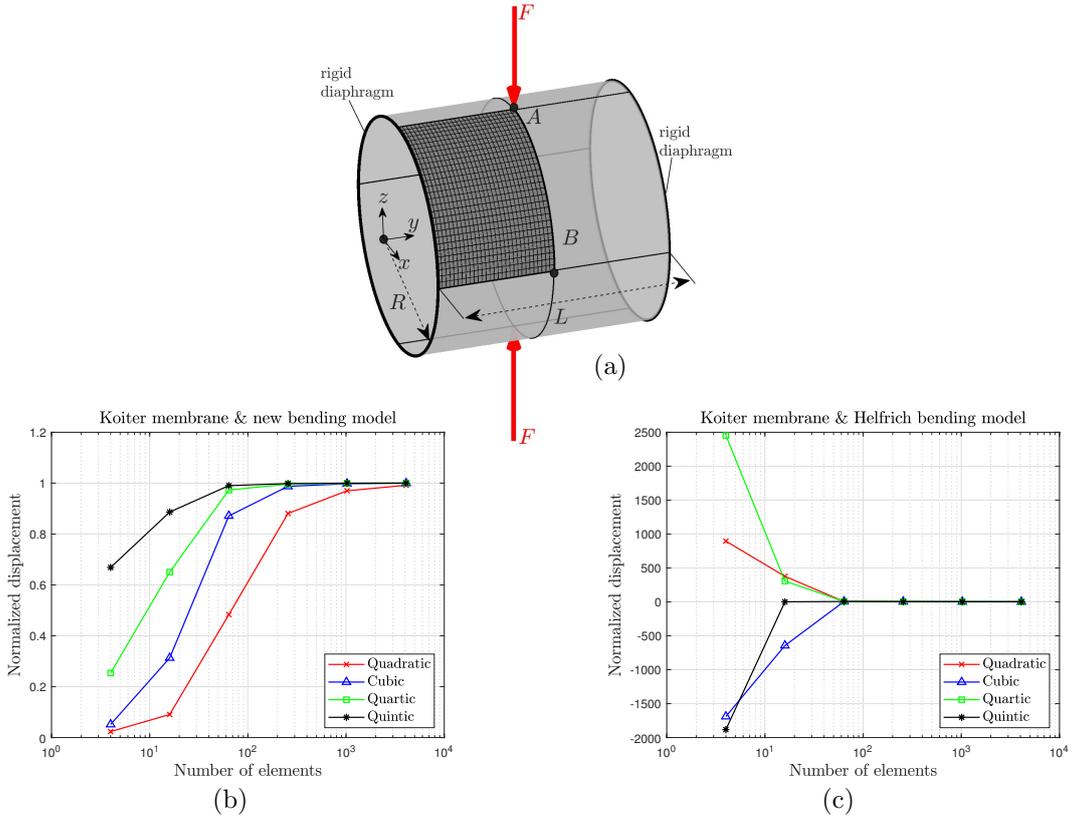
\begin{figure}[H]
	\small
	\centering
	\begin{tikzpicture}
		\node[inner sep=0pt] (a) at (-12,3) {\includegraphics[trim={6cm 2cm 0 0},clip,scale=0.14]{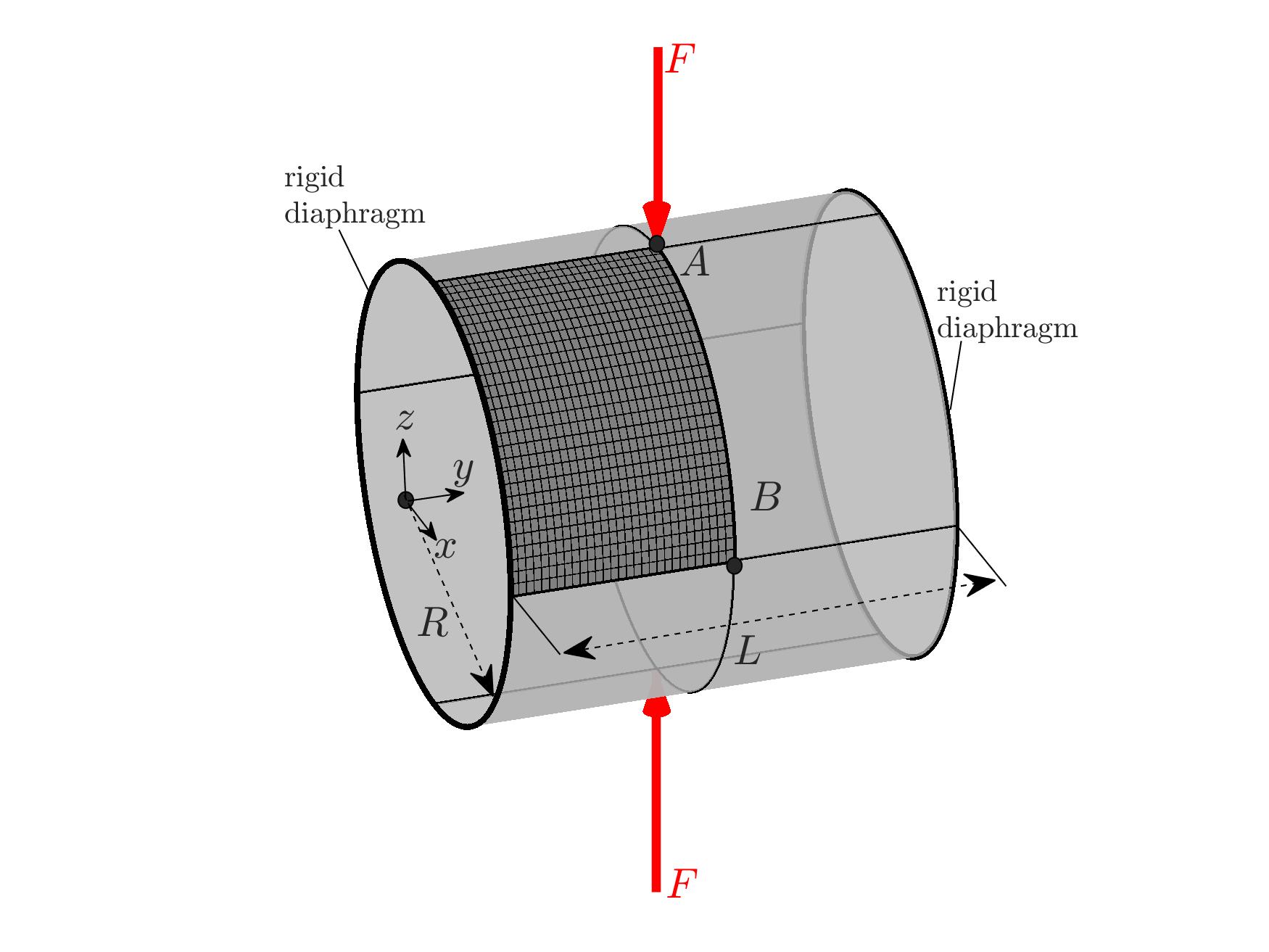}};
		\node[inner sep=0pt] (a) at (-11,1) {\small(a)};
		\node[inner sep=0pt] (a) at (-16,-2) {\includegraphics[scale=0.45]{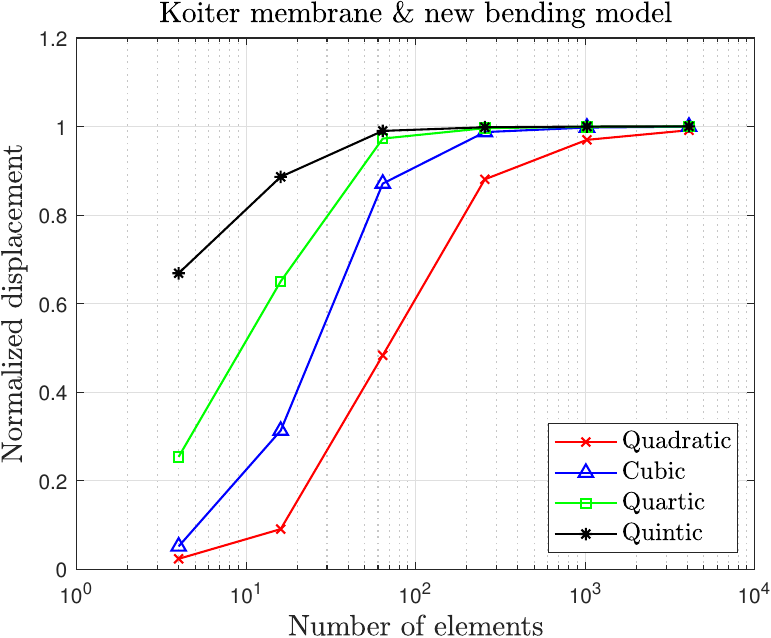}};
		\node[inner sep=0pt] (a) at (-8,-2) {\includegraphics[scale=0.45]{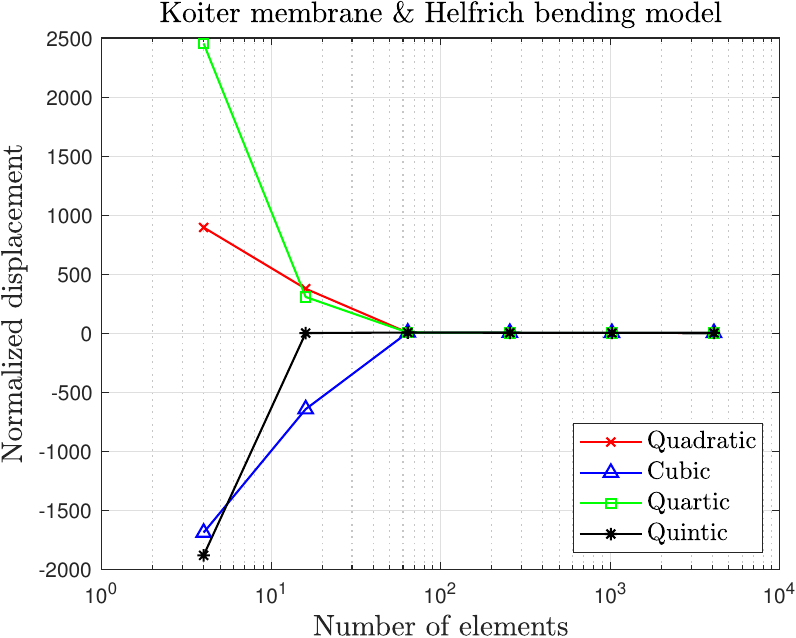}};
		\node[inner sep=0pt] (a) at (-16,-4.75) {\small (b)};
		\node[inner sep=0pt] (a) at (-8,-4.75) {\small (c)};
	\end{tikzpicture}
	\caption{Pinched cylindrical shell -- linear case: (a) Problem setup. Convergence of the displacement beneath the force for the (b) proposed new and (c) Helfrich bending model. The result obtained for the Koiter and apH bending models are similar to those of the proposed new bending model.}
	\label{fig:pinchlin}
\end{figure}

This example analyzes a cylinder with rigid diaphragms at its ends and subjected to point forces as shown in Fig.~\ref{fig:pinchlin}. The cylinder is of dimension $ R\,=\,300L_0 $, $ L\,=\,600L_0 $ and has thickness $ T\,=\,3L_0 $. The rigid diaphragm is realized by fixing the $ x$- and $ z$- degrees-of-freedom of the nodes lying at the cylinder ends. The material parameters are Young's modulus $E \,=\,3\, E_0$ and Poisson's ratio $ \nu\,=\,0.3 $. The corresponding material parameters for the different bending models are shown in Tab.~\ref{tab:8}. The magnitude of the two opposing pinching forces is $ F\,=\,10^{-6}\,E_0L^2_0 $. Owing to symmetry, the problem is solved for only $ 1/8 $ of the cylinder as shown in Fig.~\ref{fig:pinchlin}a. All symmetry conditions are enforced using a penalty method with parameter $ \epsilon\,=\,6\times10^2n^{q-1}_lE_0 L^{2}_0 $ for axial symmetry and $ \epsilon\,=\,6\times10^2n^{q-1}_tE_0 L^{2}_0 $ for circumferential symmetry, where $ n_l $ and $  n_t$ are the number of elements in axial and circumferential directions, respectively. 
\begin{table}[H]
	\small
	\centering
	\begin{tabular}{|c|c|c|}
		\hline
		Bending model & \multicolumn{2}{c|}{Parameter set} \\ \hline
		& & \\[-3.5mm]
		3D linear elasticity & $ E\, = \,3\,E_0$                 &    $ \nu\,=\,0.3 $             \\ [0.5mm]\hline
		& & \\[-3.5mm]
		Koiter               & \begin{tabular}{c}$\mu\,=\,3.4615\,E_0L_0$\\  $ \Lambda\,=\,2.967\,E_0\,L_0 $ \end{tabular}                 & $ \nu\,=\,0.3 $                 \\ [0.5mm]\hline
		& & \\[-3.5mm]
		Helfrich             & \begin{tabular}{c}$ k\,=\,2\,c_i\,=\,14.83\,E_0L^3 $ \\$ \bar{k}\,=\,-5.1923\,E_0L^3 $\\ $ H_0\,=\,-0.0016\,L_0^{-1} $\end{tabular}                 & $ \nu\,=\,0.3 $                 \\ \hline
		& & \\[-3.5mm]
		apH                  & $ \mu\, =\,4.5\,E_0L_0 $                  & $ \nu\,=\,0.5 $                 \\ [0.5mm]\hline
		& & \\[-3.5mm]
		new                  & \begin{tabular}{c} $ c_{1}\,=\,c_{1}\,=\,7.4176\,E_0\,L_0^3 $\\$ c_{12}\,=\,2.2253\,E_0\,L_0^3 $\\$ c_{3}\,=\,2.5962\,E_0\,L_0^3 $\end{tabular}                 & $ \nu\,=\,0.3 $                \\ [0.5mm]\hline
	\end{tabular}
	\caption{Pinched cylindrical shell -- linear case: Parameter set used for the different bending models according to Eqs.~\eqref{eq:surflam}, \eqref{eq:HelP} and \eqref{eq:newp}.}
	\label{tab:8}
\end{table}
The finite element solution is verified by the analytical solution for the displacement beneath the force from \cite{flugge2013stresses} and \cite{duong2017new}. The proposed bending model converges to the accurate solution as shown in Fig.~\ref{fig:pinchlin}b. Identical convergence behavior is observed for the Koiter and apH model. As the Helfrich bending model is not initially stress-free due to the rear term in Eq.~(\ref{eq:helfeq}.2), highly inaccurate results are obtained for coarse meshes (Fig.~\ref{fig:pinchlin}c).  
%----------------------------------------------------------------------------------------%
%----------------------------------------------------------------------------------------%
\subsection{Pinched cylindrical shell -- nonlinear case}\label{subsec:5.3}

\begin{figure}[H]
	\small
	\centering
	\begin{tikzpicture}
	\node[inner sep=0pt] (a) at (-11,3) {\includegraphics[scale=0.1]{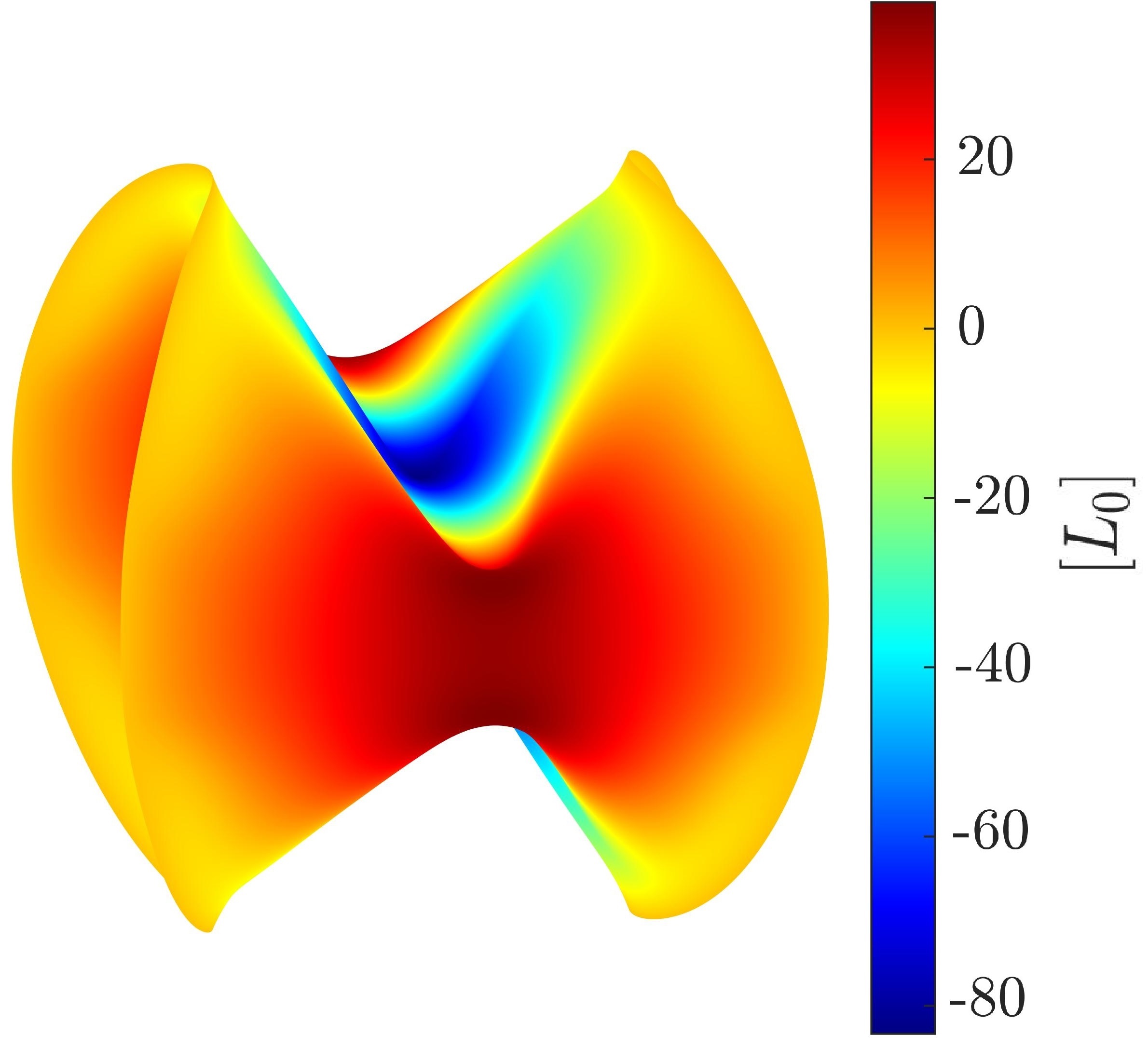}};
	\node[inner sep=0pt] (b) at (-3.5,3) {\includegraphics[scale=0.5]{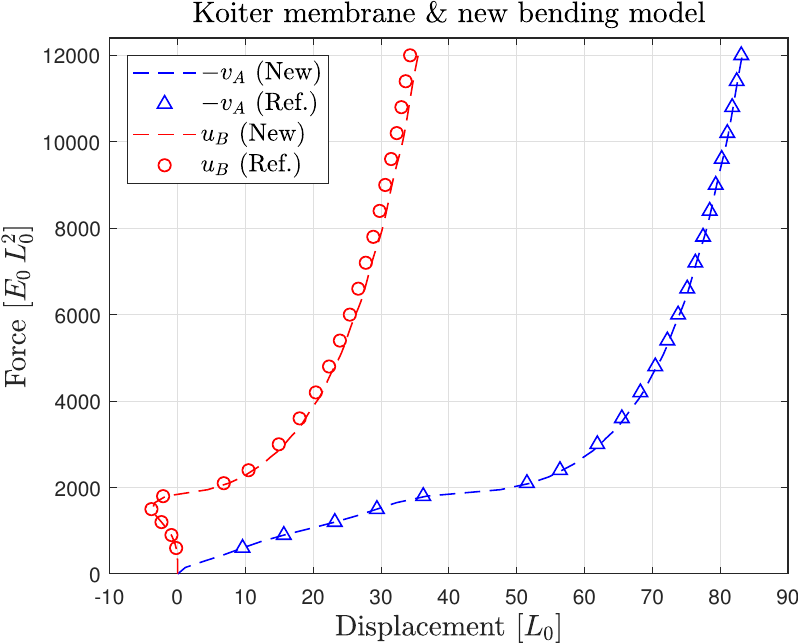}};
	\node[inner sep=0pt] (a) at (-11,0.) {(a)};
	\node[inner sep=0pt] (a) at (-3.5,0) {(b)};
	\end{tikzpicture}
	\caption{Pinched cylindrical shell -- nonlinear case: (a) Final configuration colored by the radial displacement. (b) Force-displacement curve for the new bending model. The result obtained for the Koiter and Helfrich bending models are similar to those of the new bending model and hence are omitted here.}
	\label{fig:pinch_comp}
\end{figure}

We consider the same problem as in the previous example (see Fig.~\ref{fig:pinchlin}a), but now undergoing large deformations. The radius, length, and thickness of the cylinder now is $ R\,=\,100\,L_{0} $, $ L\,=\,200\,L_{0} $, and $T\,=\,1.0\,L_{0} $, respectively. The cylinder is considered to have a Young's modulus $ E\,=\,30\,E_{0} $, Poisson's ratio $ \nu\,=\,0.3 $. The corresponding bending parameters following from this are tabulated in Tab.~\ref{tab:9}. Fig.~\ref{fig:pinch_comp}a shows the deformed configuration obtained with the point force of $ F\,=\,12\,E_{0}L_{0}^2 $ applied in 40 loading steps. By employing Lagrange multiplier-based symmetric boundary conditions as in \cite{duong2017new}, only $ 1/8 $ of the cylinder is used in the FE computations. Fig.~\ref{fig:pinch_comp}c presents the force vs.~displacement curve (measured at points A and B shown in Fig.~\ref{fig:pinchlin}a) obtained with $ 50\times50 $ quadratic NURBS elements. The reference solution from \cite{sze2004popular} is based on Reissner-Mindlin shell elements. The Koiter, Helfrich and new bending models are all in good agreement with the reference solution and thus only the numerical result of the new bending model is shown.
\begin{table}[H]
	\small
	\centering
	\begin{tabular}{|c|c|c|}
		\hline
		Bending model & \multicolumn{2}{c|}{Parameter set} \\ \hline
		& & \\[-3.5mm]
		3D linear elasticity & $ E\, = \,30\,E_0$                 &    $ \nu\,=\,0.3 $             \\ [0.5mm]\hline
		& & \\[-3.5mm]
		Koiter               & \begin{tabular}{c}$\mu\,=\,11.538\,E_0L_0$\\  $ \Lambda\,=\,9.8901\,E_0\,L_0 $ \end{tabular}                 & $ \nu\,=\,0.3 $                 \\ [0.5mm]\hline
		& & \\[-3.5mm]
		Helfrich             & \begin{tabular}{c}$ k\,=\,5.4945\,E_0L^3 $ \\$ \bar{k}\,=\,-1.9231\,E_0L^3 $\\ $ H_0\,=\,-0.005\,L^{-1}_0 $\end{tabular}                 & $ \nu\,=\,0.3 $                 \\ [0.5mm]\hline
			& & \\[-3.5mm]
		new                  & \begin{tabular}{c} $ c_{1}\,=\,c_{2}\,=\,2.7473\,E_0\,L_0^3 $\\$ c_{12}\,=\,0.8242\,E_0\,L_0^3 $\\$ c_{3}\,=\,0.9615\,E_0\,L_0^3 $\end{tabular}                 & $ \nu\,=\,0.3 $                \\ [0.5mm]\hline
	\end{tabular}
	\caption{Pinched cylindrical shell -- nonlinear case: Parameter set used for the different bending models according to Eqs.~\eqref{eq:surflam}, \eqref{eq:HelP} and \eqref{eq:newp}.}
	\label{tab:9}
\end{table}

\begin{figure}[H]
	\begin{subfigure}{.49\textwidth}\centering
		\includegraphics[trim={14cm 4cm 3cm 2cm},clip,scale=0.12]{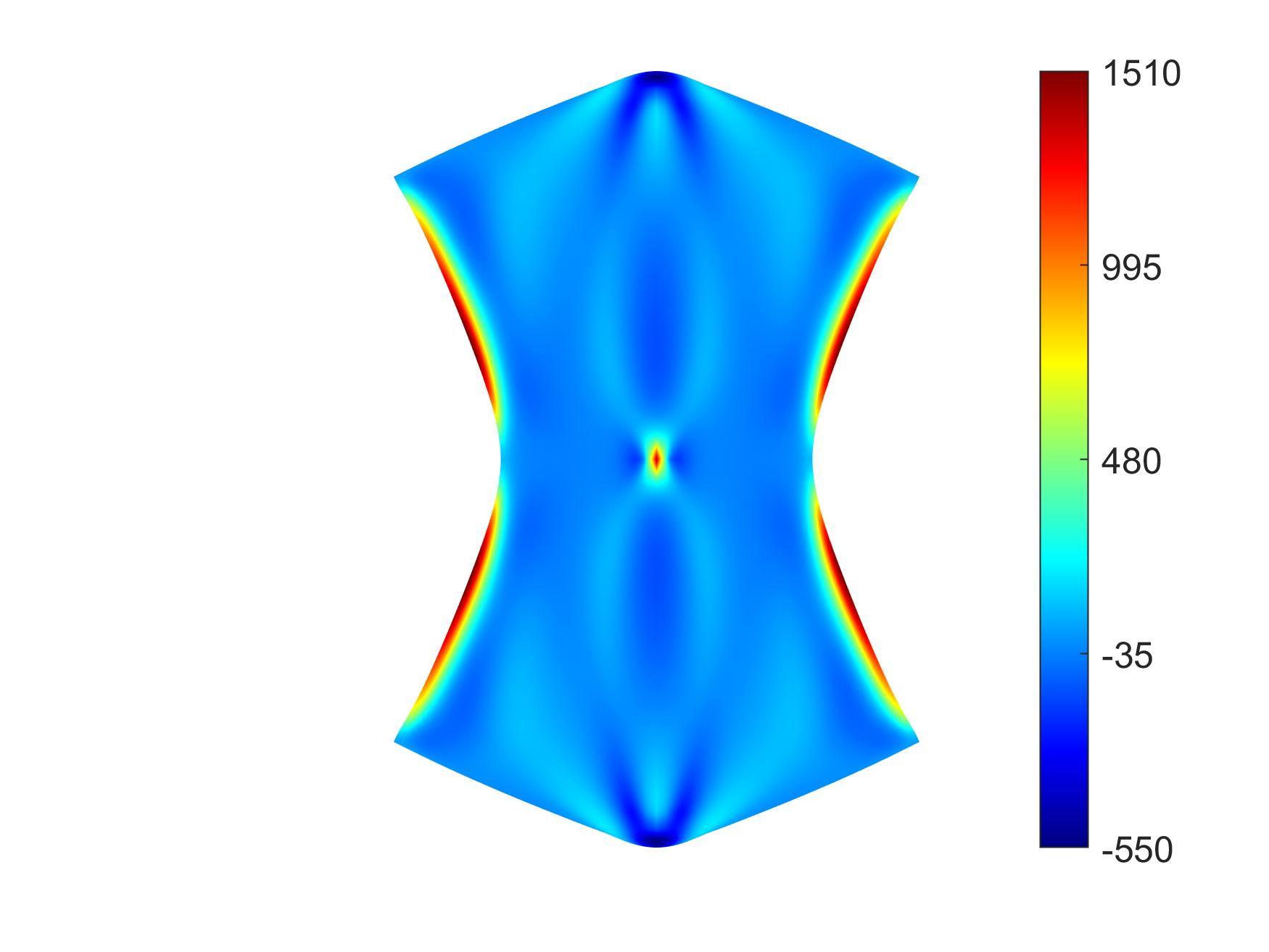}
		\subcaption{new $ N^{1}_1 $}
	\end{subfigure}
	\begin{subfigure}{.49\textwidth}\centering
		\includegraphics[trim={14cm 4cm 3cm 2cm},clip,scale=0.12]{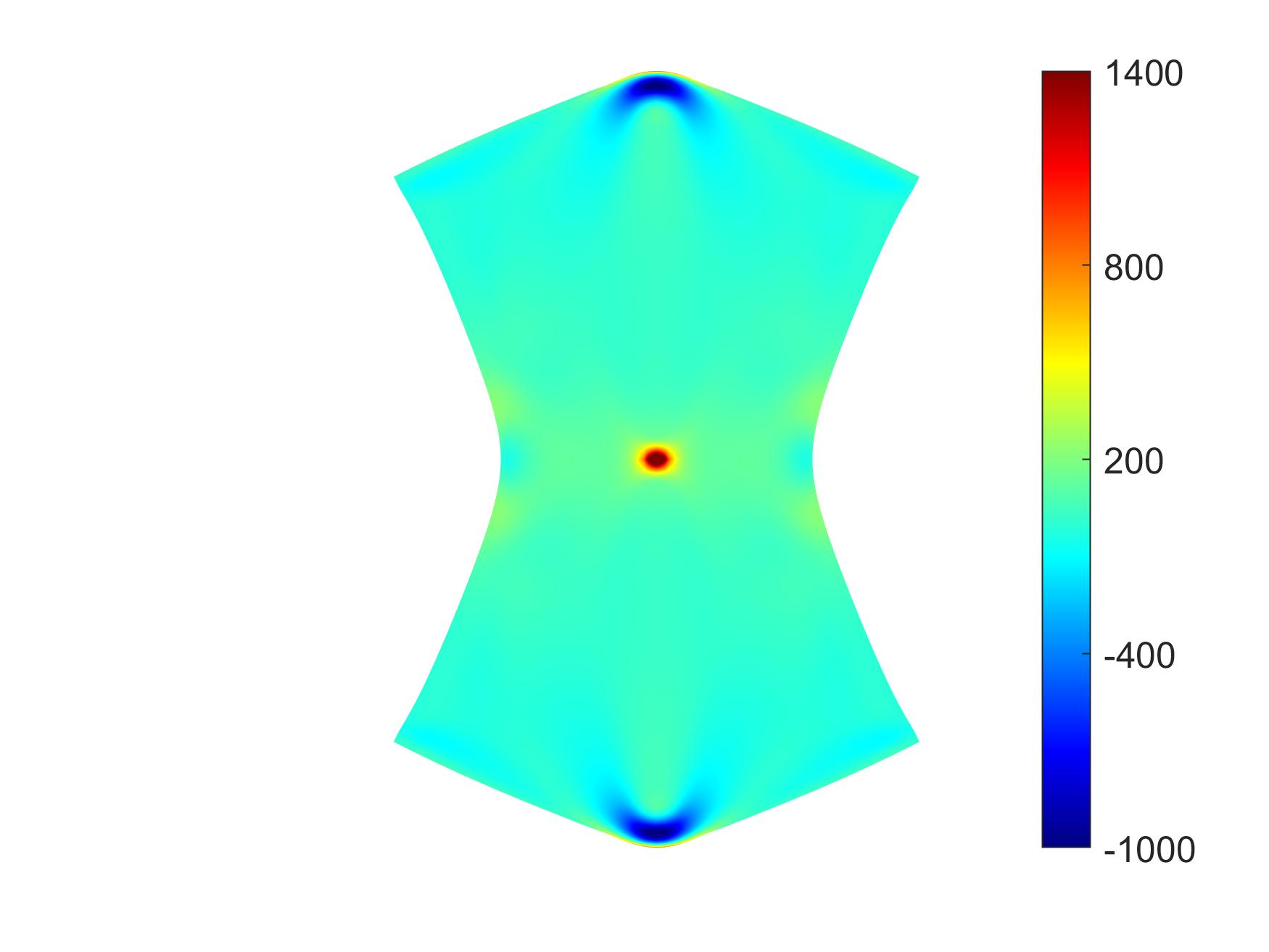}
		\subcaption{new $ N^{2}_2 $}
	\end{subfigure}
	\vspace*{2mm}
	\caption{Pinched cylindrical shell -- nonlinear case: Physical membrane stress components $ N^{\alpha}_{\beta}\, [\text{N/m}] $ for the new bending model.}
	\label{fig:totres}
\end{figure}

\begin{figure}[H]
	\begin{subfigure}{.3\textwidth}\centering
		\includegraphics[trim={14cm 4cm 22cm 3cm},clip,scale=0.12]{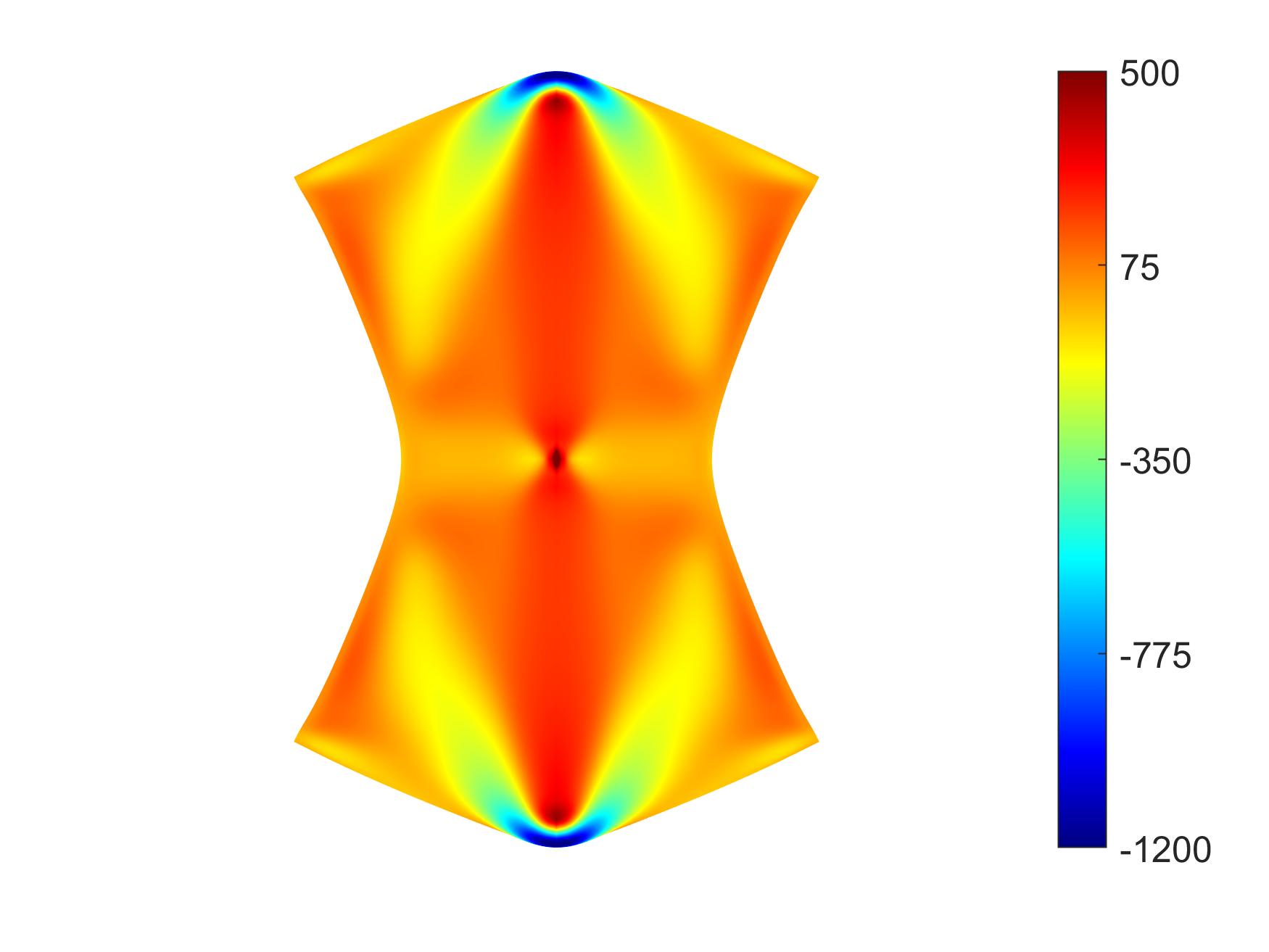}
		\subcaption{Koiter $ M^{1}_1 $}
	\end{subfigure}
	\begin{subfigure}{.3\textwidth}\centering
		\includegraphics[trim={14cm 4cm 22cm 3cm},clip,scale=0.12]{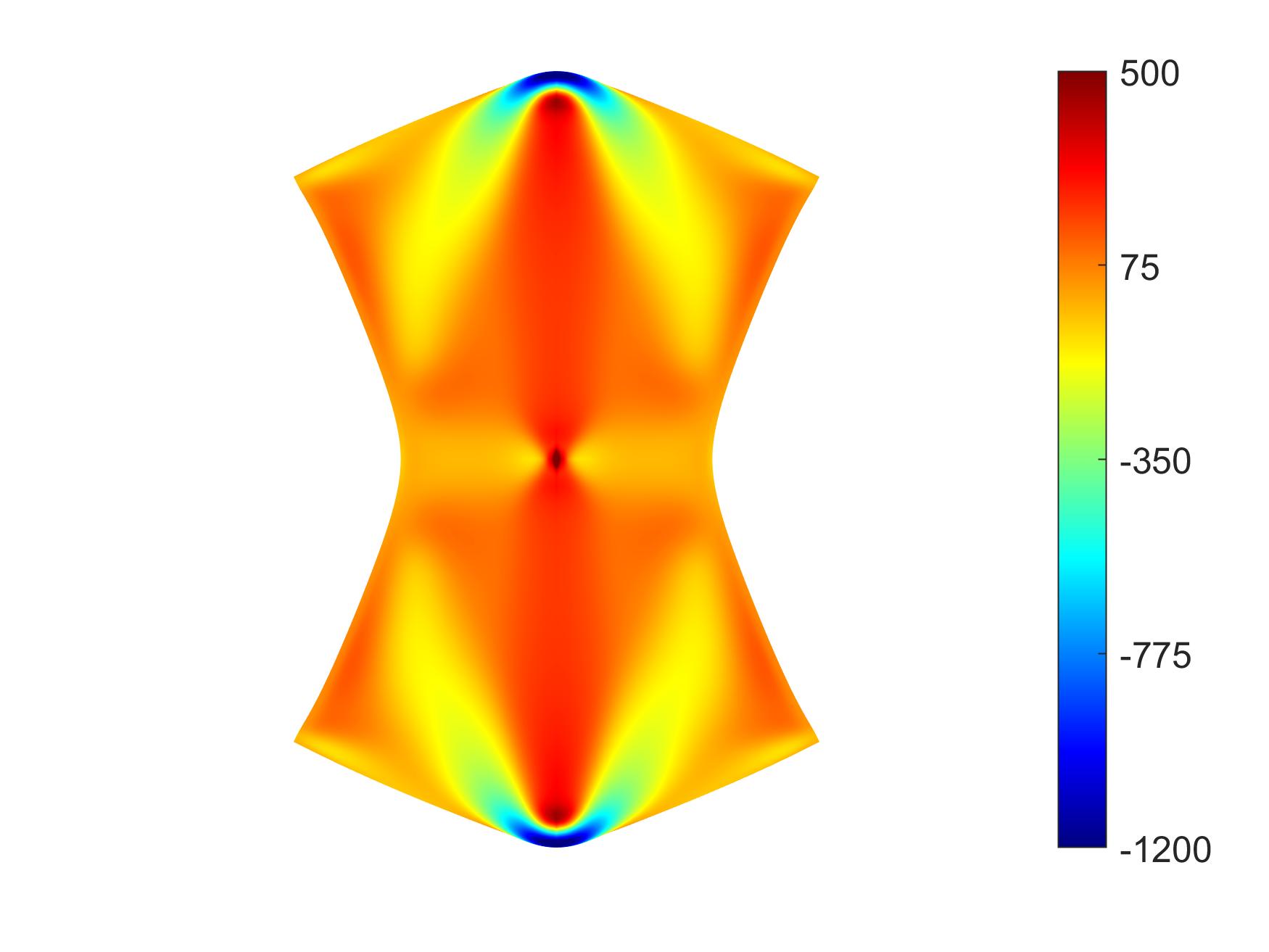}
		\subcaption{new $ M^{1}_1 $}
	\end{subfigure}
	\begin{subfigure}{.3\textwidth}\centering
		\includegraphics[trim={14cm 4cm 6.7cm 2.5cm},clip,scale=0.12]{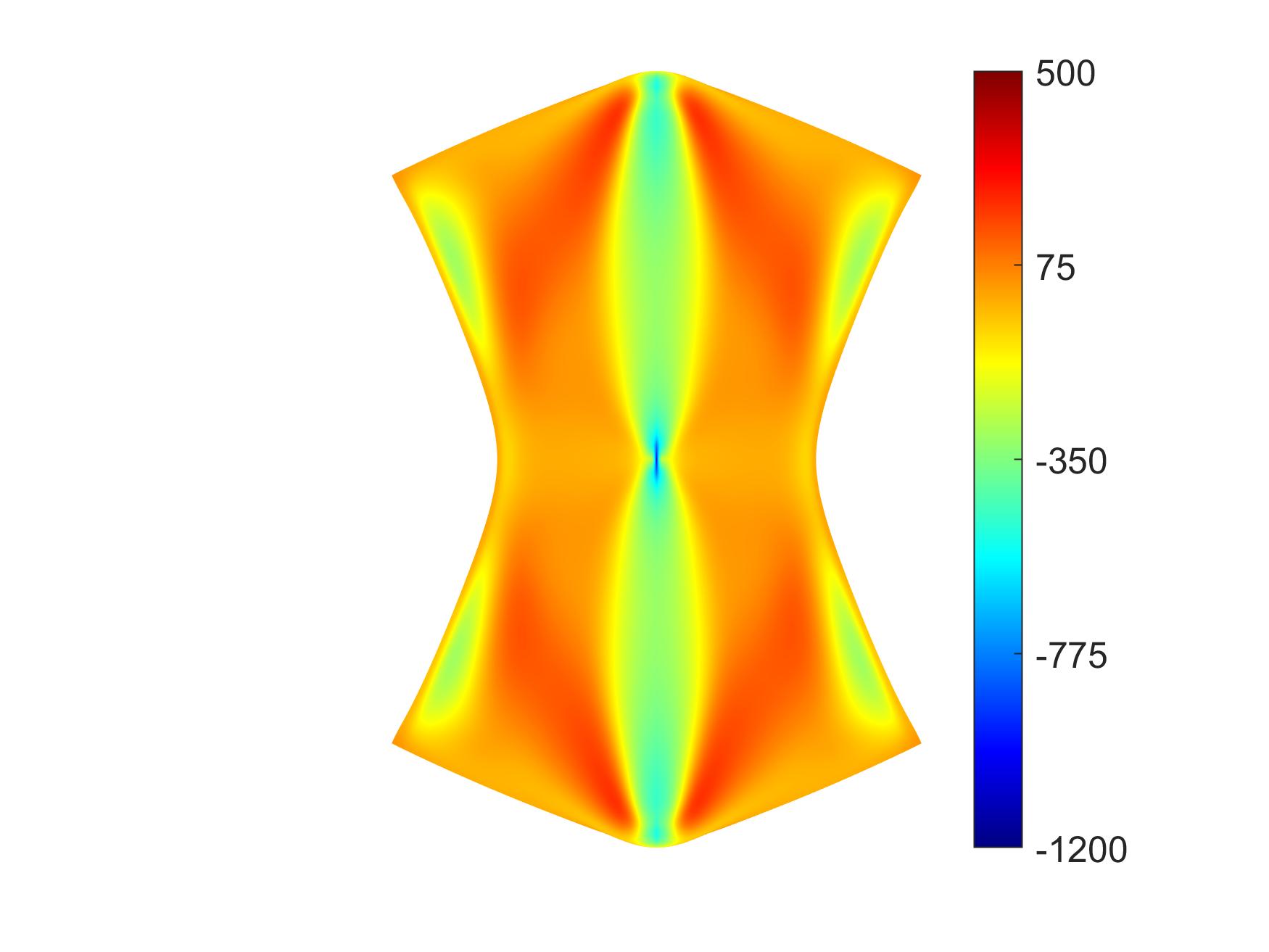}
		\subcaption{Helfrich $ M^{1}_1 $}
	\end{subfigure}\\ [5mm]
	\begin{subfigure}{.3\textwidth}\centering
		\includegraphics[trim={14cm 4cm 22cm 3cm},clip,scale=0.12]{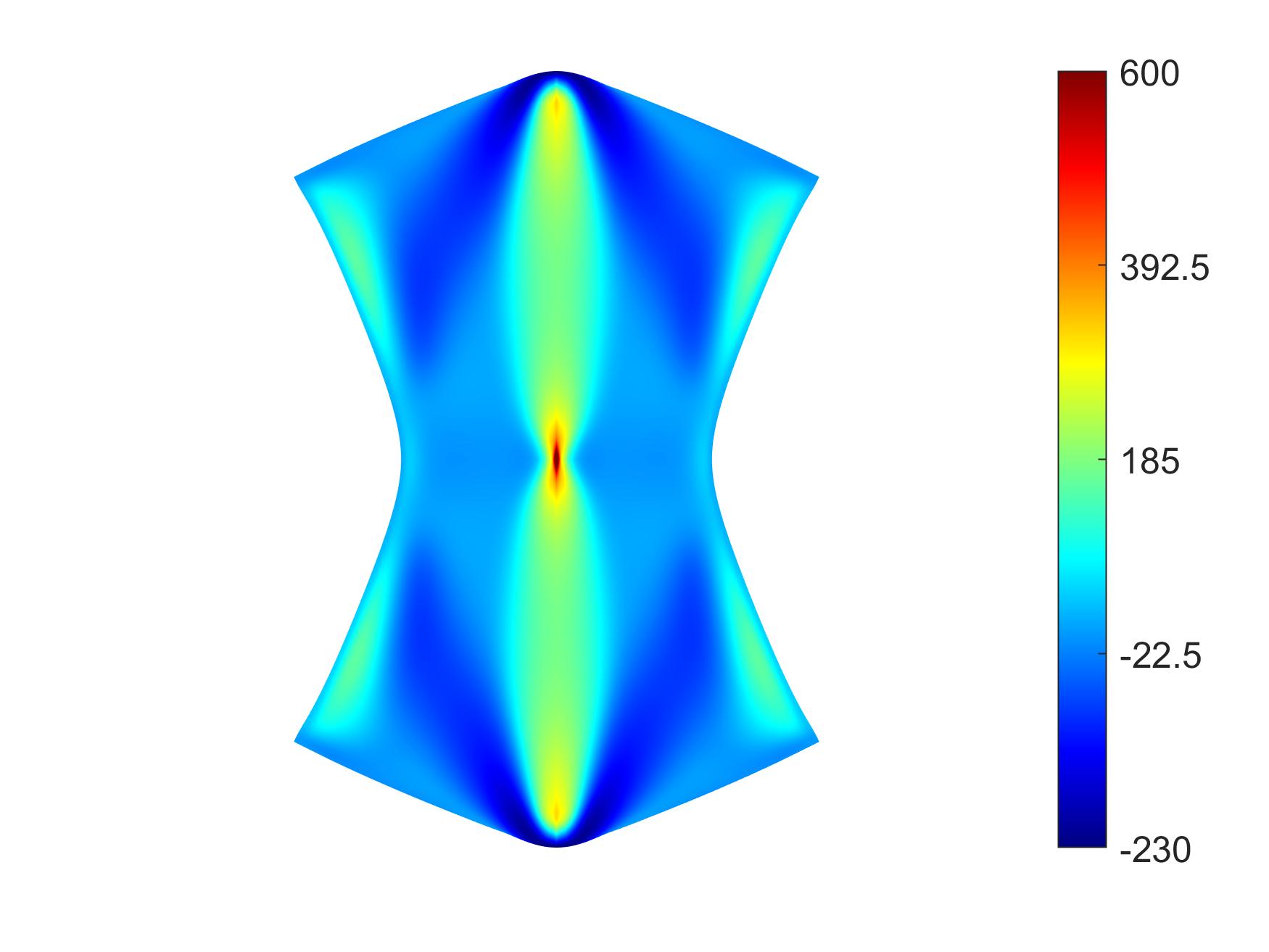}
		\subcaption{Koiter $ M^{2}_2 $}
	\end{subfigure}
	\begin{subfigure}{.3\textwidth}\centering
		\includegraphics[trim={14cm 4cm 22cm 3cm},clip,scale=0.12]{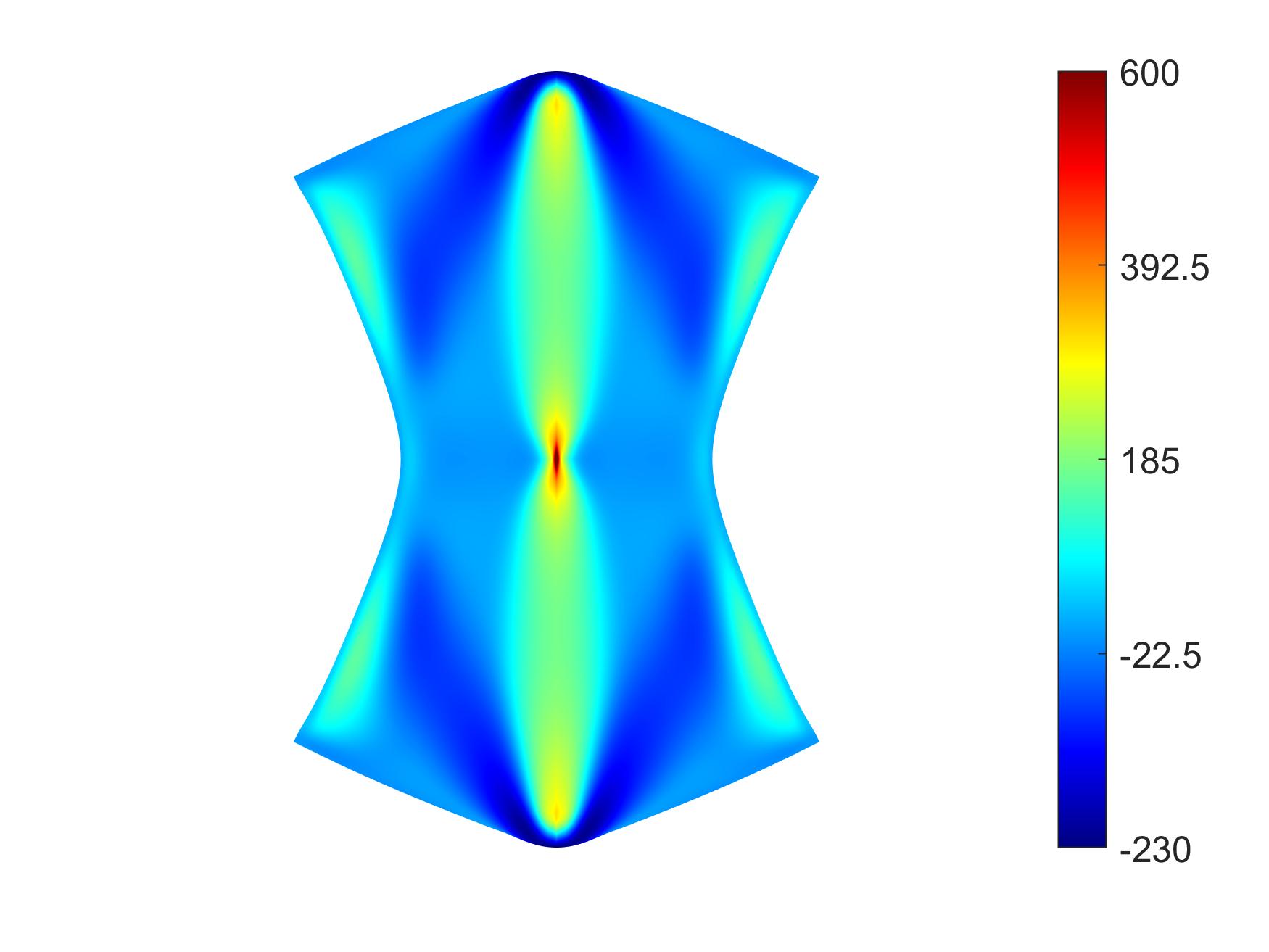}
		\subcaption{new $ M^{2}_2 $}
	\end{subfigure}
	\begin{subfigure}{.3\textwidth}\centering
		\includegraphics[trim={14cm 4cm 6.7cm 2.5cm},clip,scale=0.12]{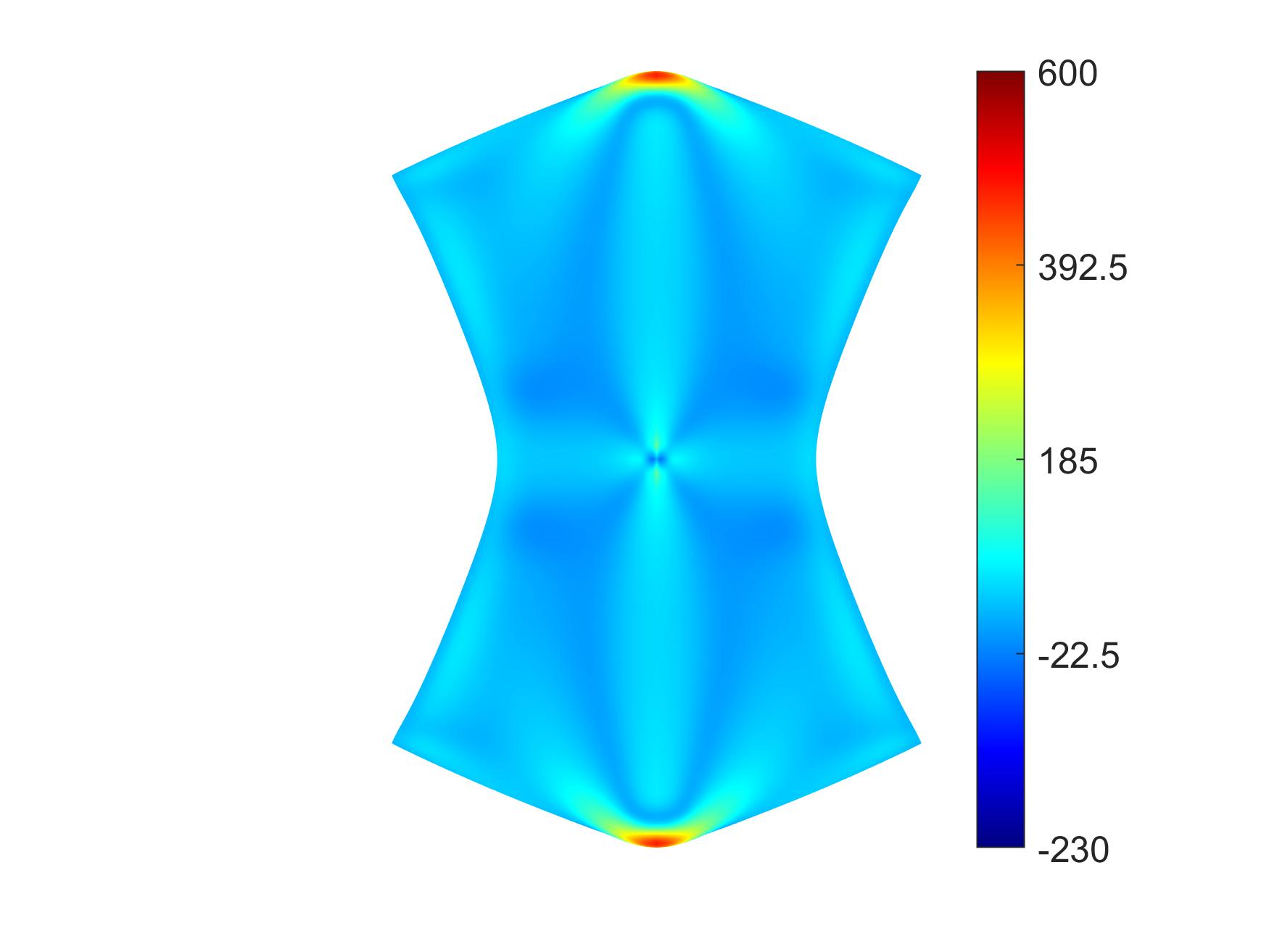}
		\subcaption{Helfrich $ M^{2}_2 $}
	\end{subfigure}
	\vspace*{2mm}\caption{Pinched cylindrical shell -- nonlinear case: Bending stress components $ M^{\alpha}_{\beta}\, [\text{N}]$}
	\label{fig:benres}
\end{figure}

The following figures show the Cauchy stresses $ N^\alpha_\beta \,=\,N^{\alpha\gamma}\,a_{\gamma\beta}$ (Fig.~\ref{fig:totres}) and bending stresses $ M^{\alpha}_{\beta}\,=\,M^{\alpha\gamma}_{0}\,a_{\gamma\beta}/J$ (Fig.~\ref{fig:benres}). In order to smooth-out stress oscillations, L2-projection \citep{stressCalc} is used with lumped mass matrix. For $ N^{\alpha}_{\beta} $, only the result of the new bending model is shown as the two other models give similar results. However, for $ M^{\alpha}_{\beta} $ the Helfrich model shows distinct differences to both the Koiter and new model. This again shows that the Helfrich model is not able to capture initially curved stress-free shells properly.   

%----------------------------------------------------------------------------------------%
\subsection{Pure bending}\label{subsec:5.x}
	
\begin{figure}[H]
		\small
		\centering
		\begin{tikzpicture}
			\node[inner sep=0pt] (a) at (-12,-3) {\includegraphics[scale=0.55]{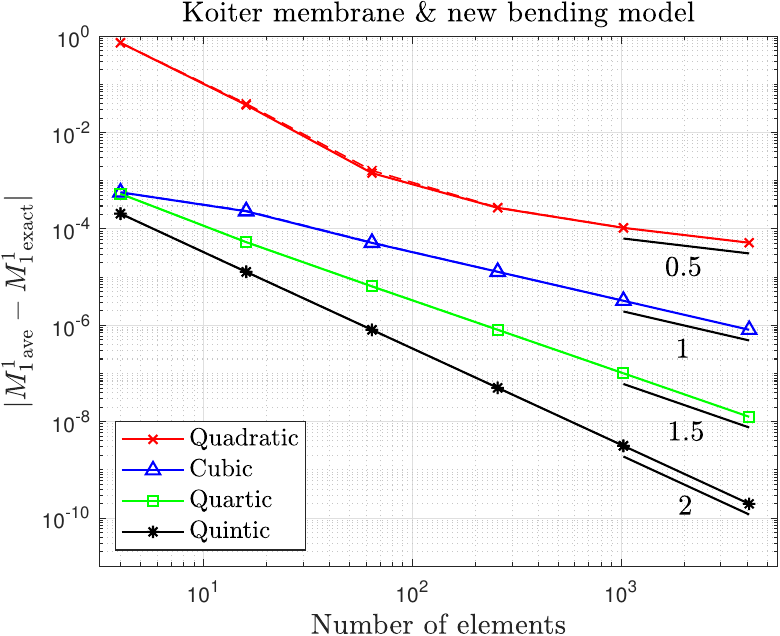}};
		\end{tikzpicture}
		\caption{Pure bending: $ M^1_1 $ error plot for the Koiter model (dashed lines) and new bending model (solid lines). The two FE results are very similar and approach the exact solution with mesh refinement and increasing shape function order.}
		\label{fig:pbx}
\end{figure}

	The second nonlinear example considers pure bending of half a cylinder as presented in Sec.~\ref{subsubsec:ben}. The radius of the cylinder is taken as $ R\,=\,L/\pi $, with length $ L\,=\,600\,L_{0} $ and thickness $ T\,=\,3\,L_{0} $. The material parameters are tabulated in Tab.~\ref{tab:x}. One straight edge of the half cylinder is clamped and a rotation is applied on the other until the angle $ 3\,\pi/4 $. The penalty approach of \cite{duong2017new} with the penalty parameter $\epsilon\,=\,n^{q-1} \,E\,L_{0} $ (cf.~Sec.~\ref{subsec:5.1}) is used for this. Fig.~\ref{fig:pbx} shows the error between the computed bending component $ M^1_1 $ and its exact analytical counterpart from Tab.~\ref{tab:4}. The computed stress component is therefore averaged over the surface, i.e.,
	\begin{equation}
		M^1_{1\,\mathrm{ave}}\,=\,\dfrac{\int_{\Omega^{e}_0}\,M^1_1\,\dif A}{\int_{\Omega^{e}_0}\,\dif A}\,.
	\end{equation}
	The error converges with increasing shape function order and  number of elements, thus verifying the FE formulation. The convergence rates are approximately $ 0.5 $, $ 1 $, $ 1.5 $ and $ 2 $ for quadratic, cubic, quartic and quintic shapes function orders, respectively.   
	\begin{table}[H]
		\small
		\centering
		\begin{tabular}{|c|c|c|}
			\hline
			Bending model & \multicolumn{2}{c|}{Parameter set} \\ \hline
			& & \\[-3.5mm]
			3D linear elasticity & $ E\, = \,300\,E_0$                 &    $ \nu\,=\,0 $             \\ [0.5mm]\hline
			& & \\[-3.5mm]
			Koiter               & \begin{tabular}{c}$\mu\,=\,450\,E_0L_0$\\  $ \Lambda\,=\,0 $ \end{tabular}                 & $ \nu\,=\,0 $                 \\ [0.5mm]\hline
			& & \\[-3.5mm]
			new                  & \begin{tabular}{c} $ c_{1}\,=\,c_{2}\,=\,675\,E_0\,L_0^3 $\\$ c_{12}\,=\,0 $\\$ c_{3}\,=\,337.5\,E_0\,L_0^3 $\end{tabular}                 & $ \nu\,=\,0 $                \\ [0.5mm]\hline
		\end{tabular}
		\caption{Pure bending: Parameter set used for the Koiter and new bending models according to Eqs.~\eqref{eq:surflam} and \eqref{eq:newp}.}
		\label{tab:x}
	\end{table}

%----------------------------------------------------------------------------------------%
\subsection{Cylindrical shell spreading}\label{subsec:5.4}

\begin{figure}[H]
	\begin{subfigure}{.49\textwidth}\centering
		\includegraphics[trim={6cm 3cm 0 2cm},clip,scale=0.15]{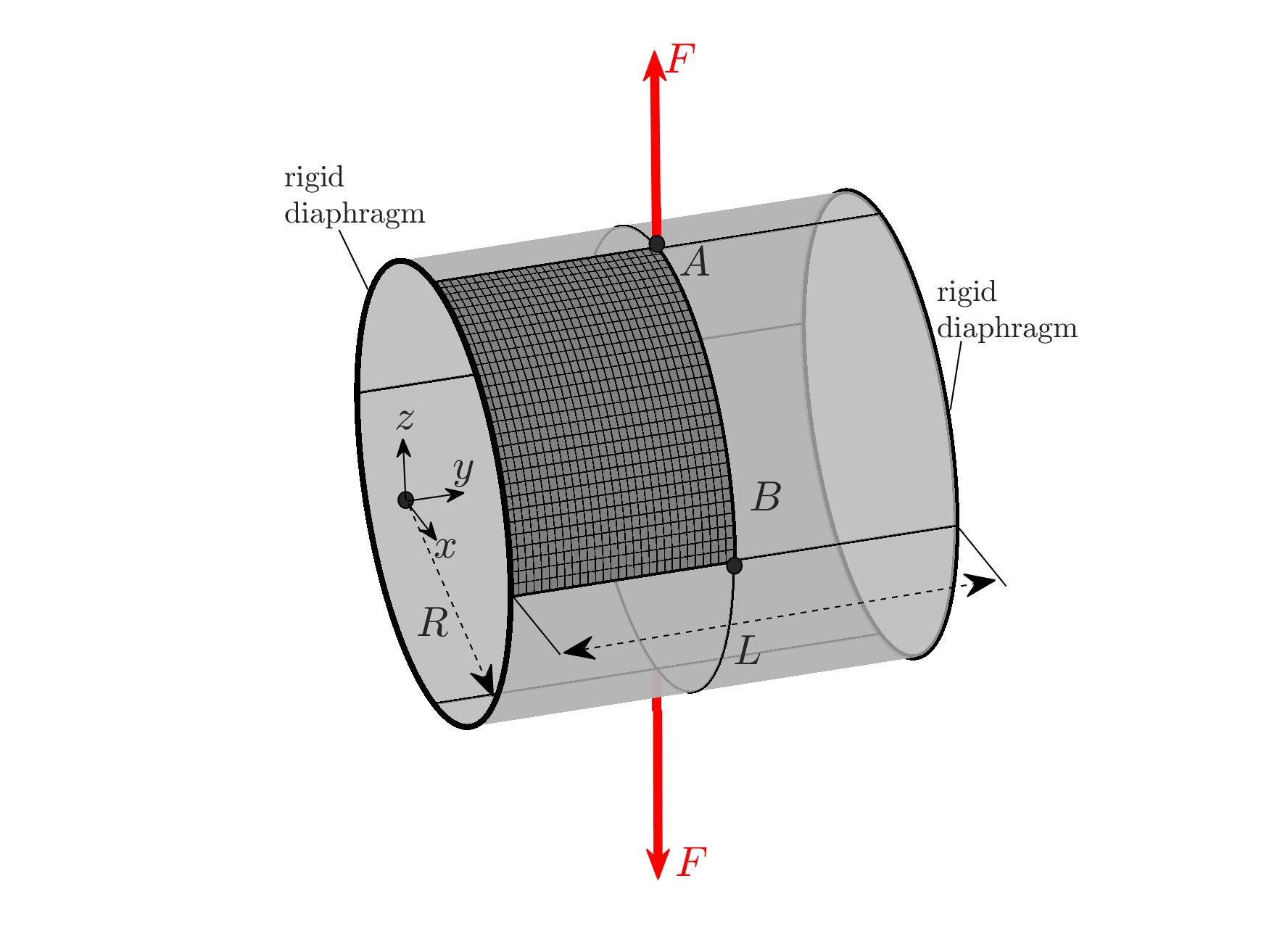}
		\subcaption{}
	\end{subfigure}
	\begin{subfigure}{.49\textwidth}\centering
		\includegraphics[scale=0.1]{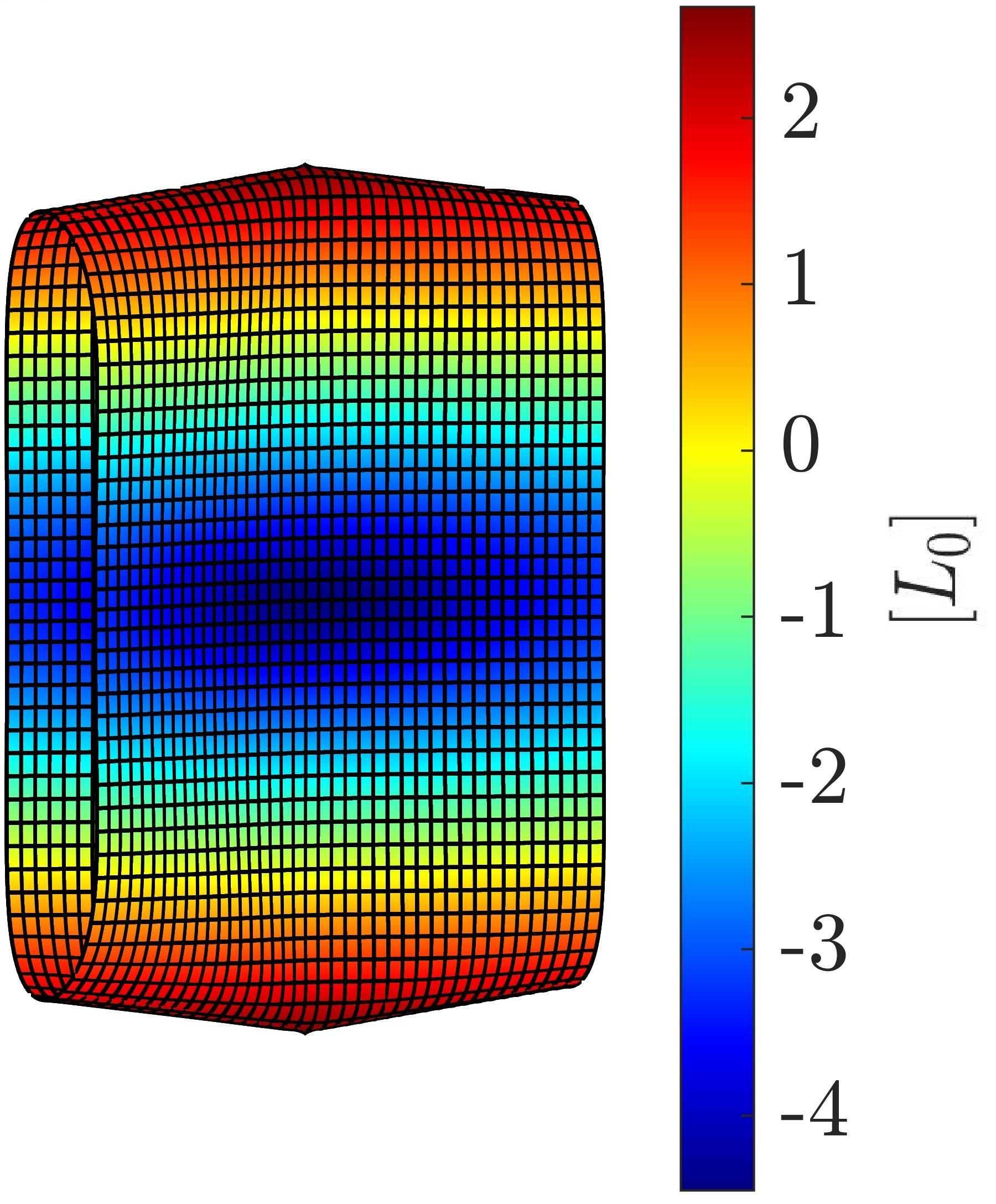}
		\subcaption{}
	\end{subfigure}\\ \\ \\
	\begin{subfigure}{.49\textwidth}\centering
		\includegraphics[scale=0.5]{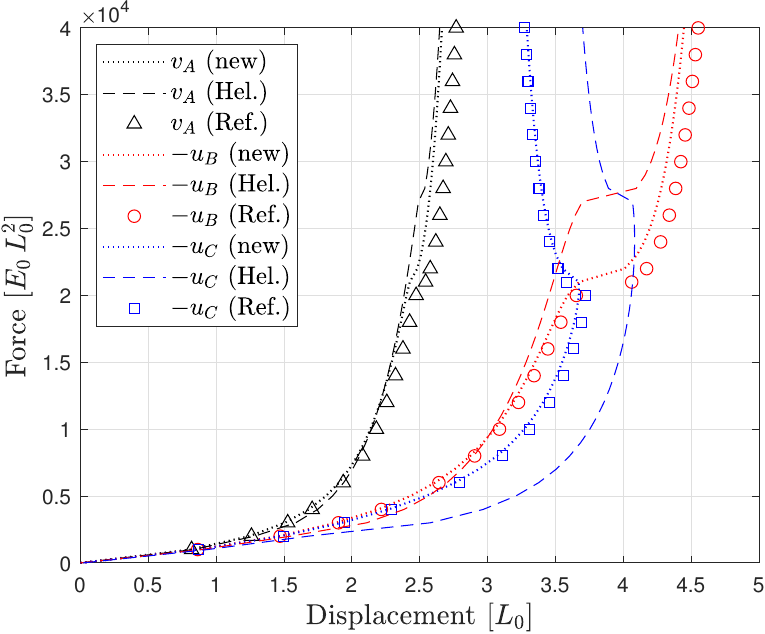}
		\subcaption{$ q\,=\,2 $}
	\end{subfigure}\centering
	\begin{subfigure}{.49\textwidth}\centering
		\includegraphics[scale=0.5]{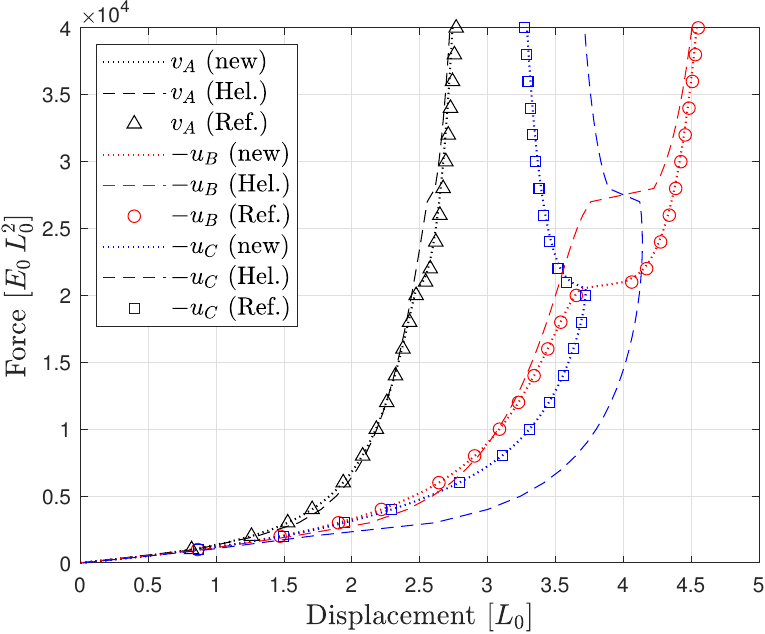}
		\subcaption{$ q\,=\,3 $}
	\end{subfigure}
	\vspace*{2mm}
	\caption{Cylindrical shell spreading: (a) problem setup and (b) deformed configuration colored by the radial displacement. Force-displacement curve for FE order (c) $ q\,=\,2 $ and (d) $ q\,=\,3 $ according to the new model, Helfrich model and reference results of \cite{sze2004popular}.}
	\label{fig:pull}
\end{figure}

Next, we consider an open ended cylinder that is being pulled apart by a pair of opposite forces. The cylinder has dimensions $ R\times L\times T\,=\,4.953\,L_{0}\,\times\,10.35\,L_{0}\,\times\, 0.094\,L_{0} $. Only $ 1/8 $ of the cylinder is discretized using 20 x 20 NURBS elements as shown in Fig.~\ref{fig:pull}a. The symmetry boundary conditions are enforced via the Lagrange multiplier method of \cite{duong2017new}. The bending parameters shown in Tab.~\ref{tab:10} are obtained from Young's modulus $ E\,=\,10.5\,\times 10^{3}\,E_{0} $ and Poisson's ratio $ \nu\,=\,0.3125 $. The magnitude of the spreading force is $ F\,=\,40\,E_{0}L_{0}^2 $ applied in 40 loading steps. 

\begin{table}[H]
	\small
	\centering
	\begin{tabular}{|c|c|c|}
		\hline		
		Bending model & \multicolumn{2}{c|}{Parameter set} \\ \hline
		&& \\ [-3.5mm]
		3D linear elasticity & $ E\, = \,10.5\times10^3\,E_0$                 &    $ \nu\,=\,0.3125 $             \\ [0.5mm]\hline
		&& \\ [-3.5mm]
		Koiter               & \begin{tabular}{c}$\mu\,=\,376\,E_0L_0$\\  $ \Lambda\,=\,341.82\,E_0\,L_0 $ \end{tabular}                 & $ \nu\,=\,0.3123 $                 \\[0.5mm] \hline
		&& \\ [-3.5mm]
		Helfrich             & \begin{tabular}{c}$ k\,=\,1.6108\,E_0L^3 $ \\$ \bar{k}\,=\,-0.5537\,E_0L^3 $\\ $ H_0\,=\,-0.1009\,L^{-1}_0 $\end{tabular}                 & $ \nu\,=\,0.3123 $                 \\[0.5mm] \hline
		&& \\ [-3.5mm]
		new                  & \begin{tabular}{c} $ c_{1}\,=\,c_{2}\,=\,0.8054\,E_0\,L_0^3 $\\$ c_{12}\,=\,0.2517\,E_0\,L_0^3 $\\$ c_{3}\,=\,0.2769\,E_0\,L_0^3 $\end{tabular}                 & $ \nu\,=\,0.3125 $                \\[0.5mm] \hline
	\end{tabular}
	\caption{Cylindrical shell spreading: Parameter set used for the different bending models according to Eqs.~\eqref{eq:surflam}, \eqref{eq:HelP} and \eqref{eq:newp}.}
	\label{tab:10}
\end{table}

Fig.~\ref{fig:pull}c-d shows the force-deflection curve for the new and Helfrich bending models in comparison to the reference solution from \cite{sze2004popular}. It is evident from these two plots that the Helfrich model again fails to capture the reference solution accurately. As in the previous examples, the solution for the Koiter bending model is similar to that of the new bending model.  
%----------------------------------------------------------------------------------------%
\subsection{Angioplasty simulation}\label{subsec:5.6}
The last example considers contact between an expanding balloon and a tube, which mimics the angioplasty procedure used to clear blocked coronary arteries. The tube has the dimensions $ R_{\mathrm{t}}\times L_{\mathrm{t}} \times T_{\mathrm{t}} \,=\,5\,\mathrm{mm}\times\,30\,\mathrm{mm}\,\times\,0.5\,\mathrm{mm}$ and represents a portion of an artery. The balloon within the artery is initially spherical and has the initial radius $ R_{\mathrm{b}}\,=\,R_{\mathrm{t}} $ and volume $ V_0 = 4\pi/3 R_\mathrm{b}^3 $. The tube is modeled using the incompressible Neo-Hookean membrane model from Eq.~(\ref{eq:wbapH}.1) together with either the Koiter or the proposed new bending model using the parameters $ \mu_{\mathrm{t}}\,=\, 8\,\mathrm{N/m}$ (corresponding to $ E\,=\,48\,\mathrm{kPa} $ and $ \nu\,=\,0.5 $ according to Eq.~\eqref{eq:surflam}). The balloon is also modeled by Eq.~(\ref{eq:wbapH}.1) but with $ \mu_{\mathrm{b}}\,=\,5\,\mu_{\mathrm{t}}$. The tube and balloon are discretized using $ n_{\mathrm{el}}\,=\,60\times55 $ and $ n_{\mathrm{el}}\,=\,64\times32 $ quadratic NURBS elements, respectively. The tube and balloon are supported by fixing all dofs normal to the $ x $--, $ y $-- and $ z $-- symmetry planes. This leaves the tube free to contract longitudinally. The balloon is inflated up to the volume $ V = 3V_0 $. Frictionless contact is enforced using the two-half-pass penalty contact formulation of \cite{sauer2013computational} with penalty parameter $ \epsilon\,=\,5\,E/\mathrm{mm}$. A similar angioplasty example was studied in \cite{roohbakhshan2017efficient,roohbakhshan2018simulation} using the material model of \cite{Gasser}.
\begin{figure}[H]
	\small
	\centering
	\begin{tikzpicture}
		\node[inner sep=0pt] (a) at (-9,0) {\includegraphics[trim={23cm 4cm 15cm 2cm},clip,scale=0.15]{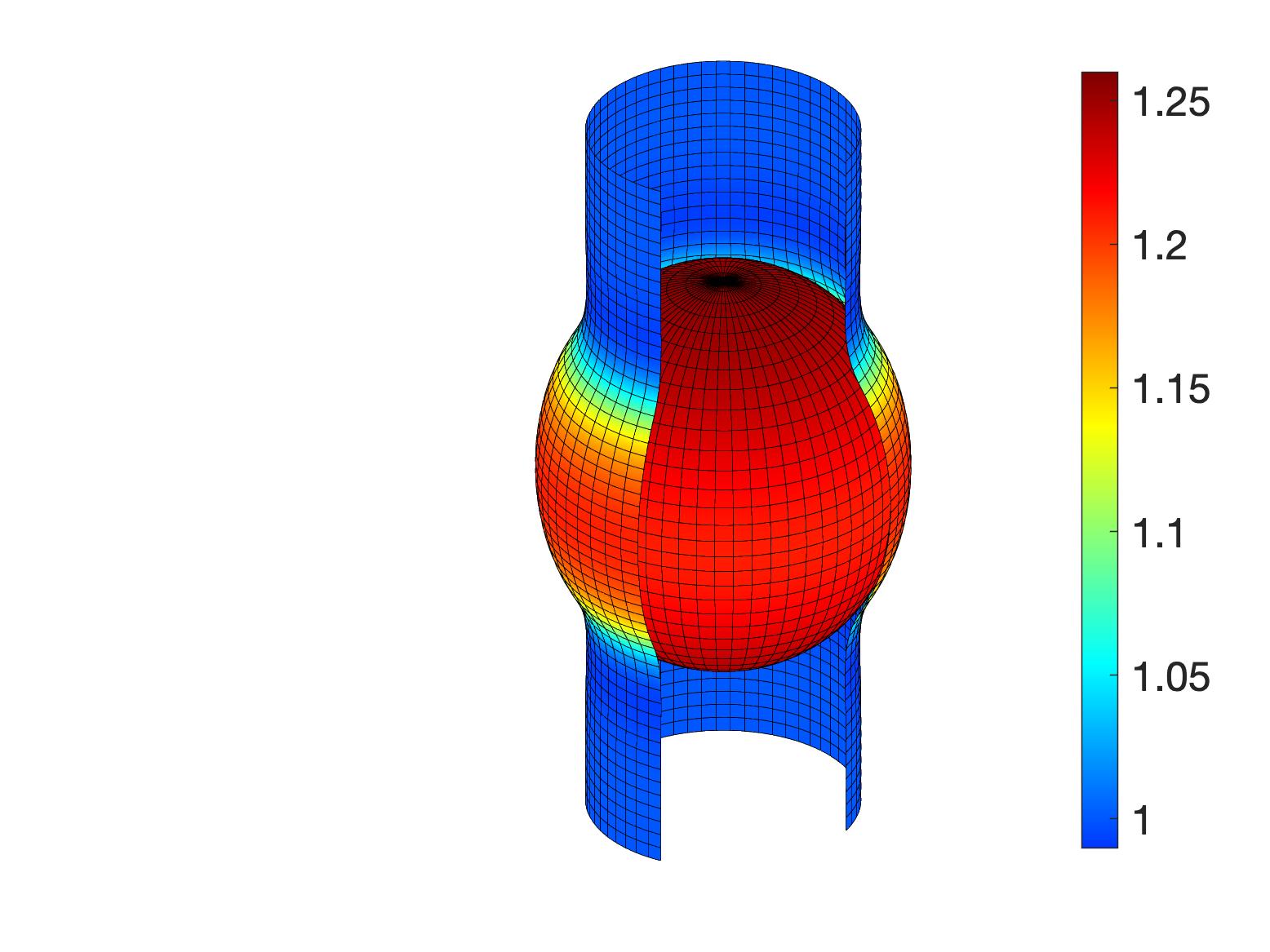}};
		\node[inner sep=0pt] (a) at (-6,0) {\includegraphics[trim={22cm 4cm 15cm 2cm},clip,scale=0.15]{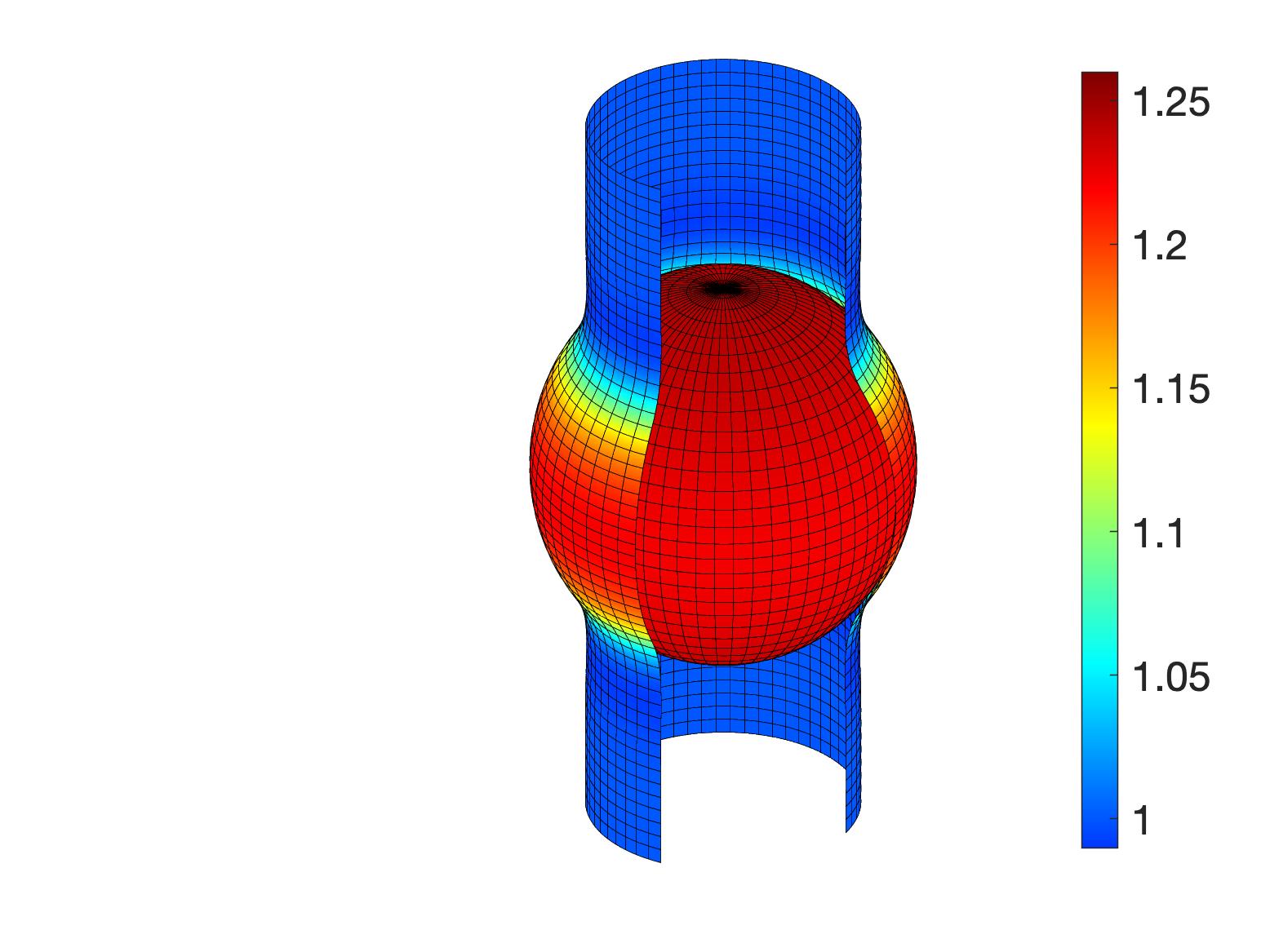}};
		\node[inner sep=0pt] (a) at (-4,0) {\includegraphics[trim={46cm 4cm 1.5cm 2cm},clip,scale=0.15]{fig/angi/Defo_Angio_SaSa_mc5_bc0_l1}};
		\node[inner sep=0pt] (a) at (-0.75,0) {\includegraphics[trim={23cm 4cm 15cm 2cm},clip,scale=0.15]{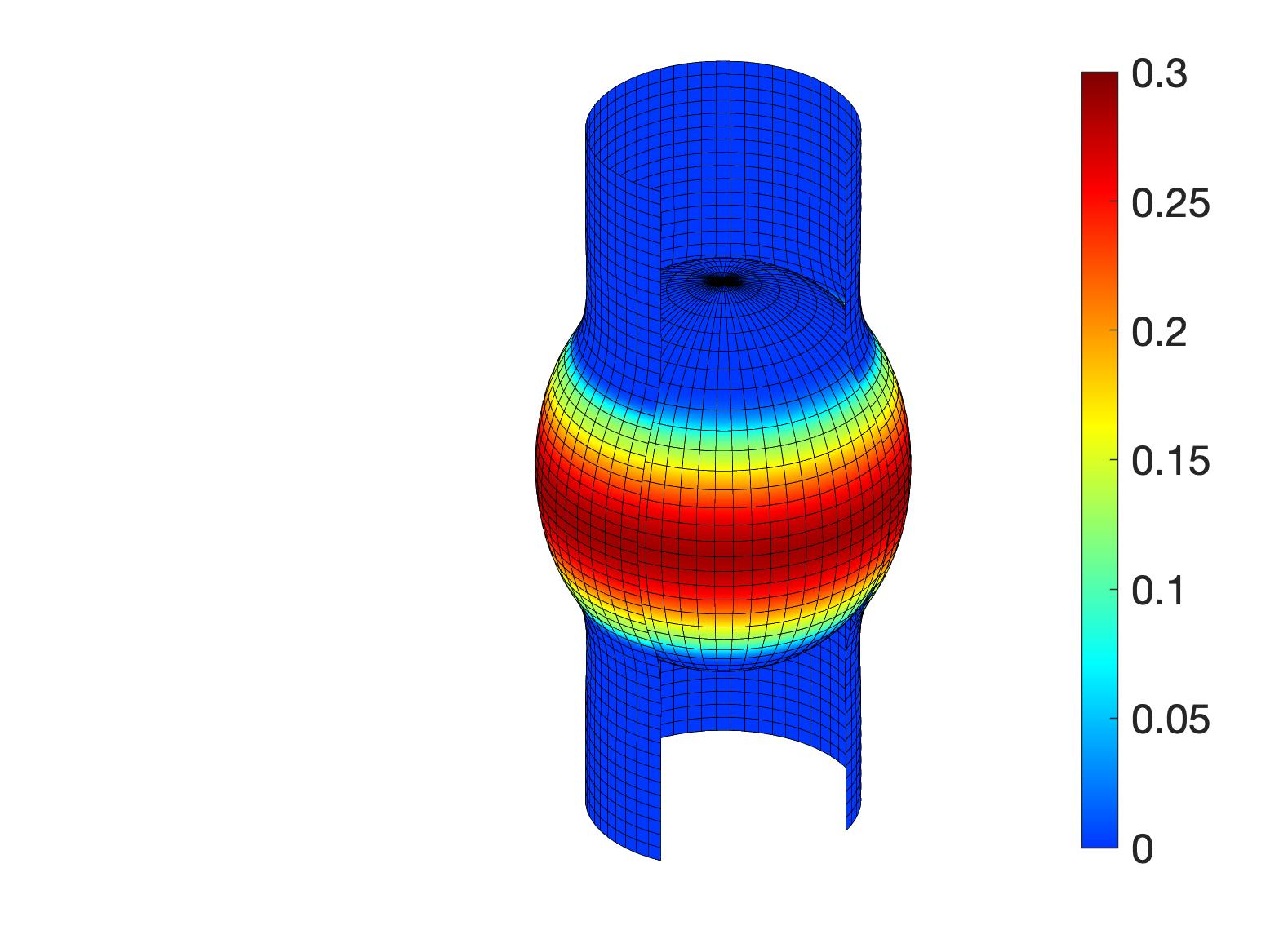}};
		\node[inner sep=0pt] (a) at (2.1,0) {\includegraphics[trim={22cm 4cm 15cm 2cm},clip,scale=0.15]{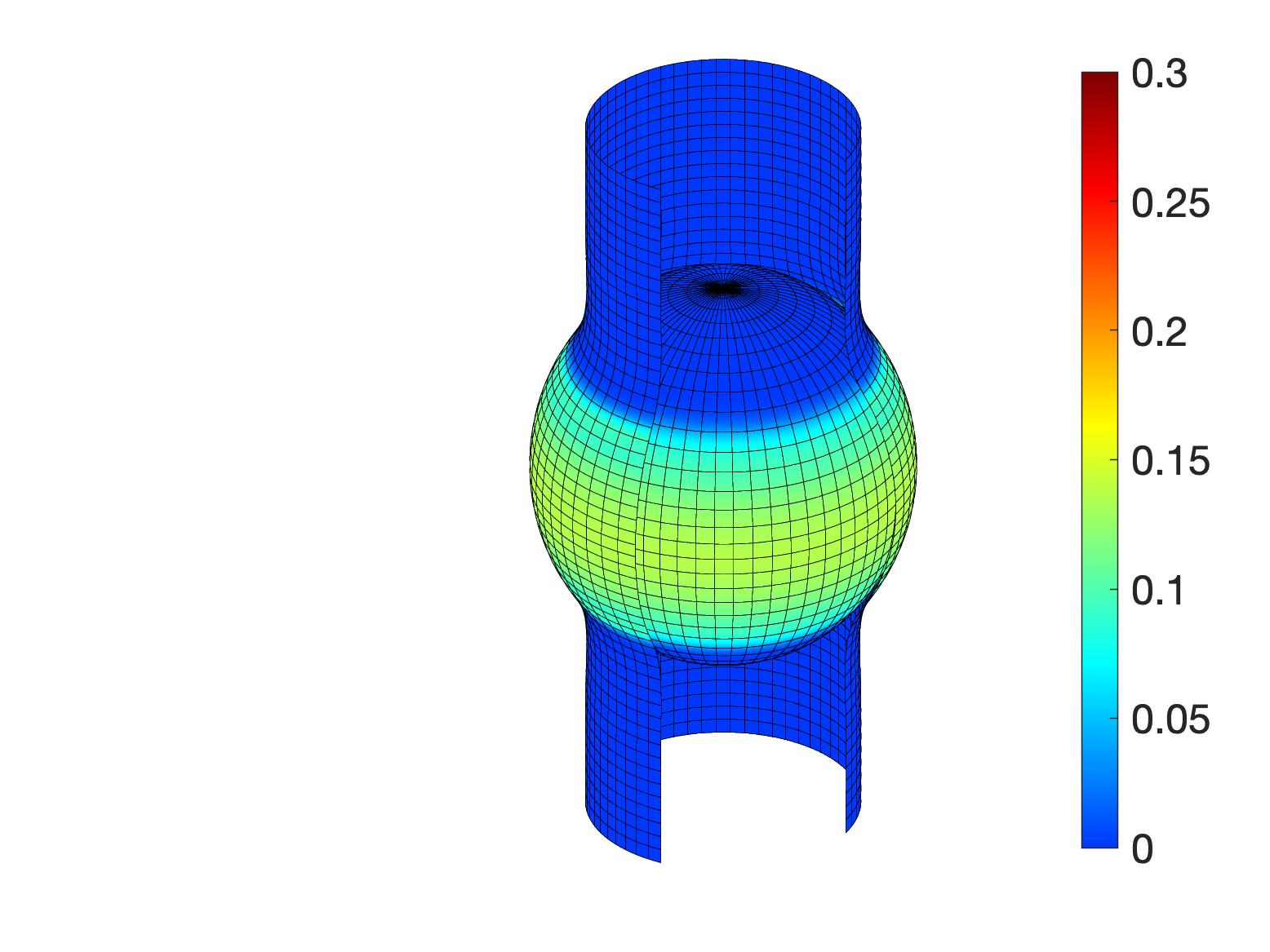}};
		\node[inner sep=0pt] (a) at (4.1,0) {\includegraphics[trim={46cm 4cm 1.5cm 2cm},clip,scale=0.15]{fig/angi/Defo_Angio_Koiter_mc5_bc0_pc}};
		\node[inner sep=0pt] (a) at (-9,-6) {\includegraphics[trim={22cm 4cm 15cm 2cm},clip,scale=0.15]{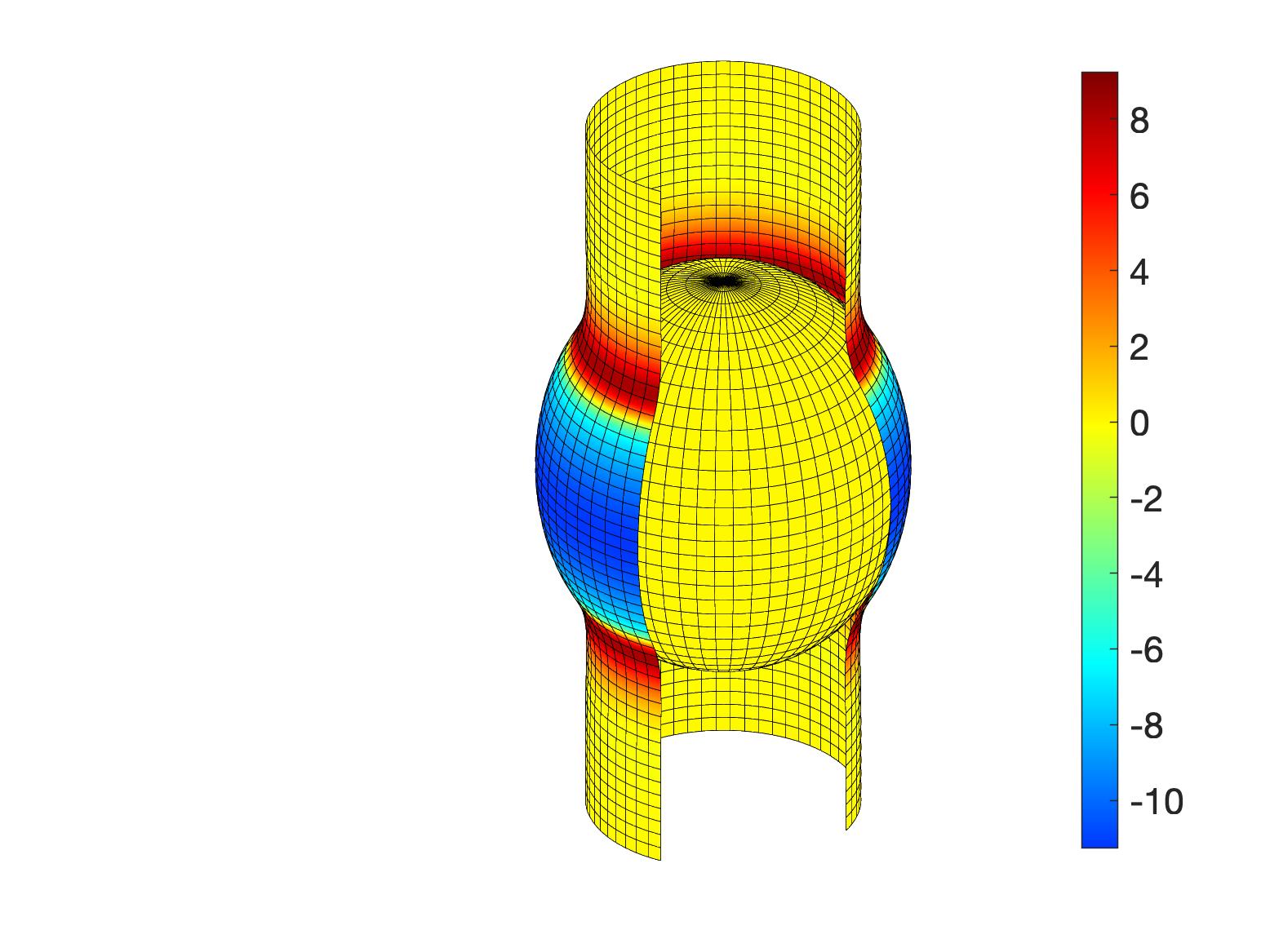}};
		\node[inner sep=0pt] (a) at (-6,-6) {\includegraphics[trim={22cm 4cm 15cm 2cm},clip,scale=0.15]{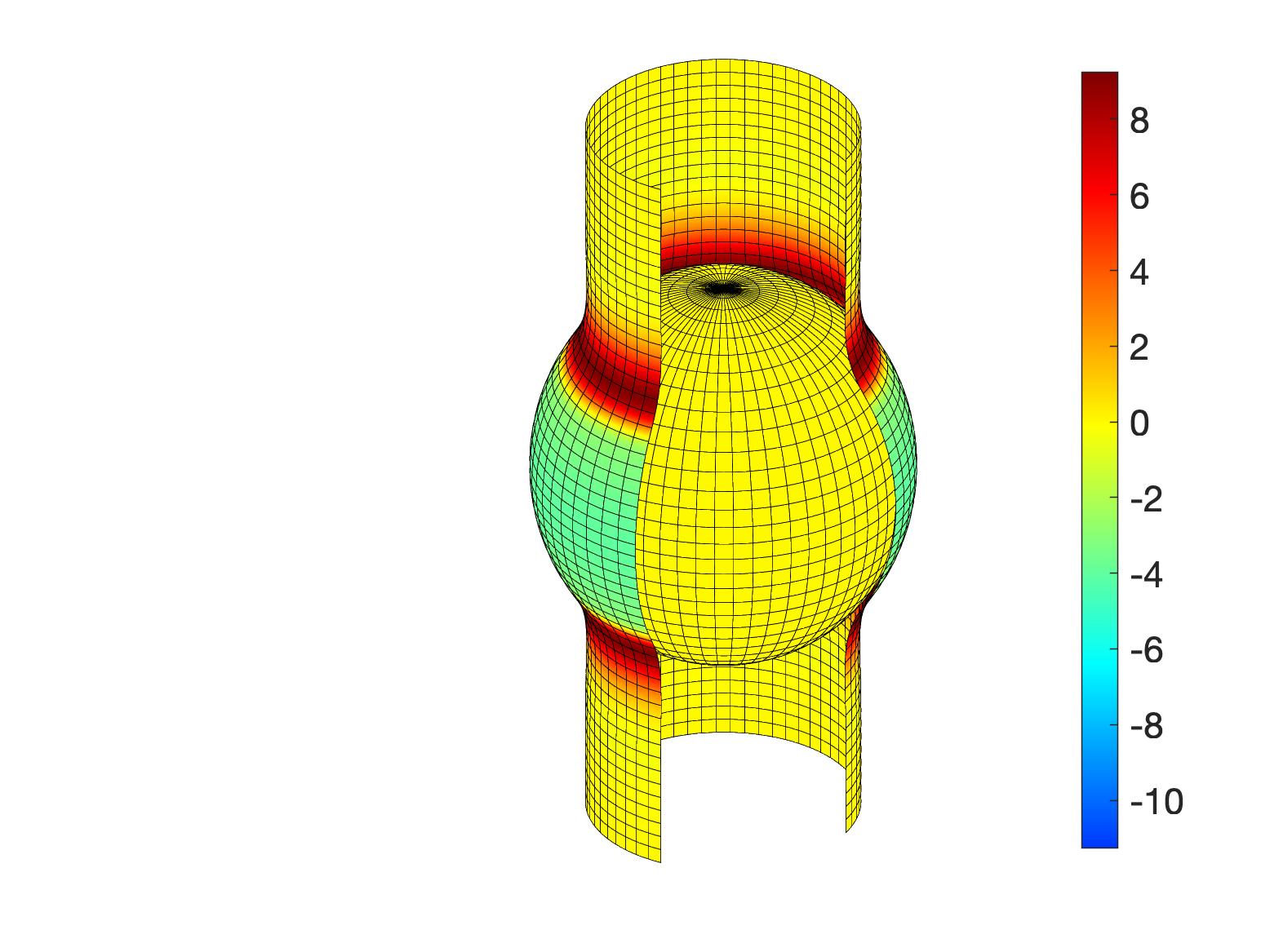}};
		\node[inner sep=0pt] (a) at (-4,-6) {\includegraphics[trim={46cm 4cm 1.5cm 2cm},clip,scale=0.15]{fig/angi/Defo_Angio_Koiter_mc5_bc0_M1_1}};
		\node[inner sep=0pt] (a) at (-0.75,-6) {\includegraphics[trim={22cm 4cm 15cm 2cm},clip,scale=0.15]{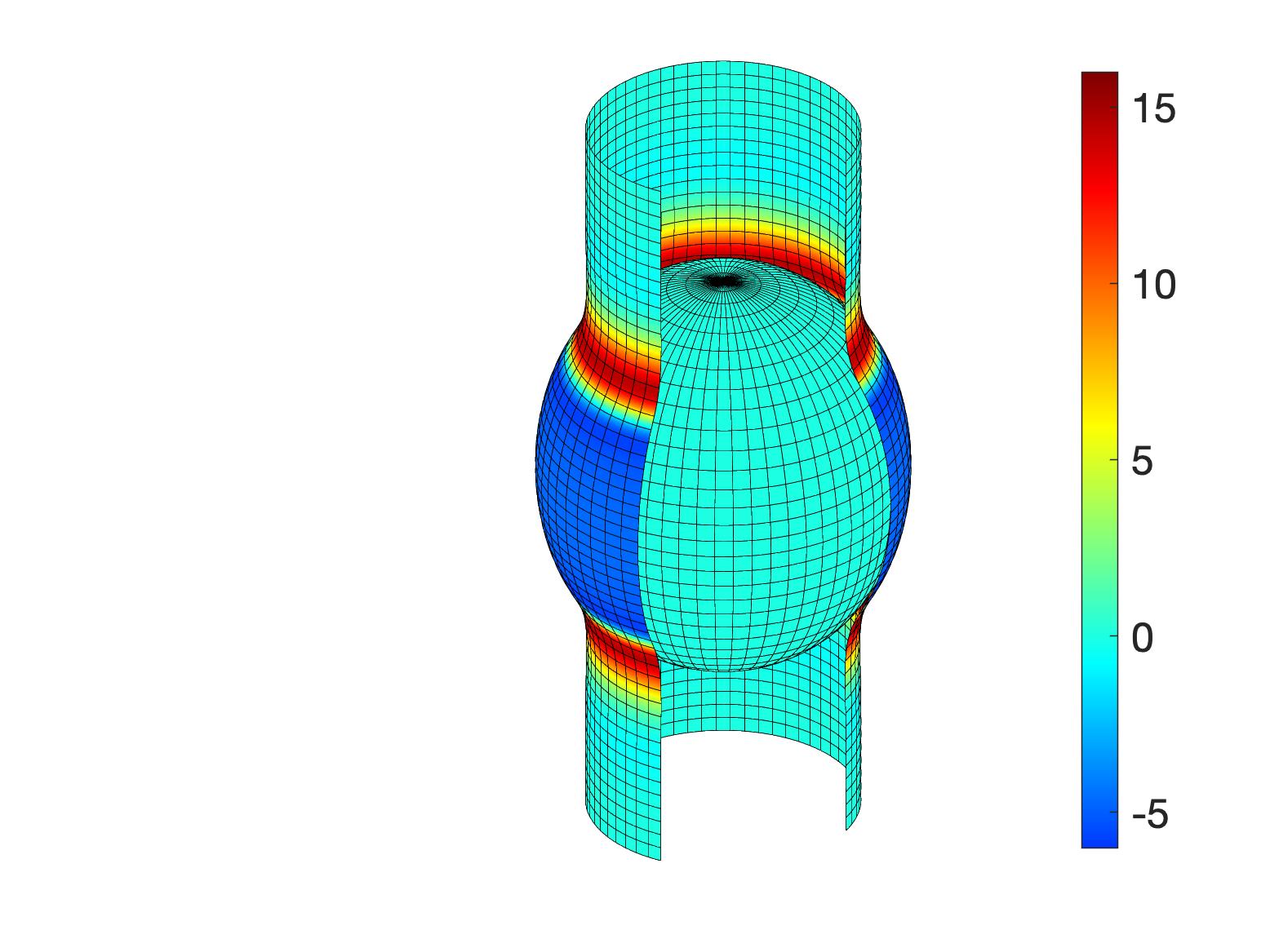}};
		\node[inner sep=0pt] (a) at (2.1,-6) {\includegraphics[trim={22cm 4cm 15cm 2cm},clip,scale=0.15]{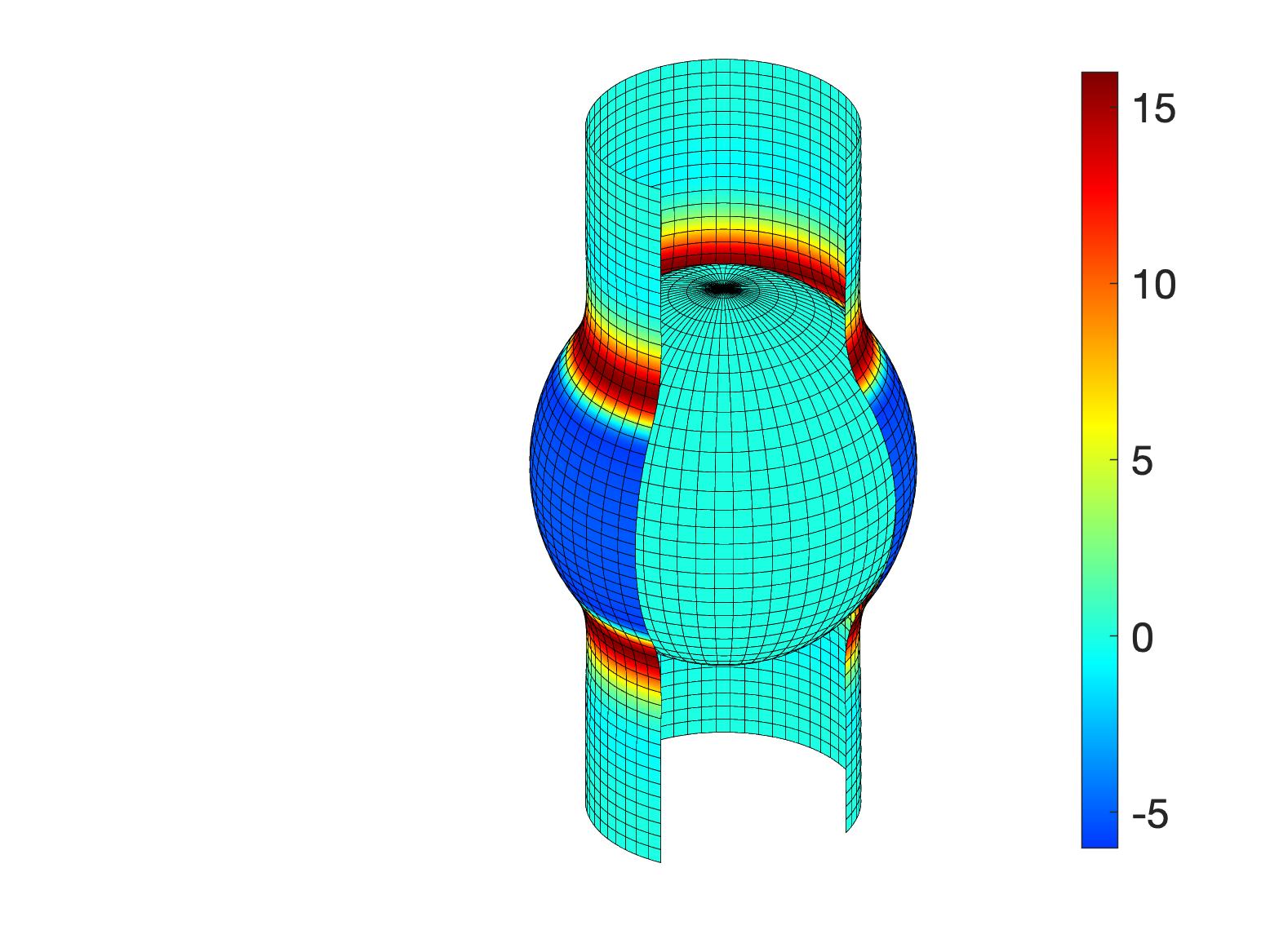}};
		\node[inner sep=0pt] (a) at (4.1,-6) {\includegraphics[trim={46cm 4cm 1.5cm 2cm},clip,scale=0.15]{fig/angi/Defo_Angio_Koiter_mc5_bc0_M2_2}};
		\node[inner sep=0pt, rotate=90] (a) at (-3.5,-0) {$\lambda_1$};
		\node[inner sep=0pt, rotate=90] (a) at (4.8,-0) {$p_{\mathrm{c}}$ [$\mathrm{kPa}$]};
		\node[inner sep=0pt, rotate=90] (a) at (-3.5,-6) {$M^1_1$ [$ \mu\mathrm{N} $]};
		\node[inner sep=0pt, rotate=90] (a) at (4.8,-6) {$M^2_2$ [$ \mu\mathrm{N} $]};
		\node[inner sep=0pt] (a) at (-7.45,-2.75) {(a)};
		\node[inner sep=0pt] (a) at (0.8,-2.75) {(b)};
		\node[inner sep=0pt] (a) at (-7.45,-8.6) {(c)};
		\node[inner sep=0pt] (a) at (0.8,-8.6) {(d)};
	\end{tikzpicture}
	\caption{Angioplasty simulation: Distribution of (a) Circumferential stretch $\lambda_{1} $, (b) distribution of contact pressure $ p_{\mathrm{c}} $ and, bending stresses (c) $ M^{1}_1 $ and (d) $ M^{2}_2 $ at balloon volume $ V \,= \,3V_0 $. In all cases the left figures belong to the Koiter model while the right figures belong to the new bending model. Significant difference between both models appear for $ p_{\mathrm{c}} $ and $ M^1_1 $, but not for $ \lambda_1 $ and $ M^2_2 $}
	\label{fig:angi}
\end{figure}
\begin{figure}[H]
	\centering
	\begin{tikzpicture}
		\node[inner sep=0pt] (a) at (-7.5,-5.2) {\includegraphics[trim={0cm 0cm 0cm 0cm},clip,scale=0.35]{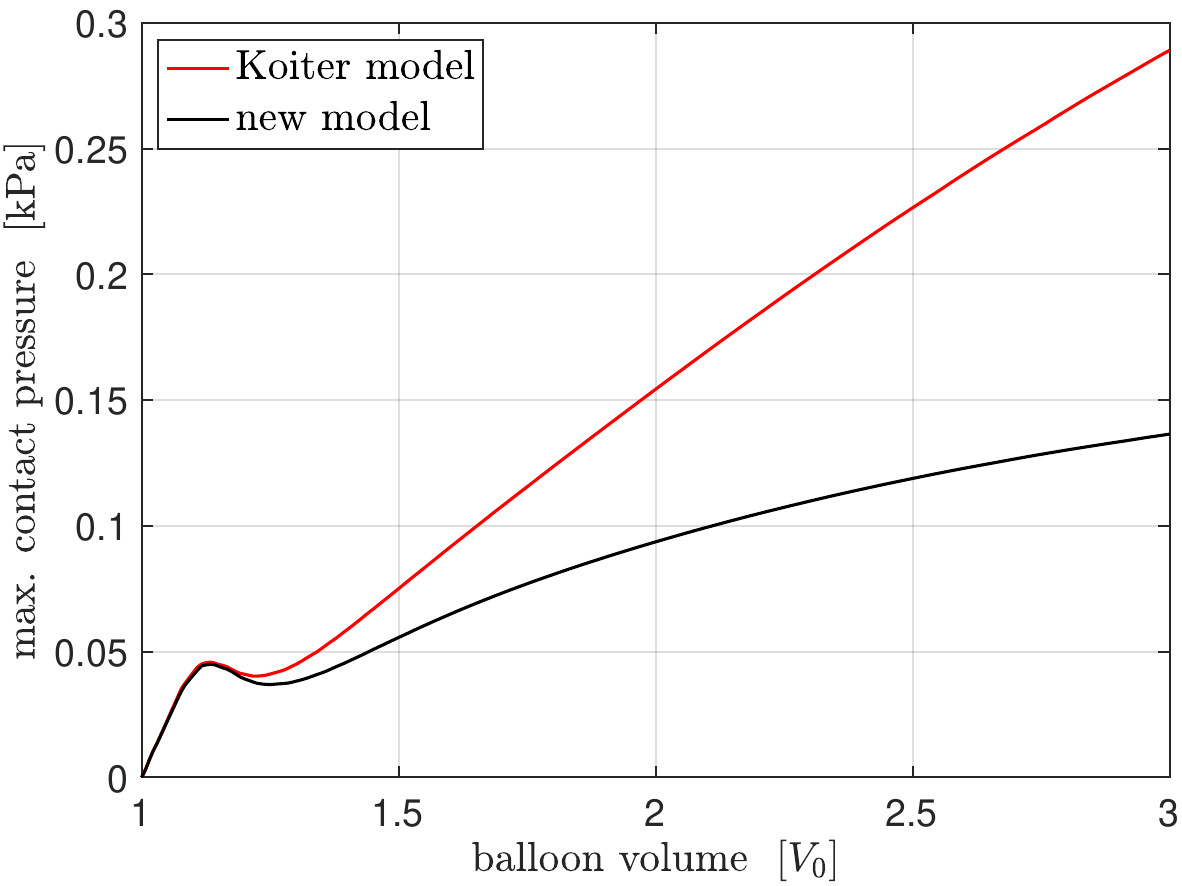}};
		\node[inner sep=0pt] (a) at (0,-5.2) {\includegraphics[trim={0cm 0cm 0cm 0cm},clip,scale=0.35]{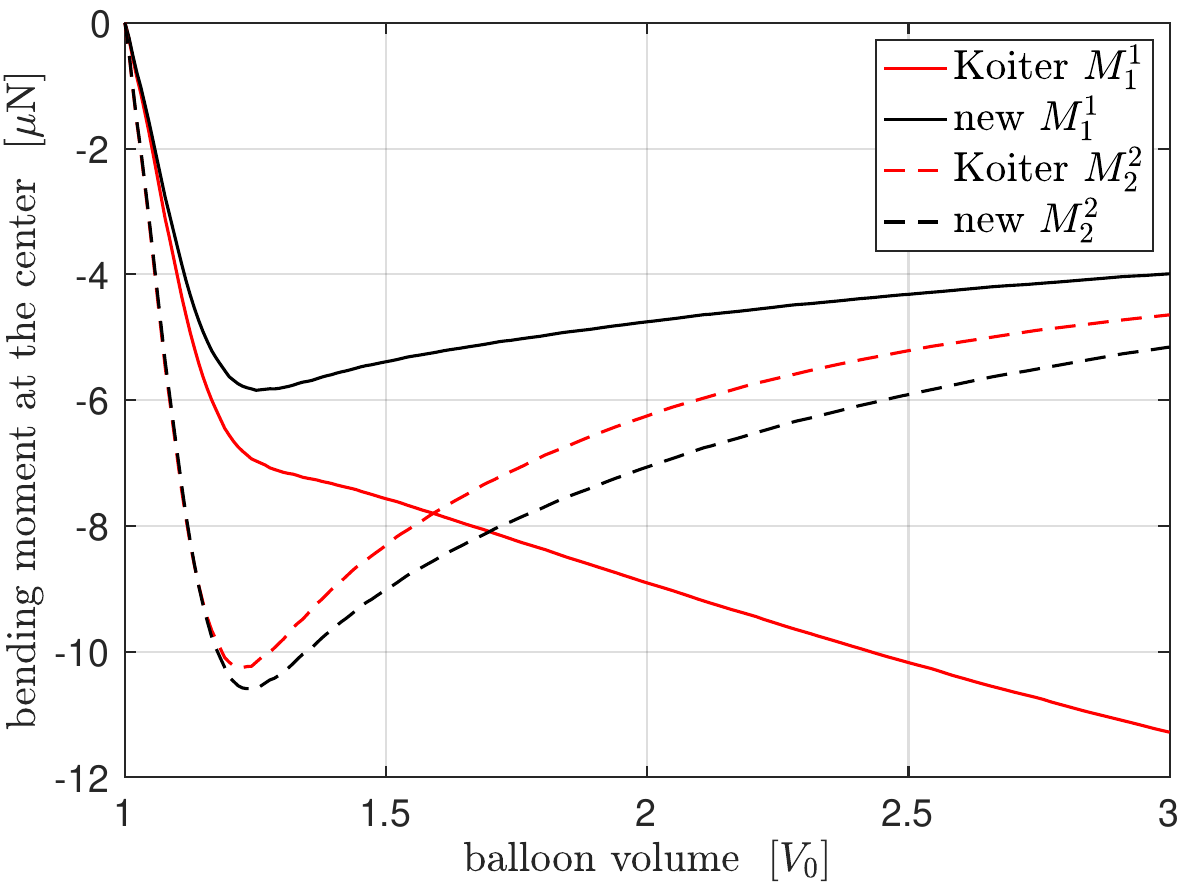}};
		\node[inner sep=0pt] (a) at (-7.2,-8) {\small (a)};
		\node[inner sep=0pt] (a) at (0.4,-8) {\small (b)};
	\end{tikzpicture}
	\caption{Angioplasty simulation: Change in (a) contact pressure ($ p_{\mathrm{c}} $) and (b) bending stresses ($ M^{1}_{1} $ and $ M^{2}_2 $) measured at the center of the tube as a function of increasing balloon volume.}
	\label{fig:angi_res}
\end{figure}	

Fig.~\ref{fig:angi} shows the deformed configurations of the Koiter (left) and the new (right) bending models colored by different quantities. In Fig.~\ref{fig:angi}a, the deformed configurations are colored by the circumferential stretch $ \lambda_{1} $. Though this distribution looks similar, the balloon has slightly higher longitudinal deformation for the Koiter bending model than for the new model, which indicates that the Koiter model is stiffer. This is confirmed by the larger contact pressure $ p_{\mathrm{c}} $ and circumferential bending stress component $ M^1_1 $ of the Koiter model shown in Fig.~\ref{fig:angi}b and c, respectively. The longitudinal bending stress component $ M^2_2 $, on the other hand, is again similar for both models as Fig.~\ref{fig:angi}d shows. The large differences between the contact pressure $ p_{\mathrm{c}} $ and bending stress component $ M^1_1 $ can also be seen in Fig.~\ref{fig:angi_res}. At $ V \,= \,3V_0 $, $ p_{\mathrm{c}} $ is almost twice as large and $ M^1_1 $ almost three times as large for the Koiter than for then new bending model. Up to $ V \,\approx \,1.2 V_0 $, however, the differences are small, illustrating once more the equivalency of the two models for small deformations.

Contrary to the previous examples, the angioplasty example thus shows that there are major differences between the Koiter bending model and the proposed new bending model. Those are due to the circumferential stretch in the tube caused by the expanding balloon. In the Koiter model this stretch contributes to the bending stress and stiffness, while it does not in the proposed new model. The effect of bending and stretching can therefore be properly separated in the new model. This was also seen in test case~2 (Sec.~\ref{subsec:t2}). Since the Koiter bending model reacts to stretches it can be expected to overestimate the bending stresses.
%%%%%%%%%%%%%%%%%%%%%%%%%%%%%%%%%%%%%%%%%%%%%%%%%%%%%%%%%%%%%%%%%%%%%%%%%%%%%%%%%%%%%%%%%%
\section{Conclusion}\label{sec:6}

This work proposes a new bending model for Kirchhoff-Love shells based on the direct surface approach. The proposed bending model eliminates the spurious influence of membrane strains on bending at large deformation. It is objective, can handle initially curved surfaces and large deformations.

The new bending model is introduced in Sec.~\ref{sec:3} along with several existing bending models. The relation between the material parameters of the different bending models is provided, and it is shown that all the bending models become equivalent for initially planar shells at small deformations. In contrast to existing bending models, the proposed new model passes an essential set of five elementary bending test cases as was seen in Sec.~\ref{sec:4}. Test case~1(a) (Sec.~\ref{subsec:t1}) shows that the Canham and Helfrich models are not initially stress-free. The Helfrich model will be initially stress-free only when one of its bending parameter ($ \bar{k} $) is zero. Test case~1(b) considers a final configuration similar to that of a rigid rotation but obtained by counter bending. This test case highlights the anisotropic nature of the new bending model in addition to providing accurate stresses. The Koiter and apH models only partially pass this test due to errors in the Cauchy stresses. The Helfrich model fails due to incorrect bending stresses and the Canham model fails due to incorrect Cauchy as well as bending stresses. In test case~2(a) (Sec \ref{subsec:t2}), a cylinder is inflated, which is a pure membrane action, and hence stresses shouldn't arise from the bending models. The Koiter and apH models partially pass this test with small errors in the bending stresses. The Canham and Helfrich models fail again due to incorrect bending and Cauchy stresses. In test case~2(b), pure bending is considered and all the bending models except the proposed model fail to give zero Cauchy stresses. In test case~3 (Sec \ref{subsec:t3}), torsion is considered for which only the Koiter and new bending model are able to provide accurate bending stresses to pass the test.

These observations are confirmed by the six numerical test cases in Sec.~\ref{sec:5}. In the simply supported linear plate problem (Sec.~\ref{subsec:5.1}), all the bending models considered are identical. This problem is also solved with skew meshes to exhibit that the proposed model gives accurate results also when the principal curvature directions are not aligned with the curvilinear coordinates. The Helfrich model fails in the second linear problem -- the pinched cylindrical shell (Sec.~\ref{subsec:5.2}) -- due to non-zero initial bending stresses. The Helfrich model also deviates from the reference solution for cylinder spreading (Sec.~\ref{subsec:5.4}) for the same reason. Though the Helfrich model gives an accurate force vs. displacement curve for the nonlinear pinching problem (Sec.~\ref{subsec:5.3}), its bending stress components ($ M^{\alpha}_{\beta} $) are different from those obtained for the Koiter and the new bending model. Finally in the angioplasty example, the difference between the Koiter and the new model is illustrated. The example shows that bending and stretching can be properly separated in the new model, which is not the case for the Koiter model. The latter thus shows much larger bending stresses.

The proposed new model is able to circumvent problems in existing bending models by using stretch-invariant bending quantities. It serves as a bending model with an union of desired features for shell models in the direct surface approach. As the model is based on the principal curvature directions and has four different material parameters, it provides flexibility in modeling new materials. In future, it would be interesting to combine the new bending model with anisotropic membrane models based on the Mooney–Rivlin \citep{mooney1940theory,Rivlin} and Gasser–Ogden–Holzapfel model for biological tissues \citep{Gasser}. 
\bigskip

{\Large{\bf Acknowledgements}}\\[2mm]
The authors are grateful to the Deutsche Forschungsgemeinschaft (DFG, German Research Foundation) – 333849990/GRK2379 (IRTG Modern Inverse Problems) for supporting this research. The authors also wish to thank Farshad Roohbakhshan for his help with the numerical examples.

%%%%%%%%%%%%%%%%%%%%%%%%%%%%%%%%%%%%%%%%%%%%%%%%%%%%%%%%%%%%%%%%%%%%%%%%%%%%%%%%%%%%%%%%%%
\appendix
\section{Material tangents of the proposed model}\label{ap:tan}

For the proposed model, the fourth order material tensor components defined in Eq.~\eqref{eq:1.31} are
\begin{align}
\cabgd\,&=\,\ell_{11}^{\alpha\beta}\,\ell_{11}^{\gamma\delta}\left[c_1\,\kappa_{1}\,\left(4\,\kappa_{1}\,-\,\dfrac{3\,\kappa_{01}}{\lambda_{1}}\right)\,+\,\dfrac{3\,c_{12}\,\kappa_{1}}{\lambda_1}\,k_2\,+\,c_3\,\left(\dfrac{3\,\kappa_{12}^2\,\lambda_{2}}{2\,\lambda_{1}}\,-\,\dfrac{5\,\kappa_{012}\,\kappa_{12}\,\sqrt{\lambda_{2}}}{4\,\lambda_{1}^{3/2}}\right)\right]\nonumber\\&+\,\ell^{\alpha\beta}_{22}\,\ell^{\gamma\delta}_{22}\,\left[c_2\,\kappa_{2}\,\left(4\,\kappa_{2}\,-\,\dfrac{3\,\kappa_{02}}{\lambda_{2}}\right)\,+\,\dfrac{3\,c_{12}\,\kappa_{2}}{\lambda_2}\,k_1\,+\,c_3\,\left(\dfrac{3\,\kappa_{12}^2\,\lambda_{1}}{2\,\lambda_{2}}\,-\,\dfrac{5\,\kappa_{012}\,\kappa_{12}\,\sqrt{\lambda_{1}}}{4\,\lambda_{2}^{3/2}}\right)\right]\nonumber\\&+\left(\ell^{\alpha\beta}_{11}\,\ell^{\gamma\delta}_{22}\, +\, \ell^{\alpha\beta}_{22}\,\ell^{\gamma\delta}_{11}\right)\left[c_{12}\,\kappa_{1}\,\kappa_{2}\,+\,c_3\left(\dfrac{\kappa_{12}^2}{2}\,-\,\dfrac{\kappa_{012}\,\kappa_{12}}{4\sqrt{\lambda_{1}\lambda_{2}}}\right)\right]\,,\\
\dabgd\,&=\,\ell^{\alpha\beta}_{11}\,\ell^{\gamma\delta}_{11}\,\left[c_1\,\left(\dfrac{\kappa_{01}}{\lambda_{1}}\,-\,\kappa_{1}\right)\,+\,\dfrac{c_{12}}{\lambda_{1}}\left(\kappa_{02}\,-\,\lambda_{2}\,\kappa_{2}\right)\right]\,-\,c_{12}\,\ell^{\alpha\beta}_{11}\,\ell^{\gamma\delta}_{22}\,\kappa_{1}\nonumber\\&+\,\ell^{\alpha\beta}_{22}\,\ell^{\gamma\delta}_{22}\,\left[c_2\,\left(\dfrac{\kappa_{02}}{\lambda_{2}}\,-\,\kappa_{2}\right)\,+\,\dfrac{c_{12}}{\lambda_{2}}\left(\kappa_{01}\,-\,\lambda_{1}\,\kappa_{1}\right)\right]\,-\,c_{12}\,\ell^{\alpha\beta}_{22}\,\ell^{\gamma\delta}_{11}\,\kappa_{2}\nonumber\\&+\,\dfrac{c_3}{2\sqrt{\lambda_{1}\lambda_{2}}}\,\left(\dfrac{\kappa_{012}}{\sqrt{\lambda_{1}\lambda_{2}}}\,-\,2\,\kappa_{12}\right)\left(\lambda_{2}\,\ell^{\alpha\beta}_{11}\,+\,\lambda_{1}\,\ell^{\alpha\beta}_{22}\right)\,\ell^{\gamma\delta}_{12}\,,\\
\eabgd\,&=\,d^{\gamma\delta\alpha\beta},\\
\fabgd\,&=\,\ell^{\alpha\beta}_{11}\,\ell^{\gamma\delta}_{11}\,c_1\,+\,\ell^{\alpha\beta}_{22}\,\ell^{\gamma\delta}_{22}\,c_2\,+\,\left(\ell^{\alpha\beta}_{22}\,\ell^{\gamma\delta}_{11}\,+\ell^{\alpha\beta}_{11}\,\ell^{\gamma\delta}_{22}\,\right)\,c_{12}\,+\,\ell^{\alpha\beta}_{12}\,\ell^{\gamma\delta}_{12}\,c_{3}\,.
\end{align}
The tensor components corresponding to the bending models of Koiter, projected Neo-Hooke, and Canham and Helfrich can be found in \cite{duong2017new}, \cite{roohbakhshan2017efficient} and \cite{sauer2017theoretical}, respectively.

%%%%%%%%%%%%%%%%%%%%%%%%%%%%%%%%%%%%%%%%%%%%%%%%%%%%%%%%%%%%%%%%%%%%%%%%%%%%%%%%%%%%%%%%%%
\section{Efficient FE implementation}\label{ap:fe}

If the initial principal curvature directions align with surface tangents ($ \bA_{\alpha} $), then the quantity defined in Eq.~\eqref{eq:auxK} will be 
\begin{align}
L^{\alpha}_{i}\,=\,\frac{\delta_{i}^{\alpha}}{\sqrt{A_{ii}}}\,,\quad (\text{no summation on}\,\, i)\,.
\end{align}
The stretches along those direction will then simply be,
\begin{align}
\lambda_{1}\,=\,\sqrt{\frac{a_{11}}{A_{11}}}\,,\quad\lambda_{2}\,=\,\sqrt{\frac{a_{22}}{A_{22}}}\,.
\end{align}
Plugging this into the quantities of Eq.~\eqref{eq:auxL} leads to 
\begin{gather}
[l^{\alpha\beta}_{11}]\,=\,\dfrac{1}{\mathcal{A}_1}\,\left[ {\begin{array}{cc}
	1 & 0 \\
	0 & 0 \\
	\end{array} }\right]\,,\nonumber
\quad[l^{\alpha\beta}_{22}]\,=\,\dfrac{1}{\mathcal{A}_2}\,\left[ {\begin{array}{cc}
	0 & 0 \\
	0 & 1 \\
	\end{array} }\right],\nonumber\\
\hspace*{-10mm}[l^{\alpha\beta}_{12}]\,=\,\dfrac{1}{\sqrt{\mathcal{A}_1\,\mathcal{A}_2}}\,\left[ {\begin{array}{cc}
	0 & 1 \\
	1 & 0 \\
	\end{array} }\right]\,, \label{eq:sim}
\end{gather}
where
\begin{align}
	\mathcal{A}_1\,:=\,\sqrt{A_{11}\,a_{11}}\,,\quad\mathcal{A}_2\,:=\,\sqrt{A_{22}\,a_{22}}\,.
\end{align}
Using this simplification, the membrane and bending stresses can be directly calculated as
\begin{gather}
\tau^{11}\,=-\left(c_1\,k_1\,+\,c_{12}k_2\,+\,\dfrac{c_3\,\sqrt{\lambda_{2}}\,\kappa_{12}\,k_{12}}{2\,\sqrt{\lambda_{1}}\,\kappa_1}\right)\,\dfrac{\kappa_{1}}{\mathcal{A}_1}\,, \\
\tau^{22}\,=-\left(c_{12}\,k_1\,+\,c_{2}k_2\,+\,\dfrac{c_3\,\sqrt{\lambda_{1}}\,\kappa_{12}\,k_{12}}{2\,\sqrt{\lambda_{2}}\,\kappa_2}\right)\,\dfrac{\kappa_{2}}{\mathcal{A}_2}\,,\\
\tau^{12}\, =\, \tau^{21}\,=\,0\,,\\
M^{11}_{0}\,=\,\left(c_1\,k_1\,+\,c_{12}\,k_{2}\right)\dfrac{1}{\mathcal{A}_1}\,, \\
M^{22}_{0}\,=\,\left(c_{12}\,k_1\,+\,c_{2}\,k_{2}\right)\dfrac{1}{\mathcal{A}_2}\,,\\ M^{12}_{0}\,=\,M^{21}_{0}\,=\,\dfrac{c_3\,k_{12}}{\sqrt{\mathcal{A}_1\,\mathcal{A}_2}}\,.
\end{gather}
%The only non zero auxiliary terms will be
%\begin{align}
%l_{1111}^{1111}\,=\,\frac{1}{(a_{11})^2},\quad	l_{2222}^{2222}\,=\,\frac{1}{(a_{22})^2},\quad	l_{1122}^{1122}\,=\,l_{2211}^{2211}\,=\,\frac{1}{a_{11}\,a_{22}}.
%\label{eq:auxLLL}
%\end{align}   
As described in \cite{duong2017new}, for the efficient computation of the FE stiffness matrices we can exploit the symmetries to rearrange the fourth order tensor, $ f^{\alpha\beta\gamma\delta} $ as
\begin{equation}
\mathbf{F}:=\left[\begin{array}{ccc}
f^{1111} & f^{1122} & f^{1112} \\
f^{2211} & f^{2222} & f^{2212} \\
f^{1211} & f^{1222} & f^{1212}
\end{array}\right]\,.
\end{equation} 
Based on the simplification in Eq~\eqref{eq:sim}, this rearrangement simplifies to
\begin{equation}
\mathbf{F}:=\left[\begin{array}{ccc}
\dfrac{c_{1}}{(\mathcal{A}_1)^2} & \dfrac{c_{12}}{\mathcal{A}_1\,\mathcal{A}_2} & 0 \\[1em]
\dfrac{c_{12}}{\mathcal{A}_1\,\mathcal{A}_2} & \dfrac{c_{2}}{(\mathcal{A}_2)^2} & 0 \\[1em]
0 & 0 & \dfrac{c_{3}}{\mathcal{A}_1\,\mathcal{A}_2}\vspace*{1mm}
\end{array}\right]\,.
\end{equation} 
Similar rearrangement can be applied to $ \Cabgd $, $ \Dabgd $ and $ \Eabgd $. Further, we define the auxiliary terms
\vspace*{-5mm}
\begin{align}
\mathbf{L}_{\alpha \beta}^{a} &:=\mathbf{N}_{, \alpha}^{\mathrm{T}} \boldsymbol{a}_{\beta}\,,\nonumber \\
\mathbf{L}_{\alpha}^{n} &:=\mathbf{N}_{, \alpha}^{\mathrm{T}} \boldsymbol{n}\,,\nonumber \\
\mathbf{G}_{\alpha \beta}^{n} &:=\tilde{\mathbf{N}}_{; \alpha \beta}^{\mathrm{T}} \boldsymbol{n}\,,
\end{align}
where each term is an array of size $ (3n\times1) $, with $ n $ being the number of control points per element. This is further reorganized as  
\begin{align}
\hat{\mathbf{L}}_{a}&\,=\,\left[\mathbf{L}_{11}^{a}\,,\, \mathbf{L}_{22}^{a}\,,\, \mathbf{L}_{12}^{a}\,+\,\mathbf{L}_{21}^{a}\right]\,,\\
\hat{\mathbf{G}}_{n}&\,=\,\left[\mathbf{G}_{11}^{n}\,,\, \mathbf{G}_{22}^{n}\,,\, \mathbf{G}_{12}^{n}\,+\,\mathbf{G}_{21}^{n}\right]\,.
\end{align}
We can then rewrite the equations for force,
\begin{align}
\mathbf{f}_{\mathrm{int} \tau}^{e}&\,=\,\int_{\Omega_{0}^{e}} \,\hat{\mathbf{L}}_{a}\, \hat{\boldsymbol{\tau}}\, \mathrm{d} A\,, \\
\mathbf{f}_{\mathrm{int} M}^{e}\,&\,=\,\int_{\Omega_{0}^{e}} \,\hat{\mathbf{G}}_{n} \,\hat{\mathbf{M}}_{0} \,\mathrm{~d} A\,,
\end{align}
material stiffness,
\begin{align}
\mathbf{k}_{\tau \tau}^{e}&\,=\,\int_{\Omega_{0}^{e}}\, \hat{\mathbf{L}}_{a}\, \mathbf{C}\, \hat{\mathbf{L}}_{a}^{\mathrm{T}}\, \mathrm{d} A\,,\\
\mathbf{k}_{\tau M}^{e}&\,=\,\int_{\Omega_{0}^{e}}\, \hat{\mathbf{L}}_{a}\, \mathbf{D}\, \hat{\mathbf{G}}_{n}^{\mathrm{T}}\, \mathrm{d} A\,,\\
\mathbf{k}_{M \tau}^{e}&\,=\,\int_{\Omega_{0}^{e}}\, \hat{\mathbf{G}}_{n}\, \mathbf{E}\, \hat{\mathbf{L}}_{a}^{\mathrm{T}}\, \mathrm{d} A\,,\\
\mathbf{k}_{M M}^{e}&\,=\,\int_{\Omega_{0}^{e}}\, \hat{\mathbf{G}}_{n}\, \mathbf{F}\, \hat{\mathbf{G}}_{n}^{\mathrm{T}}\, \mathrm{d} A\,,
\end{align}
and geometric stiffness,
\begin{align}
\mathbf{k}_{\mathrm{M} 1}^{e}&\,=\,-\int_{\Omega_{0}^{e}}\, b_{M}\,\left(a^{11}\, \mathbf{L}_{1}^{n}\, \mathbf{L}_{1}^{n \mathrm{T}}\,+\,a^{22}\, \mathbf{L}_{2}^{n}\, \mathbf{L}_{2}^{n \mathrm{T}}\,+\, a^{12}\,\left( \mathbf{L}_{1}^{n}\, \mathbf{L}_{2}^{n\mathrm{T}}\,+\,\mathbf{L}_{2}^{n}\, \mathbf{L}_{1}^{n\mathrm{T}}\right)\right)\, \mathrm{d} A\,,\\
\mathbf{k}_{\mathrm{M} 2}^{e}&\,=\,-\int_{\Omega_{0}^{e}}\,\left(\mathbf{L}_{1}^{n}\, \boldsymbol{a}^{1 \mathrm{T}}\,+\,\mathbf{L}_{2}^{n}\, \boldsymbol{a}^{2 \mathrm{T}}\right)\,\left(M_{0}^{11}\, \tilde{\mathbf{N}}_{; 11}\,+\,M_{0}^{22}\, \tilde{\mathbf{N}}_{; 22}\,+\,2\, M_{0}^{12}\, \tilde{\mathbf{N}}_{; 12}\right)\, \mathrm{d} A\,.
\end{align}
Where
\begin{equation}
\begin{aligned}
\hat{\boldsymbol{\tau}} &:=\left[\tau^{11}, \tau^{22}, \tau^{12}\right]^{T}\,, \\
\hat{\mM}_{0} &:=\left[M_{0}^{11}, M_{0}^{22}, M_{0}^{12}\right]^{T}\,, \\
\hat{\mathbf{b}} &:=\left[b_{11}, b_{22}, 2 b_{12}\right]^{T}\,,\\
b_{M}&:=\hat{\mathbf{b}}^{\mathrm{T}} \hat{\mathbf{M}}_{0}\,.
\end{aligned}
\end{equation}

%%%%%%%%%%%%%%%%%%%%%%%%%%%%%%%%%%%%%%%%%%%%%%%%%%%%%%%%%%%%%%%%%%%%%%%%%%%%%%%%%%%%%%%%%%
\section{Bending moduli extraction}\label{ap:2}

In order to calculate the bending moduli of the new bending model we directly compare the values of $ f^{\alpha\beta\gamma\delta} $ in \eqref{eq:1.42} for the Koiter model and the new bending model. For the Koiter model, the fourth order tensor is given by
\begin{align}
f^{\alpha\beta\gamma\delta}_{\textrm{Koi}}\,=\,\dfrac{T^2}{12}\,\left(\Lambda\,\Aab\,\Agd\,+\,\mu\,\left(A^{\alpha\gamma}\,A^{\beta\delta}\,+\,A^{\alpha\delta}\,A^{\beta\gamma}\right)\right)\,.
\end{align}
Choosing a parametrization with $ A_{12}\,=\,A_{21}\,=\,0 $ and noting that $ f^{\alpha\beta\gamma\delta} $ has major and minor symmetries, the tensor can be represented only by the following elements
\begin{align}
&f^{1111}_{\textrm{Koi}}\,=\,\dfrac{T^2\,\left(\Lambda\,+\,2\,\mu\right)}{12\,\norm{\bA_1}^4}\,,&&f^{2211}_{\textrm{Koi}}\,=\,\dfrac{T^2\,\Lambda}{12\,\left(\norm{\bA_1}^2\,\norm{\bA_2}^2\right)}\,,&&f^{1211}_{\textrm{Koi}}\,=\,0\,,\nonumber\\
&f^{1122}_{\textrm{Koi}}\,=\,\dfrac{T^2\,\Lambda}{12\,\left(\norm{\bA_1}^2\,\norm{\bA_2}^2\right)}\,,&&f^{2222}_{\textrm{Koi}}\,=\,\dfrac{T^2\,\left(\Lambda\,+\,2\,\mu\right)}{12\,\norm{\bA_1}^4}\,,&&f^{1222}_{\textrm{Koi}}\,=\,0\,,\label{eq:fkoi}\\
&f^{1112}_{\textrm{Koi}}\,=\,0\,,&&f^{2212}_{\textrm{Koi}}\,=\,0\,,&&f^{1212}_{\textrm{Koi}}\,=\,\dfrac{T^2\,\mu}{12\,\left(\norm{\bA_1}^2\,\norm{\bA_2}^2\right)}\nonumber\,.
\end{align}
The corresponding fourth order tensor components for the new model are of the form
\begin{align}
f^{\alpha\beta\gamma\delta}_{\mathrm{new}}\,=&\,\sum_{i = 1}^{2}\,\dfrac{c_i}{\lambda_{i}^2}L^{\alpha}_i\,L^{\beta}_i\,L^{\gamma}_i\,L^{\delta}_i\,+\,\dfrac{c_{12}}{\lambda_{1}\,\lambda_{2}}\,\left(L^{\alpha}_1\,L^{\beta}_1\,L^{\gamma}_2\,L^{\delta}_2\,+\,L^{\alpha}_2\,L^{\beta}_2\,L^{\gamma}_1\,L^{\delta}_1\right)\nonumber\\
&+\,\dfrac{c_3}{\lambda_{1}\,\lambda_{2}}\,\left(L^{\alpha}_1\,L^{\beta}_2+\,L^{\alpha}_2\,L^{\beta}_1\right)\,\left(L^{\gamma}_1\,L^{\delta}_2+\,L^{\gamma}_2\,L^{\delta}_1\right)\,.
\end{align}
The way $ L_i^{\alpha} $ is calculated in the numerical examples is by using
\begin{gather}
L_i^{\alpha}\,=\,\dfrac{\bA_i}{\norm{\bA_i}}\,\cdot\,\bA^{\alpha}\,,
\end{gather}
for a particular choice of parametrization. Thus only $ L_1^{1} $ and $ L_2^{2} $ will be non-zero. So by considering $\lambda_{i}\,=\,1$,
\begin{align}
&f^{1111}_{\textrm{new}}\,=\,\dfrac{c_1}{\norm{\bA_1}^4}\,,&&f^{2211}_{\textrm{new}}\,=\,\dfrac{c_{12}}{\norm{\bA_1}^2\,\norm{\bA_2}^2}\,,&&f^{1211}_{\textrm{new}}\,=\,0\,,\nonumber\\
&f^{1122}_{\textrm{new}}\,=\,\dfrac{c_{12}}{\norm{\bA_1}^2\,\norm{\bA_2}^2}\,,&&f^{2222}_{\textrm{new}}\,=\,\dfrac{c_{2}}{\norm{\bA_1}^4}\,,&&f^{1222}_{\textrm{new}}\,=\,0\,,\label{eq:fnew}\\
&f^{1112}_{\textrm{new}}\,=\,0\,,&&f^{2212}_{\textrm{new}}\,=\,0\,,&&f^{1212}_{\textrm{new}}\,=\,\dfrac{c_3}{\norm{\bA_1}^2\,\norm{\bA_2}^2}\nonumber\,.
\end{align}
Comparing Eqs.~\eqref{eq:fkoi} and Eqs.~\eqref{eq:fnew}, then leads to Eq.~\eqref{eq:newp}.

\bibliographystyle{apalike}
\bibliography{citations}

\end{document}